\documentclass[12pt]{article}
\usepackage{color}
\usepackage{amssymb}
\usepackage{graphicx}
\usepackage{epstopdf}
\usepackage[hang,small,bf]{caption}
\usepackage{hyperref}

\def\gsim{\raisebox{-4pt}{$\,\stackrel{\textstyle{>}}{\sim}\,$}}
\def\simlt{\stackrel{<}{{}_\sim}}

\newcommand{\zpr}{\mbox{$Z^{\prime}$}}
\newcommand{\upr}{\mbox{$U(1)^{\prime}$}}

\newcommand{\mz}{\mbox{$M_Z$}}

\newcommand{\x}{\mbox{$\times$}}

\begin{document}
\DeclareGraphicsExtensions{.jpg,.pdf,.mps,.png}

\begin{flushright}
\baselineskip=12pt
EFI-09-10, MIFP-09-20 \\
\end{flushright}


\begin{center}
\vglue 0.5 cm

{\Large\bf Electroweak Baryogenesis, CDM and Anomaly-free Supersymmetric
$U(1)^{\prime}$ Models } \vglue 0.6cm {\Large
Junhai Kang, Paul Langacker$^{a}$, Tianjun Li$^{b,c}$ and
Tao Liu$^{d}$} \vglue 0.4cm {
$^a$ School of Natural Science, Institute for Advanced Study,  \\
             Einstein Drive, Princeton, NJ 08540, USA\\
$^b$ Key Laboratory of Frontiers in Theoretical Physics, \\
  Institute of Theoretical Physics, 
Chinese Academy of Sciences, \\
  Beijing 100190, P. R. China \\
$^c$ George P. and Cynthia W. Mitchell Institute for
Fundamental Physics, \\ Texas A$\&$M University, College Station, TX
77843, USA \\
$^d$ Enrico Fermi Institute,
University of Chicago, \\Chicago, IL 60637, USA \\
}
\end{center}

\vglue 0.5cm


\begin{abstract}
We construct two anomaly-free supersymmetric \upr \ models with a
secluded $U(1)'$-breaking sector. For the one with $E_6$
embedding we 
show that there exists a strong enough first 
order electroweak phase transition for electroweak
baryogenesis (EWBG) because of the large soft trilinear terms 
in the Higgs potential. Unlike the Minimal
Supersymmetric Standard Model (MSSM), the stop masses can be very
heavy.  We then discuss possible $CP$ violation in the Higgs sector,
which can be both spontaneous and explicit, even at tree level. 
The spontaneous violation provides a direct source for baryogenesis, while its magnitude is mediated by an
explicit phase from the secluded sector. These new $CP$ sources do
not introduce significant new contributions to electric dipole
moments. EWBG in the thin wall ($\tau$ leptons) and thick wall regimes (top squarks, charginos and top quarks) 
are systematically discussed. We find that the $CP$-violating stop and chargino currents are very different from those obtained
in the MSSM. Due to the space-dependence of the relevant $CP$ phases, they do not require a variation of $\tan \beta$ 
in the bubble wall to have a non-trivial structure at the lowest order of Higgs insertion. 
In addition to $\tau$ leptons, top squarks and charginos, we find that top quarks can also play a significant role. 
Numerical results show that the baryon asymmetry 
is large enough to explain the cosmological observation today. We illustrate that EWBG and neutralino 
cold dark matter can be accommodated in the same framework, {\em i.e.},  there exists parameter space where a strong enough first order EWPT, 
large $CP$ phase variations across the bubble wall, a reasonable baryon asymmetry  
as well as an acceptable neutralino dark matter relic density can be achieved simultaneously.

\end{abstract}

\newpage
\baselineskip=14pt

\newpage
\tableofcontents
\newpage

\section{Introduction}

The baryon asymmetry of the Universe (BAU) has been precisely
measured by WMAP. Combining their five year results with other CMB
and large scale structure results, they obtain~\cite{Komatsu:2008hk}
\begin{eqnarray}
\Omega_b h^2 &=& 0.02265\pm 0.00059~,~\,
\end{eqnarray}
in excellent agreement with the 95\% C.L. range $0.017-0.024$ obtained from 
big bang
nucleosynthesis~\cite{Amsler:2008zzb}. This implies 
\begin{eqnarray}
{{n_B}\over {n_{\gamma}}}& ~=~( 6.21\pm 0.16) \times
10^{-10}~,~\, \\
{{n_B}\over {s} }& ~=~( 8.82\pm 0.23)\times 10 ^{-11}~,~\,
\end{eqnarray}
for the ratios
of baryon density ($n_B$)  to photon density ($n_{\gamma}$)
and entropy ($s$).

To generate the baryon asymmetry, the Sakharov criteria~\cite{Sak}
must be satisfied: (1) Baryon number ($B$) violation; (2) $C$ and
$CP$ violation; (3) a departure from thermal equilibrium (or $CPT$
violation). The first criterion is obvious. The second is required
because if $C$ and $CP$ are exact symmetries,
 no net baryon number can be produced. Moreover,
 the equilibrium average of
$B$ is zero in thermal equilibrium by $CPT$, so the third
criterion is necessary.

There are several baryogenesis scenarios: Grand Unified Theory
(GUT) baryogenesis~\cite{Yoshimura, SDLS, Weinberg},
Affleck-Dine baryogenesis~\cite{IAMD}, leptogenesis~\cite{LG86,
WBMP}, electroweak baryogenesis (EWBG)~\cite{krs,ACDKAN, VARMES},
etc. (For reviews,
 see~\cite{Trodden, ARMT, Werner, MDINE}.)
We will concentrate on EWBG, where the Sakharov criteria can be
satisfied. Baryon number violation is induced by the anomalous
sphaleron process, which violates $B+L$ but preserves $B-L$ at
 temperatures higher than the critical temperature $T_c$
of the electroweak (EW) phase transition (EWPT) ~\cite{GTH,
klma,KSHT}. In addition, $C$ is violated
 because the left-
and right-handed fermions transform differently under
$SU(2)_L\times U(1)_Y$. $CP$ can be violated in the
Standard Model (SM) and Minimal Supersymmetric Standard Model
(MSSM) and their extensions by the Yukawa interactions, $\mu$
term, and soft supersymmetry breaking terms. Non-equilibrium
processes may be obtained through the expansion of bubbles
containing the true vacuum after the EWPT.

However, to preserve the generated baryon asymmetry after the
EWPT, one needs a strong first order EWPT so that the sphaleron
effects in the bubbles of true vacuum are shut down quickly. This
requirement  can be expressed as\footnote{Some
authors impose the weaker constraint $\sqrt{2}{{v(T_c)}/ T_c}
\gsim 1$.}
\begin{eqnarray}
{{v(T_c)}\over T_c} \gsim 1~,~\,
\end{eqnarray}
where in the SM $v(T)$ is the vacuum expectation value of the
Higgs field $H^0$ at temperature $T$ and $T_c$ is the critical
temperature of the EWPT. Similarly, in the MSSM
\begin{eqnarray}
v(T) = {\sqrt {v_1(T)^2 +v_2(T)^2 }} ~,~\,
\end{eqnarray}
where $v_1(T)$ and $v_2(T)$ are respectively the VEVs of the two neutral Higgs fields $H_d^0$ and $H_u^0$. Our
normalization is such that $v(0) \sim 174$ GeV.

In the SM, the EWPT can be strongly first order only if the Higgs
mass is lighter than 40 GeV, which is inconsistent with the lower
bound of 114.4 GeV from LEP~\cite{higgs} (recently, it was noticed that this 
tension can be relaxed in the SM, with an extra singlet introduced~\cite{Profumo:2007wc}). Furthermore, the $CP$
violation from the CKM matrix is too small to generate a
sufficient baryon asymmetry during the EWPT. In the MSSM there may
be a small parameter space for EWBG~\cite{JRE1}-\cite{KPSW2}.
There are additional sources of $CP$ violation associated with the 
phases of the $\mu$ and supersymmetry breaking
parameters. However, a strong enough first order EWPT requires
that the lightest stop quark mass be smaller than the top quark
mass $\sim 173$ GeV, called the light stop
scenario~\cite{Carena:1996wj, Carena:1997gx, MCQCW, CMQSW1, CQSW2} 
(for recent work, also see~\cite{Carena:2008rt, Cirigliano:2006dg, Chung:2008aya, Li:2008ez, Menon:2009mz, Funakubo:2009eg}). Also, the mass of the
lightest $CP$-even Higgs must be smaller than 125 GeV, which
leaves a small window above the current limit. In the Next-to-Minimal Supersymmetric Standard Model (NMSSM)~\cite{NMSSM} there
is a trilinear term $A_h h S H_u H_d$ in the Higgs potential,
where $S$ is a singlet under the SM gauge group. This allows a
strong enough first order EWPT~\cite{Pietroni, ADFM, SJHMGS, Cline:2009sn}. In
the best-motivated versions of the NMSSM (motivated by the $\mu$ problem~\cite{Kim:1983dt}), the effective $\mu$
parameter is given by $ h \langle S \rangle$ (for recent studies, see~\cite{Liu:2008pa}). However, the
original versions involve a discrete $Z_3$ symmetry and serious
cosmological domain wall problems~\cite{domain}. These can be
avoided by introducing an elementary bilinear $\mu$ term in the
superpotential (or linear or quadratic terms in $S$), but at the
cost of reintroducing
 the $\mu$ (or analogous) problems~\cite{ADFM, SJHMGS}. The domain
 wall problem can also be avoided in an alternative version of the NMSSM~\cite{altnmssm},
 and in the nearly Minimal
 Supersymmetric Standard Model (nMSSM)~\cite{nmssm1, nmssm2, Balazs:2007pf},
 using a different discrete  symmetry broken by a loop-induced tadpole.
 (For reviews, see~\cite{xhiggsrev}.)

On the other hand, the possibility of an extra \upr \ gauge
symmetry~\cite{Langacker:2008yv} is well-motivated in superstring
constructions~\cite{string},
in grand unified theories~\cite{review},
in models of dynamical symmetry breaking~\cite{DSB}, and in little
Higgs models~\cite{littlehiggs}. In supersymmetric models, an
extra \upr \ can provide an elegant solution to the $\mu$
problem~\cite{muprob1,muprob2}, with an effective $\mu$ parameter
generated by the VEV of the Standard Model singlet field $S$ which
breaks the \upr \ symmetry. This is somewhat similar to the
effective $\mu$ parameter in the NMSSM~\cite{NMSSM}. However, with
a \upr \ there are no extra discrete symmetries or domain wall
problems. A closely related feature is that the MSSM upper bound
of \mz \ on the tree-level mass of the lightest Higgs scalar is
relaxed, both in models with a \upr \ and in the NMSSM, because of
the $h S H_u H_d$ term in the superpotential and the
$U(1)'$ $D$-term~\cite{Higgsbound}. Light Higgs masses lower than
those allowed in the MSSM are also allowed experimentally because
of possible mixings among the Higgs doublets and
singlets~\cite{lowmass,Barger:2006dh}. For specific \upr \ charge assignments
for the ordinary and exotic fields in some theories one can
simultaneously ensure the absence of anomalies; that all fields of
the TeV-scale effective theory are chiral, avoiding a generalized
$\mu$ problem; consistency with gauge coupling unification;
 and the absence of dimension-4 proton decay
operators~\cite{general}.

There are stringent limits on an extra $Z'$ from direct searches
from the Tevatron~\cite{explim} and from indirect precision tests
at the $Z$-pole, at LEP 2, and from
weak neutral current experiments~\cite{indirect}.
The constraints depend on the particular \zpr \ couplings, but in
typical models one requires $M_{Z^{\prime}} > (800-1000) $ GeV and
the $Z-\zpr$ mixing angle $\alpha_{Z-Z^{\prime}}$ to be smaller
than a few $\x 10^{-3}$.
(The specific parameters considered here  yield a  $Z'$ mass of around
1 TeV.)
Thus, explaining the $Z-Z^{\prime}$
mass hierarchy without fine-tuning is non-trivial.
Two of us with J. Erler
proposed a supersymmetric $U(1)^{\prime}$
model with a secluded $U(1)^{\prime}$-breaking sector (sMSSM) in which the
squark and slepton spectra can mimic those of the MSSM, the EW
symmetry breaking is driven by relatively large $A$ terms, and a
large \zpr \ mass can be generated by the VEVs of additional SM
singlet fields that are charged under the \upr \ but do not
directly contribute to the effective $\mu$ parameter~\cite{ELL}.
If these fields are only weakly coupled to the SM fields, {\it
i.e.}, by \upr \ interactions and possibly soft SUSY-breaking
terms, then the scale of VEVs in this sector is only weakly linked
to the EW scale. In particular, we consider the situation in which
there is an almost $F$ and $D$ flat direction involving these
secluded fields, with the flatness lifted by a small Yukawa
coupling $\lambda$. For a sufficiently small value for $\lambda$,
the \zpr \ mass can be arbitrarily large.

In this paper, we systematically study EWBG and its correlation with neutralino CDM in the sMSSM 
(for a discussion on EWBG in the thin wall regime in the sMSSM, also see~\cite{letter}). We construct
two explicit anomaly-free models. In Model I, we embed the
$SU(3)_C\times SU(2)_L \times U(1)_Y \times U(1)'$ gauge symmetry
into a subgroup of $E_6$. We consider three families of the SM
fermions as arising from three fundamental ${\bf 27}$
representations of $E_6$. The ${\bf 27}$'s also contain candidates
for the Higgs doublets, SM singlets (needed to break the \upr),
and additional exotic fields. Additional vector-like pairs from
${\bf 27+27^*}$ representations can also exist at low energy
without introducing anomalies. Consistency with gauge coupling
unification is achieved by introducing one pair of vector-like
Higgs-type doublets from ${\bf
27+27^*}~\cite{PLJW}$\footnote{These are in addition to those in
the three ${\bf 27}$'s. It is not important for our purposes which
ones correspond to the two MSSM Higgs doublets.}, while the
necessary \upr \ breaking and the generation of masses for the
exotic fermions is achieved by introducing three pairs of
vector-like SM singlets from two pairs of ${\bf 27+27^*}$.
The spectrum of exotics is rather complicated in Model I. In
contrast, in Model II we  add the minimal number of exotic
particles with rational $U(1)'$ charges to cancel the $U(1)'$
anomalies. This model has the minimal exotic particle content but
does not respect the simplest form of gauge coupling unification.
We present the general superpotential and the supersymmetry
breaking soft terms for both models.

For concreteness, we will work in Model I (the $E_6$ 
embedding). First, we discuss the one-loop effective potential at finite temperature
in the 't~Hooft-Landau gauge in the $\overline{MS}$-scheme. We
calculate the gauge boson, Higgs,
 neutralino, chargino, squark, and slepton
mass matrices, and those for the scalar components of the exotic
chiral superfields. We also calculate the temperature dependent
masses for the Higgs fields, the longitudinal components of the
gauge bosons, squarks, sleptons, and the scalar components of
the exotic chiral superfields. There are two relevant phase transitions as the temperature decreases: the \upr\ symmetry is
broken at about 1 TeV, and the EW symmetry at the weak scale.
We show that there exists strong enough first order EWPT because of the large
trilinear term $A_h h S H_u H_d$ in the tree-level Higgs
potential. Thus, unlike the MSSM, the stop masses can be very
heavy compared to the top quark mass. The EWPT features in 
model II or the other possible anomaly-free embeddings are similar, 
because the exotic particles' contributions to the one-loop effective potential at finite
temperature are suppressed by  their heavy masses.

Subsequently, we discuss possible $CP$ violation
introduced by the extended Higgs sector of the sMSSM. 
As first pointed out in~\cite{letter}, unlike the MSSM in which
there is no $CP$-violation in the Higgs sector at tree level, 
both explicit and spontaneous violation (denote them as ECPV and SCPV) can occur.
There are five complex parameters in the supersymmetry breaking
soft terms and only four gauge-independent Higgs phase degrees of
freedom, which implies that one of these complex phases cannot be
removed by field redefinition. The $CP$-symmetry of this tree-level potential
can therefore be explicitly broken. In addition, as a result of balancing
different terms in the minimization of the tree-level neutral Higgs potential,  
the phases of the Higgs fields can obtain non-trivial VEVs, which implies that $CP$ symmetry 
can also be spontaneously broken. With loop corrections included, there may coexist vacua at finite temperature with broken and unbroken EW symmetry. 
The values of the spontaneous $CP$ phases are usually different in these vacua. 

In this work SCPV provides a direct source for baryogenesis, while its magnitude is mediated by an
explicit phase from the secluded sector\footnote{We turn off all $CP$-violating sources beyond the Higgs sector, e.g., 
the ones from soft gaugino and sfermion masses, which are generally used for EWBG 
in supersymmetric models. Though they may provide non-trivial contributions to
EWBG, these $CP$ phases usually suffer strong constraints from electric dipole moments. }. 
These new $CP$ sources do not introduce significant new contributions to the electric dipole
moments (EDMs) of the electron and neutron. After proper field redefinitions, the
$CP$-violation phases will only appear in the Higgs mass matrix
through soft masses associated with singlet components. Numerical estimates show that, for typical
parameter values, their contributions to EDMs will be about six or
seven orders of magnitude smaller than the experimental upper limits. These contributions disappear completely in the
limit with a trivial explicit $CP$ phase, where the spontaneous $CP$ phases are absent in the true vacuum but not where 
EWBG occurs.

In the early Universe the first order EWPT is realized by
nucleating bubbles of the broken phase. The dynamical properties of these bubbles,
such as the wall profile and expansion velocity, can have important 
influences on the production of the baryon asymmetry. We also 
study their physics in detail. The VEVs of the Higgs fields (including both their magnitudes and phases) are space-dependent, as one crosses the bubble wall. 
The bubble wall thickness is estimated by minimizing the action under the
kink ansatz~\cite{MQS,P.John,CMS}. Numerical results show that it
varies from $3~T_c^{-1}$ to $30~T_c^{-1}$ as an approximately monotonically
increasing function of the phase changes of the Higgs fields. In addition, we argue that the
wall velocity in the sMSSM cannot be larger than that
in the MSSM under the same phase transition condition and thus
should be non-relativistic. This fact implies that  EW sphaleron processes 
have more time to occur and hence will enhance the final baryon asymmetry.

We then systematically study non-local EWBG in both the thin wall and thick wall
regimes. Non-local baryogenesis has a great advantage over the
local one~\cite{CKN1}: the baryon number violation processes take
place in an effective volume extending from the wall surface to a
region in the symmetric phase, and thus we may expect an
enhancement in the generation of the baryon asymmetry. According to the 
bubble wall profile, we calculate the contribution from $\tau$ leptons in the thin wall regime, and the ones from
stop, charginos and top quarks in the thick wall regime. 
We find that the $CP$-violating currents induced by stops and charginos are very different from those obtained 
in the MSSM: (1) the current of the $v_1 \partial v_2 - v_2 \partial v_1$ type (in our formalism, the concrete expression is a little different) 
never requires $\partial \tan\beta\neq 0$ in the bubble wall due to the variance of the relevant $CP$ phases crossing the wall\footnote{In the MSSM, there also
exists a $CP$-violating current of the $v_1\partial v_2 + v_2 \partial v_1$ type for charginos which arises from a resummation of corrections associated
with higher order Higgs insertions or multiple scattering effects in the bubble wall.
Though not suppressed by $\partial \tan\beta$, the current of this type is usually subdominant unless $\partial \tan\beta$ 
is suppressed (in our model it is always subdominant). For relevant or more general discussions, see~\cite{mass insertion, Rius:Sanz, CK, CJK, Carena:1997gx, MCQCW, CMQSW1, CQSW2}. 
We will therefore not discuss the currents of this type. }; (2) there is a new type of $CP$-violating current at the leading order which is proportional 
to $v_iv_i$ with $i=1,2$ and is absent in the MSSM. The $CP$-violating current of the new type has important influence on the EWBG.  
First, the stop contribution can be quadratically enhanced by a large soft $A_{h_t}$ parameter. Second, in addition to 
$\tau$ leptons, top squarks and charginos, top quarks can also play a significant role in the EWBG.  
All of these features are results of the SCPV-driven EWBG, so they are not sensitive to the concrete embeddings of the sMSSM.
Numerical calculations show that the produced baryon asymmetry 
is large enough to explain the cosmological observation today. 
Though the EWBG is directly driven by SCPV, there is no dilution problem 
between matter and anti-matter bubbles. Such a dilution is 
caused by a $Z_2$ discrete symmetry which usually exists in the SCPV scenarios. 
In our work, however, this symmetry has been explicitly broken at tree level by the explicit $CP$-violating phase.

We also study the correlation between EWBG and cold dark matter (CDM) in the sMSSM. 
Large trilinear soft parameters (used for strong enough first order EWPT) more often arise from gravity-mediated 
SUSY breaking, where the lightest supersymmetric particle (LSP) is usually the lightest neutralino. We study the possibility of describing
the EWBG and neutralino CDM in the same framework. Though we do not scan the whole parameter
space, we find that there indeed exist some regions where strong enough first order EWPT, large $CP$ phase variations across the bubble wall, reasonable baryon asymmetry,  
as well as acceptable neutralino LSP relic density can be achieved simultaneously. The relevant neutralino mass $m_{\chi_1^0}$ is close to $M_Z/2$.
Finally, we comment briefly on possible cosmological signals: superconducting cosmic strings and gravitational waves (GWs). 
We expect that particle emission from the decays of cosmic strings and GWs from EWPT
may be observed within the foreseeable future.

This paper is organized as follows. We briefly review the
tree-level Higgs potential in the sMSSM and construct two
anomaly-free embeddings in Section 2. The one-loop effective potential
at finite temperature is considered in Section 3. In Section 4, we
show that the EWPT is strongly first order. We study $CP$ violation 
in Section 5 and the bubble wall physics in Section 6.
In Section 7, we systematically study non-local EWBG in both the thin wall and thick wall regimes, 
and discuss the contributions from leptons, squarks, charginos and quarks in detail. 
In Section 8, we illustrate that  
there exists common parameter space where the baryon asymmetry 
and the CDM can both be explained by EWBG and neutralino LSP, respectively.
Some simple comments on the cosmological signals of cosmic string decays and GWs  
are given in Section 9. The last section is our discussion and conclusions.

\section{Two Anomaly-free \upr\ Models}

Let us briefly review the sMSSM~\cite{ELL}. There are one
pair of Higgs doublets, $H_u$ and $H_d$, and four SM singlets,
$S$, $S_1$, $S_2$, and $S_3$. The $U(1)'$ charges for the Higgs
fields satisfy
\begin{eqnarray}
\label {qcharge} 
Q'_{H_d}+Q'_{H_u}+Q'_S=0, \ \ \ \  Q'_S=-Q'_{S_1} =-Q'_{S_2} ={1\over 2} Q'_{S_3} .
\end{eqnarray}
The superpotential for the Higgs is
\begin{eqnarray}
W_{H} &=& h S H_u H_d + \lambda S_1 S_2 S_3 ~,~\,
\end{eqnarray}
where the Yukawa couplings $h$ and $\lambda$ are respectively
associated with the effective $\mu$ term and with an (almost) $F$
and $D$-flat  direction. 
For simplicity, we assume that terms such as $S_1^2S_3, S_2^2S_3$ and their
associated soft terms are absent, and that there are no bilinears $SS_{1,2}$ in $W_H$.
The existence of a number of SM singlets
and the non-diagonal nature of the superpotential is in part
motivated by explicit superstring
constructions~\cite{freefermionic}. The corresponding $F$-term
scalar potential is
\begin{eqnarray}
V_F &=& h^2 \left( |H_d^0|^2 |H_u^0|^2 + |S|^2 |H_d^0|^2 +
|S|^2|H_u^0|^2\right) \nonumber\\&& +\lambda^2 \left(|S_1|^2
|S_2|^2 + |S_2|^2 |S_3|^2 + |S_3|^2 |S_1|^2\right) ~.~
\label{VFpotential}
\end{eqnarray}
The $D$-term scalar potential for the neutral fields is
\begin{eqnarray}
V_D &=& {{G^2}\over 8} \left(|H_u^0|^2 - |H_d^0|^2\right)^2
\nonumber\\&& +{1\over 2} g_{Z'}^2\left(Q'_S |S|^2 + Q'_{H_d}
|H_d^0|^2 + Q'_{H_u} |H_u^0|^2 + \sum_{i=1}^3 Q'_{S_i}
|S_i|^2\right)^2 ~,~\, \label{VDpotential}
\end{eqnarray}
where $G^2=g_1^{2} +g_2^2$; $g_1, g_2$,  and $g_{Z'}$ are the
coupling constants for $U(1)$, $SU(2)_L$ and $U(1)^{\prime}$; and
$Q'_{\phi}$ is the $U(1)^{\prime}$ charge of the field $\phi$.

In addition, one introduces the supersymmetry breaking soft terms
\begin{eqnarray}
V_{soft}^{H} &=& m_{H_d}^2 |H_d^0|^2 + m_{H_u}^2 |H_u^0|^2 + m_S^2
|S|^2 + \sum_{i=1}^3 m_{S_i}^2 |S_i|^2 \nonumber\\&&
 -\left(A_h h S H_d^0
H_u^0 + A_{\lambda} \lambda S_1 S_2 S_3 + m_{S S_1}^2 S S_1
 \right.\nonumber\\&&\left.
 + m_{S S_2}^2 S S_2 + m_{S_1 S_2}^2 S_1^{\dagger} S_2
+ {\rm H. C.} \right)~.~\, \label{vsoftH}
\end{eqnarray}
$m_{S S_i}^2, i=1,2$ are needed to break two unwanted global
$U(1)$ symmetries (for a recent discussion of such global symmetries, see~\cite{Langacker:2008dq}). There is an almost $F$ and $D$ flat direction
involving $S_i$, with the flatness lifted by a small Yukawa
coupling $\lambda$. For a sufficiently small value of $\lambda$,
the \zpr \ mass can be arbitrarily large. For example, if $ h\sim
10 \lambda$, one can generate a $Z-Z'$ mass hierarchy in which the
$Z'$ mass is of order 1 TeV. The dimensional parameters in
$V_{soft}^{H}$ are specified in arbitrary units, and then rescaled
after the (effective) potential is minimized so that the EW scale
is $v={\sqrt {v_1^2 + v_2^2 }} \sim 174$ GeV, with $v_1=\langle
H_u^0 \rangle$ and $v_2=\langle H_d^0 \rangle$.

In~\cite{ELL}, $m_{S_1 S_2}^2$ was set to 0  for simplicity.
 In that case, the fields can be defined so that $A_h, \ A_\lambda, \ m_{S S_1}^2,$
 and $m_{S S_2}^2$ are all real and positive. The minimum then occurs at a point
 at which all of the fields are real and positive, i.e.,  there is no $CP$ violation in the tree-level
 Higgs potential.  For
$m_{S_1 S_2}^2 \ne 0$, however, there will in general be $CP$
violation.

The \upr \ charges for the fermions and the anomaly cancellations
were not discussed in Ref.~\cite{ELL}, so in the following two
subsections we  construct two anomaly-free models, one of which is
consistent with minimal gauge coupling unification.

\subsection{Model I: sMSSM with $E_6$ Embedding}
$U(1)'$ models necessarily imply that new fermions are needed for anomaly
cancellations~\cite{Langacker:2008yv,general}. Since all complete representations
of $E_6$ are anomaly-free, it is convenient and conventional to
simply consider the \upr\ charge assignment and exotic particle
content of $E_6$ as an example of an anomaly-free construction. We
are not considering a full $E_6$ grand unified theory, because a
light \zpr\ would prevent a large doublet-triplet splitting,
leading to rapid proton decay if the $E_6$ Yukawa relations were enforced.
(Detailed studies of $E_6$ theories with broken Yukawa relations may be 
found in~\cite{King:2005jy}.)

The $E_6$ gauge group can be broken as~\cite{Group,Hewett:1988xc}
\begin{eqnarray}
E_6 \to\ SO(10) \x\ U(1)_{\psi} \to\ SU(5) \x\ U(1)_{\chi} \x\
U(1)_{\psi}~.~\,
\end{eqnarray}
The $U(1)_{\psi}$ and $U(1)_{\chi}$ charges for the $E_6$
fundamental ${\bf 27}$ representation are given in Table
\ref{E6charge}. The \upr\ in Model I is one linear combination of
the $U(1)_{\chi}$ and $U(1)_{\psi}$
\begin{eqnarray}
Q^{\prime} &=& \cos\theta \ Q_{\chi} + \sin\theta \ Q_{\psi}~,~\,
\label{E6MIX}
\end{eqnarray}
where the angle $\theta$ will be chosen to ensure the needed \upr
\ charges for the SM singlets. For simplicity, we assume that the
other $U(1)$ gauge symmetry from the orthogonal linear combination
of the $U(1)_{\chi}$ and $U(1)_{\psi}$ is absent or broken at a
high scale.

We assume three ${\bf 27}$s, which include
 three families of the SM fermions, one
pair of Higgs doublets ($H_u$ and $H_d$), and a number of SM
singlets, extra Higgs-like doublets, and other exotics. The
embeddings of  the SM fermions are obvious. For definiteness, we
assume that the Higgs doublets ($H_u$ and $H_d$) are the doublets
in ${\bf 10}$ (or ${\bf 5}$ and ${\bf {\bar 5}}$) in the third
${\bf 27}$. (Of course, the exact identification of the $H_u$,
$H_d$ and $S$ with specific $\bf 27$s is irrelevant.) In addition,
we assume that the four SM singlets $S$, $S_1$, $S_2$, $S_3$ are
the $S_L$, $S_L^*$, $S_L^*$ and ${\bar N}^*$, respectively, in two
(partial) pairs of $\bf 27$ and $\bf 27^*$. To cancel the $U(1)'$
anomalies, we introduce $X$ and $X_3$, which are ${S_L}$ and
${\bar N}$, respectively, in the two pairs. Thus,  ($S$, $S_1$),
($X$, $S_2$) and ($X_3$, $S_3$) are three pairs of vector-like SM
singlets and will not introduce any anomalies. To be consistent
with minimal gauge coupling unification, we introduce one pair of
vector-like doublets $H_u^{\prime}$ and $\bar H_u^{\prime}$ from a
pair of
 $\bf 27+27^*$~\cite{PLJW}. For simplicity,
we assume that the other particles in the  $\bf 27+27^*$ pairs are
absent or are very heavy and decouple at low energy. From the
requirement $Q_S = {1 \over 2} {Q_S}_3$, {\it i.e.}, $Q_{S_L}= {1
\over 2} Q_{{\bar N}^*}$, we obtain
\begin{eqnarray}
 \tan \theta &=& {\sqrt{15} \over 9} ~.~\,
\end{eqnarray}
The exotic particles in Model I are $D_i$, $\bar D_i$, $S_L^i$,
${\bar N}_i$, where $i=1,2,3$; $H_u^{\prime k}$ and $\bar
H_d^{\prime k}$, where $k=1, 2$; $X$, $X_3$; and $H_u^{\prime}$
and $\bar H_u^{\prime}$. The $U(1)'$ charges for the Standard
Model fermions and exotic particles are also given in Table
\ref{E6charge}.

\begin{table}[t]
\caption{Decomposition of the $E_6$ fundamental  ${\bf 27}$
representation under $SO(10)$, $SU(5)$, and the $U(1)_{\chi}$,
$U(1)_{\psi}$ and $U(1)'$ charges.}
\begin{center}
\begin{tabular}{|c| c| c| c| c|}
\hline $SO(10)$ & $SU(5)$ & $2 \sqrt{10} Q_{\chi}$ & $2 \sqrt{6}
Q_{\psi}$ & $2 \sqrt{15} Q$ \\
\hline
16   &   $10~ (u,d,{\bar u}, {\bar e} )$ & --1 & 1  & $-{1/2}$ \\
            &   ${\bar 5}~ ( \bar d, \nu ,e)$  & 3  & 1  & 4          \\
            &   $1 \bar N$             & --5 & 1  &--5         \\
\hline
       10   &   $5~(D,H^{\prime}_u)$    & 2  & --2 & 1          \\
            &   ${\bar 5} ~(\bar D, H^{\prime}_d)$ & --2 &--2 & $-{7/2}$ \\
\hline
       1    &   $1~ S_L$                  &  0 & 4 & $5/2$ \\
\hline
\end{tabular}
\end{center}
\label{E6charge}
\end{table}

The general superpotential is $W_H + W_Y$, where
\begin{eqnarray}
W_{Y} &=&  \alpha S X X_3 + \alpha_{ij}^N S S_L^i {\bar N}_j
+\mu^{\prime} \bar H_u^{\prime} H_u^{\prime} 
+h^u_{ij} Q_i H_u \bar u_j + h^d_{ij} H_d Q_i \bar d_j  
+h^e_{ij} H_d L_i {\bar e}_j \nonumber\\&&
+ h^N_{ij} H_u L_i {\bar N}_j  
 + \alpha_{ij}^D S D_i \bar D_j +\alpha_{kl}^{H^{\prime}} S
H_u^{\prime k} H_d^{\prime l} 
+\lambda^u_{ij} Q_i H'_u \bar u_j \nonumber\\&&
+ \lambda^N_{ij} H'_u L_i {\bar N}_j
+ \lambda^1_{ijm} D_i {\bar u}_j {\bar e}_m 
+ \lambda^2_{ijm} {\bar D}_i Q_j L_m
+ \lambda^3_{ijm} D_i {\bar d}_j {\bar N}_m \nonumber\\&&
+ \lambda^4_{ijm} D_i Q_j Q_m
+ \lambda^5_{ijm} {\bar D}_i {\bar u}_j {\bar d}_m 
+ \lambda^6_{ki} \phi S_k S_L^i + M_{\phi} \phi^2
~,~\,    \label{exoticY}
\end{eqnarray}
with $Q_i=(u_i,d_i)$, $L_i=(\nu_i,e_i)$, and $i, j, m =1, 2, 3$ while
$k, l=1,2$.  For simplicity, we will assume that the Yukawa
couplings $h^u_{i}$, $ h^d_{ij}$, $h^e_{ij}$ and $h^N_{ij}$
 and those associated
with the exotics are diagonal in the numerical calculations. 
In addition, to explain the neutrino masses and mixings, we
introduce a SM singlet $\phi$ which is neutral under $U(1)'$.
Integrating out $\phi$, we obtain
the non-renormalizable terms $S_k S_l S_L^i S_L^j/M_{\phi}$,
and then Majorana mass terms for $S_L^i S_L^j$ will appear after $U(1)'$ breaking.
Thus, the light neutrino masses and mixings in Model I can
be generated via a double-see-saw mechanism
naturally~\cite{DSS, JKPLTL}.
We see from Eq.~(\ref{exoticY}) that all of
the exotics except $\bar H_u^{\prime}$ and $ H_u^{\prime}$ can
have masses generated by the VEV of $S$. The latter can be given a
supersymmetric mass $\mu^{\prime}$. This reintroduces a form of
the $\mu$ problem, though not for the Higgs fields associated with
EW symmetry breaking, and is clearly a flaw of the specific
construction. However, this segment of the model is only loosely
connected with the issue of concern in this paper, {\it i.e.}, EWBG, so
we will tolerate it rather than going to a more complicated model.

The new scalar supersymmetry breaking soft terms (in addition to
Eq.~(\ref{vsoftH}))  are
\begin{eqnarray}
 V_{soft}^{A} &=& - A_{\alpha} \alpha S X X_3 
- A_{\alpha_{ij}^N }  \alpha_{ij}^N S S_L^i {\bar N}_j
- B_{\mu^{\prime}} \mu^{\prime} \bar H_u^{\prime} H_u^{\prime} 
- A_{h^u_{ij}}  h^u_{ij} Q_i H_u \bar u_j \nonumber\\&&
- A_{h^d_{ij}}  h^d_{ij} H_d Q_i \bar d_j   
-  A_{h^e_{ij}} h^e_{ij} H_d L_i {\bar e}_j
- A_{h^N_{ij}}  h^N_{ij} H_u L_i {\bar N}_j  
 - A_{\alpha_{ij}^D} \alpha_{ij}^D S D_i \bar D_j \nonumber\\&&
- A_{\alpha_{kl}^{H^{\prime}}} \alpha_{kl}^{H^{\prime}} S
H_u^{\prime k} H_d^{\prime l} 
- A_{\lambda^u_{ij}} \lambda^u_{ij} Q_i H'_u \bar u_j 
- A_{\lambda^N_{ij}} \lambda^N_{ij} H'_u L_i {\bar N}_j \nonumber\\&&
- A_{\lambda^1_{ijm}} \lambda^1_{ijm} D_i {\bar u}_j {\bar e}_m 
- A_{\lambda^2_{ijm}} \lambda^2_{ijm} {\bar D}_i Q_j L_m
- A_{\lambda^3_{ijm}} \lambda^3_{ijm} D_i {\bar d}_j {\bar N}_m \nonumber\\&&
- A_{\lambda^4_{ijm}} \lambda^4_{ijm} D_i Q_j Q_m
- A_{\lambda^5_{ijm}} \lambda^5_{ijm} {\bar D}_i {\bar u}_j {\bar d}_m 
- A_{\lambda^6_{ki}} \lambda^6_{ki} \phi S_k S_L^i 
\nonumber\\&&
-B_{\phi} M_{\phi} \phi^2
 +{\rm ~H.C.},
\end{eqnarray}
\begin{eqnarray}
V_{soft}^{m} &=& \sum_{i=1}^3 \left(m_{\tilde Q_i}^2 |\tilde
Q_i^2| + m_{u^i_R}^2 |{\tilde {\bar u}}_i|^2 + m_{d^i_R}^2
|{\tilde {\bar d}}_i|^2 + m_{\tilde L_i}^2 |\tilde L_i^2| +
m_{e^i_R}^2 |{\tilde {\bar e}}_i|^2
 \right.\nonumber\\&&\left.
+ m_{{\tilde {\bar N}}_i}^2 |{\tilde {\bar N}}_i|^2
 + m_{S_L^i}^2 |S_L^i|^2
+ m_{{\tilde D}_i}^2 |{\tilde D}_i|^2 + m_{{\tilde {\bar D}}_i}^2
|{\tilde {\bar D}}_i|^2 \right) \nonumber\\&& +\sum_{k=1}^2
\left(m_{H_u^{\prime k}}^2 |H_u^{\prime k}|^2 + m_{H_d^{\prime
k}}^2 |H_d^{\prime k}|^2 \right) + m_{\phi}^2 |\phi|^2 
\nonumber\\&& + m_{\tilde X}^2
|\tilde X|^2 + m_{{\tilde X}_3}^2 |{\tilde X}_3|^2 +
m_{H_u^{\prime}}^2 |H_u^{\prime}|^2 + m_{\bar H_u^{\prime}}^2
|\bar H_u^{\prime}|^2, \label{vsoft}
\end{eqnarray}
where we have assumed that the mass-squared terms are diagonal.
We assume that only $S, \ S_1, \ S_2, \ S_3, \ H_u^0$, and $H_d^0$
acquire VEVs.

The vector-like particles $H_u^{\prime}$ and ${\bar H}_u^{\prime}$
can decay via the Yukawa  $\lambda_{ij}^u$ and $\lambda_{ij}^N$ terms
in Eq.~(\ref{exoticY}). In addition, 
the charged exotic particles $D_i$ and ${\bar D}_i$ can decay through
the Yukawa $\lambda_{ijm}^n$ terms with $n=1,~2,...,~5$ 
in Eq.~(\ref{exoticY})~\cite{Hewett:1988xc, Kang:2007ib}. 
However, to avoid the proton decays via dimension-5 operators, 
we must forbid some of
the Yukawa $\lambda_{ijm}^n$ terms. For example, we can 
choose $\lambda_{ijm}^n=0$ either for $n=1,~2,~3$ or for $n=4,~5$.

\subsection {Model II: Minimal Anomaly-free Model}
 In this subsection,
we consider the minimal sMSSM in which all the particles have
rational $U(1)'$ charges. In order to cancel the
gauge and gauge-gravity mixed anomaly due to the $U(1)'$ gauge
symmetry, we introduce the exotic particles $D_1$, $\bar
D_1$, $D_2$, $\bar D_2$, $V$, $\bar V$, $X_1$, $\bar X_1$, $X_2$,
 $\bar X_2$, $X_3$, $\bar X_3$, $X_4$, and $\bar X_4$. 
The quantum numbers under $SU(3)_C\times
SU(2)_L\times U(1)_Y\times U(1)'$ for the SM fermions, 
Higgs fields 
and extra exotic particles 
are given in Table~\ref{tab:SUV3}. If one allowed irrational
$U(1)'$ charges, only
 $\bar D_1$, $D_2$, $\bar D_2$,
$V$, and $\bar V$ would be needed.

With the $U(1)'$ charge assignments for the particles
 in Table~\ref{tab:SUV3}, the conditions
for $[SU(3)_C]^2 U(1)'$ and $[{\rm Gravity}]^2 U(1)'$ are
automatically satisfied, and the condition for
$[SU(2)_L]^2 U(1)'$ is the same as that for $[U(1)_Y]^2 U(1)'$.
The anomaly-free conditions for  $[SU(2)_L]^2 U(1)'$,
$U(1)_Y[U(1)']^2$ and $[U(1)']^3$, respectively, are
\begin{eqnarray}
c_1 + c_2  + 3 (a + 3 b) = 0 ~,~\, \label{A23}
\end{eqnarray}
\begin{eqnarray}
&& -3 a^{2}+3 b^{2} + 3 (a+c_1)^2-6(b+c_2)^2+3(b+c_1)^2
-c_1^2+c_2^2\\ \nonumber && -d_1^2+(s+d_1)^2 -d_2^2+(2s+d_2)^2
-d_3^2+(2s+d_3)^2 = 0 ~,~\, \label{A4}
\end{eqnarray}
\begin{eqnarray}
 &&6 a^{3} - 3 ( a + c_2 )^3 - 3 ( a + c_1 )^3 + 18 b^{3} - 9 ( b + c_2 )^3
- 9 ( b + c_1 )^3  \\ \nonumber && + 3 d_1^3 -3(s+d_1)^3 + 3 d_2^3
-3(2s+d_2)^3 +  d_3^3 - (2s+d_3)^3  \\ \nonumber &&+ 2 d_4^3
-(d_4-s)^3 -(d_4 + s)^3 + 7 s^3 + 2 c_1^3 + 2 c_2^3= 0 ~.~\, \label{A5}
\end{eqnarray}
The simplest solution is $c_1=9/10$,
$c_2=9/10$, $s=-9/5$, $a=-18/5$, $b=1$, $d_1=-2$, $d_2=-1/5$,
$d_3=9/5$, $d_4=27/10$, $ d_5=9/10$, and $d_6=-3$.

\begin{table}[t]
\caption{ Minimal Anomaly-free Model. Quantum numbers under
$SU(3)_C\times SU(2)_L\times U(1)_Y\times U(1)'$ for the
left-chiral Standard Model fermions ($Q_i$, $\bar u_i$, $\bar
d_i$, $L_i$, $\bar N_i$, $\bar e_i$), Higgs fields ( $H_u$, $H_d$,
$S$, $S_1$, $S_2$, $S_3$), and extra exotic particles ( $D_1$,
$\bar D_1$, $D_2$, $\bar D_2$, $V$, $\bar V$,  $X_1$, $\bar X_1$,
$X_2$ and $\bar X_2$).} \vspace{0.4cm}
\begin{center}
\begin{tabular}{|c|c|c|c|}
\hline
 Particles &  Quantum Numbers & Particles &  Quantum Numbers \\
\hline $L_i$ & (1; 2; $-1/2$; $a$) &
$Q_i$ & (3; 2; 1/6; $b$) \\
\hline
$\bar N_i$ & (1; 1; 0; $-(a+c_2)$) & $\bar u_i$ & ($\bar 3$; 1; $-2/3$;  $-(b+c_2)$)\\
\hline
$\bar e_i$ & (1; 1; 1; $-(a+c_1)$) & $\bar d_i$ & ($\bar 3$; 1; 1/3;  $-(b+c_1)$)\\
\hline \hline
$H_d$ & (1; 2; $-1/2$; $c_1$) & $H_u$ & (1; 2; $1/2$; $c_2$) \\
\hline
$S$ & (1; 1; 0; $s$) & $S_3$ & (1; 1; 0; $2s$) \\
\hline
$S_1$ & (1; 1; 0; $-s$) & $S_2$ & (1; 1; 0; $-s$) \\
\hline \hline
$D_1$ & (3; 1; $-1/3$; $d_1$) & $\bar D_1$ & ($\bar 3$; 1; 1/3;  $-(s+d_1)$)\\
\hline
$D_2$ & (3; 1; $-1/3$; $d_2$) & $\bar D_2$ & ($\bar 3$; 1; 1/3;  $-(2s+d_2)$)\\
\hline
$V$ & (1; 1; $-1$; $d_3$) & $\bar V$ & (1; 1; 1;  $-(d_3+2s)$)\\
\hline
$X_1$ & (1; 1; 0; $d_4$) & $\bar X_1$ & (1; 1; 0;  $-(d_4-s)$)\\
\hline
$X_2$ & (1; 1; 0; $d_5$) & $\bar X_2$ & (1; 1; 0;  $-(d_5+s)$)\\
\hline
$X_3$ & (1; 1; 0; $d_6$) & $\bar X_3$ & (1; 1; 0;  $-(d_6-s)$)\\
\hline
$X_4$ & (1; 1; 0; $d_6$) & $\bar X_4$ & (1; 1; 0;  $-(d_6+s)$)\\
\hline
\end{tabular}
\end{center}
\label{tab:SUV3}
\end{table}

The general superpotential in Model II is
\begin{eqnarray}
W &=& \alpha_1 S D_1 \bar D_1 + \alpha_2 S_3 D_2 \bar D_2 +
\alpha_3 S_3 V \bar V  + \alpha_4 S_1 X_1 \bar X_1 +
\alpha_5 S_2 X_1 \bar X_1 \nonumber\\&& + \alpha_6 S X_2 \bar X_2 
+ \alpha_7 S_1 X_3 \bar X_3 +
\alpha_8 S_2 X_3 \bar X_3 + \alpha_9 S X_4 \bar X_4 
\nonumber\\&&
+ \alpha'_1 S_1 X_1 X_1 + \alpha'_2 S_1 {\bar X}_1 {\bar X}_1
+ \alpha'_3 S_2 X_1 X_1 + \alpha'_4 S_2 {\bar X}_1 {\bar X}_1
\nonumber\\&&
+h^u_{ij} Q_i H_u \bar u_j + h^d_{ij} H_d Q_i \bar d_j  
+h^e_{ij} H_d L_i {\bar e}_j
+ h^N_{ij} H_u L_i {\bar N}_j   \nonumber\\&&
+ \lambda^1_{ij} D_1 Q_i Q_j
+ \lambda^2_{ij} {\bar D}_2 {\bar u}_i {\bar d}_j 
+ \lambda^3_{i} V {\bar e}_i { X}_1
+ \lambda^4_{i} V {\bar e}_i {\bar X}_1~.~\,
\label{Mod-II-SP}
\end{eqnarray}
All of the exotics can be made massive by \upr \ breaking.

The charged exotic particles ($D_i$, ${\bar D}_i$), and
($V$, $\bar V$) can decay through
the Yukawa $\lambda^1_{ij}$, $\lambda^2_{ij}$, $\lambda^3_{i}$ and
$\lambda^4_{i}$ terms 
in Eq.~(\ref{Mod-II-SP})~\cite{Hewett:1988xc, Kang:2007ib}, 
where  we have assumed that $m_{V} > m_{{X}_1}$.
After the $U(1)'$ breaking, the vector-like particles 
($X_i$, ${\bar X}_i$) are neutral. There is no discrete symmetry after $U(1)'$ breaking, so
the vector-like particles ($X_i$, ${\bar X}_i$) can
decay via higher dimensional operators.


\section{One-Loop Effective Potential at Finite Temperature}

\label{effective potential}

In the 't~Hooft-Landau gauge and in the $\overline{MS}$-scheme, the
one-loop effective potential at finite temperature
is~\cite{Quiros}
\begin{eqnarray}
\label{total} V_{\rm{eff}}(\phi,T) &=& V_0(\phi) + V_1(\phi,0) +
\Delta V_1(\phi,T) +\Delta V_{\rm{daisy}}(\phi,T) ~,~\,
\end{eqnarray}
where $V_0(\phi)$ is the tree-level potential, $V_1(\phi,0)$ is
the one-loop zero temperature correction, $\Delta V_1(\phi,T)$ is
the temperature dependent one-loop correction,
 and $\Delta V_{\rm{daisy}}(\phi,T)$ is the multi-loop daisy
correction. The explicit expressions for $V_0(\phi)$,
$V_1(\phi,0)$, $\Delta V_1(\phi,T)$ and $\Delta
V_{\rm{daisy}}(\phi,T)$ are
\begin{eqnarray}
\label{v0} V_0(\phi) & = & V_F + V_D + V_{soft}^{H} ~,~\,
\end{eqnarray}
\begin{eqnarray}
\label{deltav} V_1(\phi,0) & = & \sum_i {n_i \over 64 \pi^2} m_i^4
(\phi) \left[ \log {m_i^2 (\phi) \over Q^2} - C_i \right]  ~,~\,
\end{eqnarray}
\begin{eqnarray}
\label{deltavt} \Delta V_1(\phi,T) & = & {T^4 \over 2 \pi^2}
\left\{ \sum_i n_i \, J_i \left[ { m^2_i (\phi) \over T^2 }
\right]  \right\} ~,~\,
\end{eqnarray}
\begin{eqnarray}
\label{dvdaisy} \Delta V_{\rm{daisy}}(\phi,T) & = & - {T \over 12
\pi} \sum_{i={\rm bosons}} n_i \left[ {\cal M}_i^3 (\phi, T ) -
m_i^3 (\phi) \right] ~,~\,
\end{eqnarray}
where $m_i(\phi)$ are field-dependent masses, $n_i$ are the number
of degrees of freedom, and $C_i$ are constants dependent on the
regularization scheme: in the $\overline{MS}$-scheme that we are
assuming, $C_i=5/6$ for the gauge bosons and $3/2$ for scalars and
fermions\footnote{For the $\overline{DR}$-scheme, $C_i$ in
Eq.~(\ref{deltav}) is $3/2$ for gauge bosons, scalars, and
fermions. Our discussions and conclusions do not
depend on the scheme.}. The function $J_i$ comes from the
 one-loop corrections to the effective
potential at finite temperature~\cite{Quiros}. For bosons
\begin{eqnarray}
\label{jb}
 J_B(m^2(\phi)/{T^2})=\int_0^{\infty} dx\
x^2\log\left[1-e^{-\sqrt{x^2+ {{m^2(\phi)} \over {T^2}}
}}\right]~.~\,
\end{eqnarray}
At relatively high temperature, {${{m(\phi)}
\over {T}} < 2.2$}, one can expand~\cite{Quiros}
\begin{eqnarray}
\label{jbexp} J_B(m^2(\phi)/T^2)^{HT} & = & -\frac{\pi^4}{45}+
\frac{\pi^2}{12}\frac{m^2(\phi)}{T^2} -\frac{\pi}{6}
\left(\frac{m^2(\phi)}{T^2}\right)^{3/2} \nonumber\\&&
-\frac{1}{32} \frac{m^4(\phi)}{T^4}\log\frac{m^2(\phi)}{a_b T^2}
-2\pi^{7/2}\sum_{\ell=1}^{\infty}(-1)^{\ell} \nonumber\\&&
\frac{\zeta(2 \ell+1)}
{(\ell+1)!}\Gamma\left(\ell+\frac{1}{2}\right)
\left(\frac{m^2(\phi)}{4\pi^2 T^2} \right)^{\ell+2} ~,~\,
\end{eqnarray}
where $a_b=16\pi^2\exp(3/2-2\gamma_E)$ ($\log a_b=5.4076$) and
$\zeta$ is the Riemann $\zeta$-function. At relatively low
temperature, { ${{m(\phi)} \over {T}} > 2.2$},
we can expand~\cite{GWALJH}
\begin{eqnarray}
\label{jbexpp} J_B(m^2(\phi)/T^2)^{LT} & = & -\frac{\pi^4}{45}
+\left({\pi\over 2}\right)^{1/2}
\left(\frac{m(\phi)}{T}\right)^{3/2} e^{-m(\phi)/T} \nonumber\\&&
 \left[1 + {15\over 8} {T\over {m(\phi)}} + ... \right]~.~
\end{eqnarray}
The function $J_F$ for fermion $\psi$ with mass $m$
is~\cite{Quiros}
\begin{eqnarray}
\label{jf} J_F(m^2(\psi)/{T^2})=\int_0^{\infty} dx\
x^2\log\left[1+e^{-\sqrt{x^2+ {{m^2(\psi)} \over {T^2}}
}}\right]~,~\,
\end{eqnarray}
For {${{m(\psi)} \over {T}} < 1.8$}, one finds~\cite{Quiros}
\begin{eqnarray}
\label{jfexp} J_F(m^2(\psi)/T^2)^{HT} & = & \frac{7\pi^4}{360}-
\frac{\pi^2}{24}\frac{m^2(\psi)}{T^2}-\frac{1}{32}
\frac{m^4(\psi)}{T^4}\log\frac{m^2(\psi)}{a_f T^2} \nonumber\\&&
-\frac{\pi^{7/2}}{4}\sum_{\ell=1}^{\infty}(-1)^{\ell}
\frac{\zeta(2\ell+1)}{(\ell+1)!} \left(1-2^{-2\ell-1}\right)
\nonumber\\&&
\Gamma\left(\ell+\frac{1}{2}\right)\left(\frac{m^2(\psi)}{\pi^2
T^2} \right)^{\ell+2} ~,~\,
\end{eqnarray}
where $a_f=\pi^2\exp(3/2-2\gamma_E)$ ($\log a_f=2.6351$). At low
temperature, ${{m(\psi)} \over {T}} > 1.8$~\cite{GWALJH},
\begin{eqnarray}
\label{jfexpp} J_F(m^2(\psi)/T^2)^{LT} & = & \frac{7\pi^4}{360}
+\left({\pi\over 2}\right)^{1/2}
\left(\frac{m(\psi)}{T}\right)^{3/2} e^{-m(\psi)/T} \nonumber\\&&
 \left[1 + {15\over 8} {T\over {m(\psi)}} + ... \right]~.~\,
\end{eqnarray}

The $m(\phi)$ and $m(\psi)$
are the Higgs field dependent masses from the
eigenvalues of the mass matrices. Thus, we need the mass matrices
for all the particles in the models. The tree-level mass
matrices\footnote{The use of tree-level mass matrices leads to a
slight inconsistency when they are evaluated at the minimum of the
full one-loop effective potential. In particular, the tree-level
approximations to the Goldstone Boson masses are sometimes shifted
slightly from zero. We avoid singularities by setting slightly
negative mass-squares to zero. We have checked that these small
shifts have negligible effects on the phase structure, as should
be apparent from the formula (\ref{jbexp}) relevant for small
${{m(\phi)} \over {T}} $.} for the particles in  Model I 
are presented in Appendix A. (Those for the
ordinary and exotic fermions are trivial.)

There are  small discontinuities between the
$J_B(m^2(\phi)/T^2)^{HT}$ and $J_B(m^2(\phi)/T^2)^{LT}$ at
${{m(\phi)} \over {T}} \sim 2.2$, and between the
$J_F(m^2(\phi)/T^2)^{HT}$ and $J_F(m^2(\phi)/T^2)^{LT}$ at
${{m(\phi)} \over {T}} \sim 1.8$. When we numerically calculate
the minimum of the potential, we need smooth interpolations for
$J_B(m^2(\phi)/T^2)$ and $J_F(m^2(\phi)/T^2)$. Thus, in our
numerical calculation we use the following approximation
\begin{eqnarray}
J_B(m^2(\phi)/T^2) &=&
{{\tanh(-{m_i^2(\phi)/T^2}+2.2^2)+1}\over\displaystyle 2} \times
J_B(m^2(\phi)/T^2)^{HT} \nonumber\\&& +
{{\tanh({m_i^2(\phi)/T^2}-2.2^2)+1}\over\displaystyle 2} \times
J_B(m^2(\phi)/T^2)^{LT} ~,~\,
\end{eqnarray}
\begin{eqnarray}
J_F(m^2(\phi)/T^2) &=&
{{\tanh(-{m_i^2(\phi)/T^2}+1.8^2)+1}\over\displaystyle 2} \times
J_F(m^2(\phi)/T^2)^{HT} \nonumber\\&& +
{{\tanh({m_i^2(\phi)/T^2}-1.8^2)+1}\over\displaystyle 2} \times
J_F(m^2(\phi)/T^2)^{LT} ~.~\,
\end{eqnarray}

The temperature dependent scalar mass-squared
 ${\cal M}_i^2 (\phi,T)$ is obtained from the ${m}_i^2 (\phi)$
by adding the leading temperature dependent self-energy
contributions $\Pi_{\Phi} (T)$,
\begin{eqnarray}
{\cal M}^2(\phi, T) = m^2(\phi) + \Pi_{\Phi} (T) ~,~\,
\end{eqnarray}
where $ \Pi_{\Phi} (T)$ is proportional to $T^2$~\cite{DCJRE}. For
the gauge bosons,
 only the longitudinal components
receive such contributions. The $ \Pi_{\Phi} (T)$'s for the
particles in Model I  are presented in Appendix B. The daisy corrections are of higher order, but
important at high temperature.

\section{The Electroweak Phase Transition}

\label{EWPT}

Generally, the EWPT in the sMSSM is more complicated than
those in the SM, MSSM, and NMSSM due to
 the secluded $U(1)'$-breaking sector.
The temperature dependent corrections to the effective potential
are a function of the masses ${\cal M}_i(\phi,
T)$. At low temperature, the temperature
dependent corrections are negligible, while at very high
temperature they are dominant. At the critical temperature, the
temperature dependent corrections  are comparable to the effective
potential at zero temperature.
Our one-loop effective potential is a function of 10 physical variables
(the complex VEVs of the six neutral Higgs fields\footnote{We ignore 
the possibility of charge or color breaking minima.}
 minus two gauge
degrees of freedom). There might exist local minima near the
critical temperature. In our analysis we always  choose the
 global minimum (at one-loop level plus daisy correction).

Because the $Z'$ mass is assumed to be of order 1 TeV, the VEVs
of the SM singlets $S_1$, $S_2$, and $S_3$ are about an order of
magnitude larger than those of the Higgs doublets and
$S$~\cite{ELL}. There are therefore  two phase transitions: the
\upr\ symmetry breaking at TeV scale and the EW symmetry breaking at the weak scale. 

\begin{table}[t]
\caption{Three sets of typical parameter values. The energy units
are 69 GeV, 92 GeV and 120 GeV, corresponding to $m_{H_d^0}^2 =$
5.0, 9.0 and 13.0, respectively. 
In this paper, we will denote
them as cases a, b and c. Note that the soft mass squares in the
superpotential are relatively small, {\em i.e.}, the symmetry breaking
is $A$-term dominant. $M_1'$, $M_1$ and $M_2$ denote the three neutral
gaugino soft masses. \label{parameter}}
\begin{center}
\begin{tabular}{|c|c|c|c|c|c|c|c|}
\hline $h$ & $A_h$ & $\lambda$  & $A_{\lambda}$
& $m_{SS_1}^2$& $m_{SS_2}^2$ & $m_{S_1S_2}^2$ & $m_{H_d^0}^2$  \\
\hline 0.8 & 4.2 & 0.06 & 3.3
& 0.02& 0.1 & 4.8 $\times 10^{-4}$  & 5.0 / 9.0 / 13.0 \\
\hline $m_{H_u^0}^2$  & $m_S^2$
& $m_{S_1}^2$& $m_{S_2}^2$ & $m_{S_3}^2$ &$M_1'$&$M_1$&$M_2$\\
\hline -0.1  & 0.5
& 0.03& 0.03 & 0.03 &1.5&1.5&3.0\\
\hline $m^2_{\tilde Q_3}$ & $m^2_{\tilde u_R^3}$ & $m^2_{\tilde
d_R^3}$ & $m^2_{\tilde Q_{1,2}}$ & $m^2_{\tilde u_R^{1,2}}$
& $m^2_{\tilde d_R^{1,2}}$ &$m^2_{\tilde l_L^{1,2,3}}$ & $m^2_{\tilde l_R^{1,2,3}}$\\
\hline 8  & 8
& 25 & 25 & 25 & 25 & 25 & 25 \\
\hline
\end{tabular}
\end{center}
\end{table}

\begin{figure}
\begin{center}
\includegraphics[height=3.2 in]{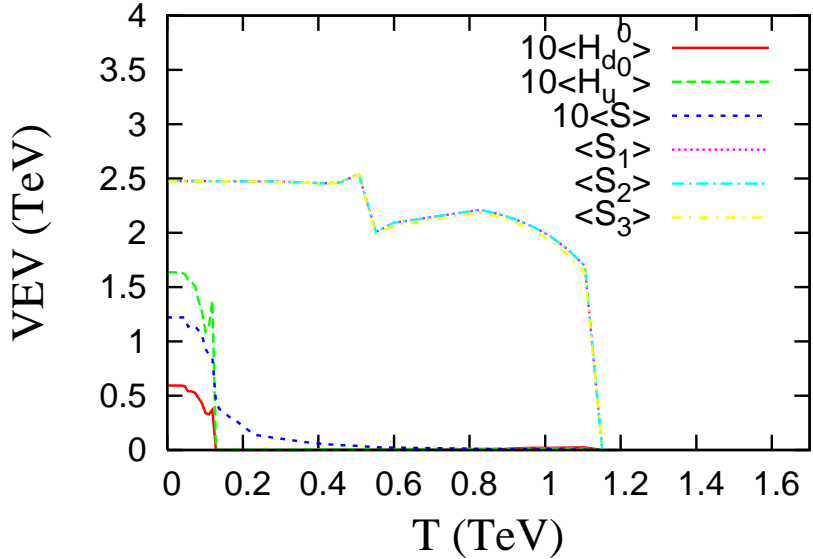}
\end{center}
\caption{The Higgs VEVs vs. temperature $T$. The VEVs of
$H_{u,d}^0$ and $S$ are multiplied by 10. The VEVs of $S_1$, $S_2$ and $S_3$ are indistinguishable.} \label{vevtt}
\end{figure}

In Figure~\ref{vevtt}, we plot the Higgs VEVs versus temperature
for the given parameters values in case b (see Table~\ref{parameter}). 
From the figure, one can see that the $U(1)'$ phase
transition occurs just below 1200 GeV, and the EWPT at about 120
GeV. We are interested in whether the EWPT is strongly first
order. From the figure, we also see that the VEVs of $H_d^0$,
$H_u^0$ and $S$ all experience a first order transition due to
their discontinuity at the weak scale. The magnitudes of the  VEVs of
the Higgs fields are plotted at the global minimum of the
potential for the given temperature. There is an abrupt change in
the $S_i$ VEVs at temperatures around 400--450 GeV because
 another local minimum becomes the global one\footnote{In a real transition
 the fields would most likely remain in the same minimum, but that 
 would have little effect on our conclusions. Detailed investigations of issues such as these
 are very model dependent and beyond the scope of this investigation.}.

After the \upr\ symmetry is broken, $S_1$, $S_2$, and $S_3$ have
non-zero VEVs of order 2 TeV, while the VEVs for $H_d^0$
and $H_u^0$ are still zero. The supersymmetry breaking soft terms
$m_{S S_1}^2 S S_1$ and $m_{S S_2}^2 S S_2$ generate a linear term
for $S$ in the potential after the \upr\ symmetry breaking, which
induces a non-zero VEV for $S$. At high temperature,
  $\langle S \rangle$ is tiny because $m_{S S_1}^2$ and $m_{S S_2}^2$
are small  and the temperature dependent corrections to the $S$
mass are large. At lower temperature, $\langle S
\rangle$ becomes
 larger, as can be seen from Figure~\ref{vevtt}.
 At the EWPT
the term $A_h h S H_u H_d$ induces a first order phase
transition in the VEV of $S$ as well.

The first order EWPT can be strong because of
the trilinear term $A_h h S H_u H_d$ in the tree-level potential.
Unlike in the MSSM~\cite{MCQCW, CMQSW1, CQSW2}, in which the phase
transition is associated with loop effects, there is no upper
bound on the lightest stop mass. This feature is independent of the
concrete embeddings of the sMSSM, though we are using an effective 
potential of finite temperature in Model I as an illustration. 
This feature is also shared by the NMSSM~\cite{Pietroni, ADFM, SJHMGS, Cline:2009sn}, the nMSSM~\cite{nmssm1,
nmssm2}, the $U(1)'$ models with no secluded sector~\cite{Ham:2007wc}, and the singlet extensions of the SM~\cite{Profumo:2007wc}.

\begin{table}[t]
\caption{$v(T_c)/T_c $ versus the stop soft mass squares
$m^2_{\tilde Q_3}$, $m^2_{\tilde u_R^3}$ and $A_h$, where the other
parameters are fixed to the values in case b (see Table~\ref{parameter}).
For simplicity, we assume
$m^2_{\tilde Q_3}=m^2_{\tilde u_R^3}$. The scales are different
for different points, but typically they are about $100$ GeV. The
slashes in the table mean that there is no first order EWPT. \label{table100}}
\begin{center}
\begin{tabular}{|c|c|c|c|c|c|}
\hline
$m^2_{\tilde Q_3}=m^2_{\tilde u_R^3} \backslash A_h$ & 2.0 & 3.0 & 4.0 & 5.0  \\
\hline
4.0 &$/$&1.29 &1.59& 2.06\\ \hline
8.0 &0.91&1.02&1.21&1.50 \\ \hline
12.0 &$/$&0.56& 1.03&1.22 \\ \hline
16.0 &$/$&$/$&$/$&1.37 \\ \hline
\end{tabular}
\end{center}
\end{table}

In Table~\ref{table100}, we calculate $v(T_c)/T_c $ versus the
stop soft mass square $m^2_{\tilde Q_3}$, $m^2_{\tilde u_R^3}$ and
$A_h$ while the other parameters are fixed to the values given by
Table~\ref{parameter} with the choice $m_{H_d^0}^2=9.0$. The
scales are different for different points, but typically they are
about $100$ GeV. 
As expected, when
$A_h$ decreases (with fixed $m^2_{\tilde Q_3}$ and $m^2_{\tilde
u_R^3}$), $v(T_c)/T_c $ decreases. The first order EWPT is not
strong enough when $A_h\leq 2 - 3$. This table also displays the
dependence of $v(T_c)/T_c $ on $m^2_{\tilde Q_3}$ and $m^2_{\tilde
u_R^3}$. For fixed $A_h$, the strength of the first order EWPT usually 
becomes weaker as $m^2_{\tilde Q_3}$ increases. Unlike the MSSM, even for heavy stop soft masses
(or physical masses) one can still have a strong enough first order
EWPT. 

For simplicity, we did not turn on any $CP$ phases for the discussion 
in this section. We will see later that the results obtained in this section 
are insensitive to these phases.

%
%

%
%
%
%
%
%
%
\section{Generation of $CP$ Violation}

\label{Generation of Large $CP$ Violation}

\subsection{$CP$ Violation at Tree Level}

One nice feature of 
the sMSSM is that both ECPV and SCPV can happen in the Higgs sector at tree level~\cite{letter}. 
This model involves six complex neutral fields. To study
the $CP$-violation, we define
\begin{eqnarray}
H_d^0 \equiv |H_d^0| e^ {i \theta_1} ~,~ H_u^0 \equiv |H_u^0| e^
{i \theta_2} ~,~ S \equiv |S| e^ {i\alpha} ~,~ \nonumber \\S_1
\equiv |S_1| e^ {i\alpha_1} ~,~ S_2 \equiv |S_2| e^ {i\alpha_2}
~,~ S_3 \equiv |S_3| e^ {i\alpha_3}~.~\, \label{Hphase}
\end{eqnarray}
Among these six phase degrees of freedom, only four of them, which
are given by
\begin{eqnarray}
\beta_1=\alpha+\alpha_1~,~\beta_2=\alpha+\alpha_2,\nonumber\\
\beta_3=\alpha+\theta_1+\theta_2~,~\beta_4=\alpha_1+\alpha_2+\alpha_3~.~\,
\label{beta definition}
\end{eqnarray}
are invariant under the $U(1)_Y$ and $U(1)'$ gauge transformations.
The two remaining degrees
of freedom
\begin{eqnarray}
\Sigma_i Y_{\phi_i} \theta_{\phi_i} ~,~ \Sigma_i Q'_{\phi_i}
\theta_{\phi_i}~,~\,
\end{eqnarray}
correspond to Goldstone bosons associated with the two
spontaneously broken gauge symmetries, and decouple from the Higgs
sector. 
Here, $Y_{\phi_i}$ and $ Q'_{\phi_i}$ are the $U(1)_Y$
and $U(1)'$ charges for $\phi_i$, $\theta_{\phi_i}$ is the phase
of $\phi_i$, and the index $``\phi_i"$ runs over all six complex
Higgs fields. 
The complex phases of the Higgs fields
$\theta_{\phi_i}$ are linear combinations of the
gauge-invariant and gauge-dependent components. 
Of course, physical consequences only depend on
the gauge-invariant components.

There are five supersymmetry breaking complex
parameters in the tree-level Higgs potential
Eq (\ref{VFpotential}-\ref{vsoftH}), $A_h h$, $A_{\lambda}
\lambda$, $m_{S S_1}^2$, $m_{S S_2}^2$ and $m_{S_1 S_2}^2$. 
Four of the five complex phases can be resolved by $\beta_{1,2,3,4}$.
Without loss of generality, we assume that
\begin{eqnarray}
m_{S_1 S_2}^2 &=& |m_{S_1 S_2}^2| e^{i\gamma} ~.~\,
\end{eqnarray}
is complex and the other parameters are real and
positive.
If $\gamma$ is not equal to zero or $\pi$, there will be ECPV at the tree level. Unlike the MSSM, SCPV can also occur at
tree level in this model. Let us rewrite the supersymmetry
breaking soft terms for the  neutral Higgs bosons as
\begin{eqnarray}
V_{soft}^{H} &=& m_{H_d^0}^2 |H_d^0|^2 + m_{H_u^0}^2 |H_u^0|^2 +
m_S^2 |S|^2 + \sum_{i=1}^3 m_{S_i}^2 |S_i|^2 \nonumber\\&&
 -2 A_h h |S| |H_d^0| |H_u^0| \cos \beta_3 - 2 A_{\lambda} \lambda |S_1| |S_2| |S_3| \cos
\beta_4 \nonumber\\
&& - 2 m_{S S_1}^2 |S| |S_1| \cos \beta_1 - 2
m_{S S_2}^2 |S| |S_2| \cos \beta_2 \nonumber\\
&& - 2 |m_{S_1 S_2}^2| |S_1| |S_2| \cos (-\beta_1 +
\beta_2+\gamma). \label{vphase}
\end{eqnarray}
It is easy to see that a mildly dominant $m_{S_1S_2}^2$ soft
term will make  $\beta_1 {\rm \ or \ } \beta_2 \neq 0, \pi$ for
$\gamma \neq 0, \pi$ and hence lead to SCPV. 
Meanwhile, the strength of SCPV can be controlled by the explicit $CP$ phase $\gamma$.

\subsection{$CP$ Violation at Finite Temperature}

\label{CP subsection2}

Explicit $CP$ phases from soft parameters beyond the Higgs sector could also contribute to EWBG. 
However, it is advantageous to turn off these phases and use the spontaneous $CP$ phases as the source for baryogenesis. 
To implement the EWPT, we need to include radiative corrections to the Higgs potential, as discussed in Section~\ref{effective potential}. At finite temperature there may coexist 
multiple vacua, in which spontaneous $CP$ phases usually have different values. If  the phase transition between two vacua is 
realized through bubble nucleation, as in a first order EWPT, these $CP$ phases are space-dependent as one crosses the bubble wall.  In the next sections we will see that 
the generated baryon asymmetry depends on these variations.

To make this more explicit, it is convenient to write the particle interactions during EWBG using the individual Higgs phases.  We define the two gauge-dependent phases as  $A$ and $B$, given by
\begin{eqnarray}
Y_{H_d^0} \theta_1 + Y_{H_u^0} \theta_2 + Y_{S} \alpha +
\Sigma_i Y_{S_i} \alpha_i=A  ~,~\, \nonumber \\
Q'_{H_d} \theta_1
+ Q'_{H_u} \theta_2 + Q'_{S} \alpha + \Sigma_i Q'_{S_i} \alpha_i=B
~.~\, \label{gauge fixing}
\end{eqnarray}
Inverting Eq.~(\ref{beta definition}) and Eq.~(\ref{gauge fixing}),
we obtain
\begin{eqnarray}
\theta_1 &=& -{\frac{1}{5}} \beta1 - {\frac{1}{5}} \beta_2 +
{\frac{7}{15}} \beta_3 + {\frac{2}{15}} \beta_4 - {\frac{11}{25}}A
- {\frac{4}{15}}B ~,~\nonumber \\
\theta_2 &=& -{\frac{1}{5}}
\beta1 - {\frac{1}{5}} \beta_2 + {\frac{7}{15}} \beta_3 +
{\frac{2}{15}} \beta_4 + {\frac{14}{25}}A
- {\frac{4}{15}}B ~,~ \nonumber \\
\alpha &=& {\frac{2}{5}} \beta1 + {\frac{2}{5}} \beta_2 +
{\frac{1}{15}} \beta_3 - {\frac{4}{15}} \beta_4 - {\frac{3}{25}}A
+ {\frac{8}{15}}B ~,~ \nonumber\\
\alpha_1 &=& {\frac{3}{5}} \beta1 - {\frac{2}{5}} \beta_2 -
{\frac{1}{15}} \beta_3 + {\frac{4}{15}} \beta_4 + {\frac{3}{25}}A
- {\frac{8}{15}}B  ~,~ \nonumber\\
\alpha_2 &=& -{\frac{2}{5}} \beta1 + {\frac{3}{5}} \beta_2 -
{\frac{1}{15}} \beta_3 + {\frac{4}{15}} \beta_4 +
{\frac{3}{25}}A - {\frac{8}{15}}B  ~,~ \nonumber \\
\alpha_3 &=& -{\frac{1}{5}} \beta1 - {\frac{1}{5}} \beta_2 +
{\frac{2}{15}} \beta_3 + {\frac{7}{15}} \beta_4 - {\frac{6}{25}}A
+ {\frac{16}{15}}B  ~,~\, \label{gauge eigenstates}
\end{eqnarray}
where we have used the rescaled $U(1)_Y$ and $U(1)'$ charges for the Higgs
particles in Model I
\begin{eqnarray}
Y_{H_d^0}=-Y_{H_u^0}=-1~,~
Y_S=Y_{S_1}=Y_{S_2}=Y_{S_3}=0
~;~\, \\
Q'_{H_d}={-7 \over {20}}~,~Q'_{H_u}={1 \over 10}~,~
Q'_S=-Q'_{S_1}=-Q'_{S_2}={1 \over 2}Q'_{S_3}={1\over 4} ~.~\, \label{charges}
\end{eqnarray}
Because $A$ and $B$ correspond to the Goldstone bosons of the $U(1)_Y$
and $U(1)'$ breaking, they can be rotated away by gauge
transformations. 

The variations of the Higgs phases crossing the bubble wall are given by 
\begin{eqnarray}
\Delta \langle \theta_{\phi_i} \rangle=\langle \theta_{\phi_i} \rangle_t-\langle \theta_{\phi_i} \rangle_f,
\end{eqnarray}
where $\langle \theta_{\phi_i} \rangle_{f,t}$ are the values of the Higgs phases in the false and true vacua, respectively. More explicitly, 
\begin{eqnarray} \Delta \theta_1&=&\Delta
\theta_2=-{\frac{1}{5}} \Delta \beta_1 - {\frac{1}{5}} \Delta
\beta_2 + {\frac{7}{15}} \Delta \beta_3 +
{\frac{2}{15}} \Delta \beta_4 \nonumber \\
\Delta\alpha &=& {\frac{2}{5}} \Delta\beta1 + {\frac{2}{5}}
\Delta\beta_2 + {\frac{1}{15}}\Delta \beta_3 - {\frac{4}{15}}
\Delta\beta_4 ~,~ \nonumber\\
\Delta\alpha_1 &=& {\frac{3}{5}} \Delta\beta1 - {\frac{2}{5}}
\Delta\beta_2 - {\frac{1}{15}} \Delta\beta_3 +
{\frac{4}{15}}\Delta \beta_4  ~,~ \nonumber\\
\Delta\alpha_2 &=& -{\frac{2}{5}} \Delta\beta1 + {\frac{3}{5}}
\Delta\beta_2 -
{\frac{1}{15}} \Delta\beta_3 + {\frac{4}{15}} \Delta\beta_4 ~,~ \nonumber \\
\Delta\alpha_3 &=& -{\frac{1}{5}} \Delta\beta1 - {\frac{1}{5}}
\Delta\beta_2 + {\frac{2}{15}}\Delta \beta_3 + {\frac{7}{15}}
\Delta\beta_4. \label{CP phase change}
\end{eqnarray}
These quantities are not uniquely defined, due to the uncertainties in the winding numbers of the
$\langle \beta_i \rangle_{f,t}$. If there is a large difference between the winding numbers of 
$\langle \beta_i \rangle_{f}$ and $\langle \beta_i \rangle_{t}$, then $\Delta \theta_{\phi_i}$
can be large, and hence the variation of $\theta_{\phi_i}$ in the bubble wall is also large. 
However, the nucleation rate for a phase transition between two such vacua is generically small~\cite{letter},
so we will require $|\beta_i| < \pi$.

Let us consider how $\Delta \theta_{\phi_i}$ is affected by
the behavior of the related field variables in the wall. For a
qualitative discussion, we neglect all loop as well as finite
temperature corrections, even though they are important for the
complete physical picture. Due to $h\gg \lambda$, the $|S_i|$ are
fixed to almost the same large VEVs in both the symmetric and
broken phases, which leads to $\langle \beta_4 \rangle_{f,t} = 0$ and $\Delta\beta_4=0$.
The change in $\beta_3$, however, is subtle. For the unbroken
phase, $v_1=v_2=0$ and hence $\beta_3$ is not fixed. But for non-local EWBG 
the relevant scattering processes mainly occur in the bubble wall where $v_{1,2}$
do not vanish, so we can understand $\beta_3$ in the unbroken phase as
its asymptotic value as $v_1$ and $v_2$ approach zero. 
From the tree-level potential Eq.~(\ref{vphase}), it is easy to see that the
non-vanishing Higgs VEVs lead to $\beta_3 =
0$ inside the bubble wall except the boundary region between the wall and the false vacuum.
We simply assume $\Delta\beta_3=0$. The phase changes of the Higgs fields therefore
are mainly due to $\Delta \beta_1$ and $\Delta \beta_2$. To
illustrate how to obtain large $\Delta \beta_1$ and $\Delta \beta_2$,
let us consider their first order differential equations from the
tree-level potential Eq.~(\ref{vphase}),
\begin{eqnarray}
\left(m_{SS_1}^2|S||S_1| \cos \beta_1 + |m_{S_1S_2}^2| |S_1||S_2|
\cos(-\beta_1+\beta_2+\gamma)\right)
\Delta \beta_1 \nonumber\\
-|m_{S_1S_2}^2| |S_1||S_2|\cos(-\beta_1+\beta_2+\gamma) \Delta
\beta_2 = -m_{SS_1}^2|S_1|\sin \beta_1 \Delta |S| \nonumber \\
\left(m_{SS_2}^2|S||S_2|\cos\beta_2+|m_{S_1S_2}^2| |S_1||S_2|\cos(-\beta_1+\beta_2+\gamma)\right)
\Delta \beta_2 \nonumber\\
-|m_{S_1S_2}^2| |S_1||S_2|\cos(-\beta_1+\beta_2+\gamma) \Delta
\beta_1 =-m_{SS_2}^2|S_2|\sin\beta_2\Delta |S|.     \label{variation}
\end{eqnarray}
Obviously, large $\Delta \theta_{\phi_i}$ or large $\Delta
\beta_1$ and $\Delta \beta_2$ requires large $\Delta |S|$, and
also depends significantly on $\gamma$.

\begin{figure}
\begin{center}
\includegraphics[height=3.2 in]{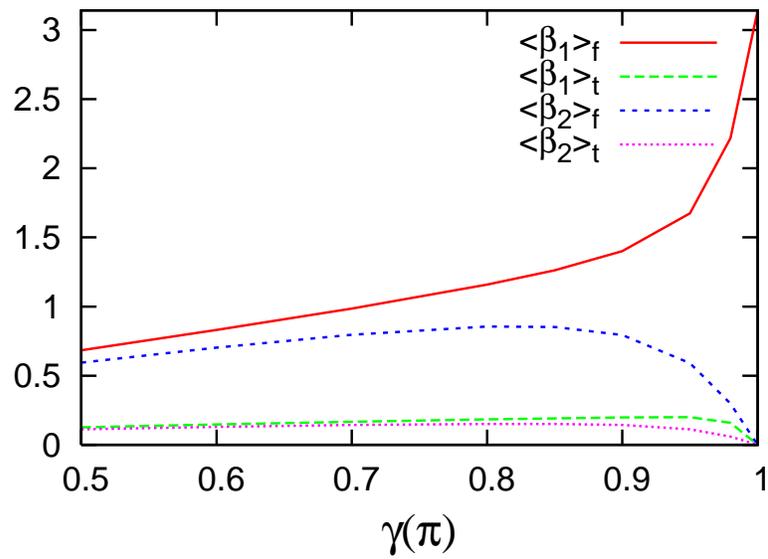}
\end{center}
\caption{ $\langle \beta_{1,2}\rangle_{f,t}$ vs. $\gamma (\pi)$, in case b. } 
\label{CPphase1}
\end{figure}

\begin{figure}
\begin{center}
\includegraphics[height=3.2 in]{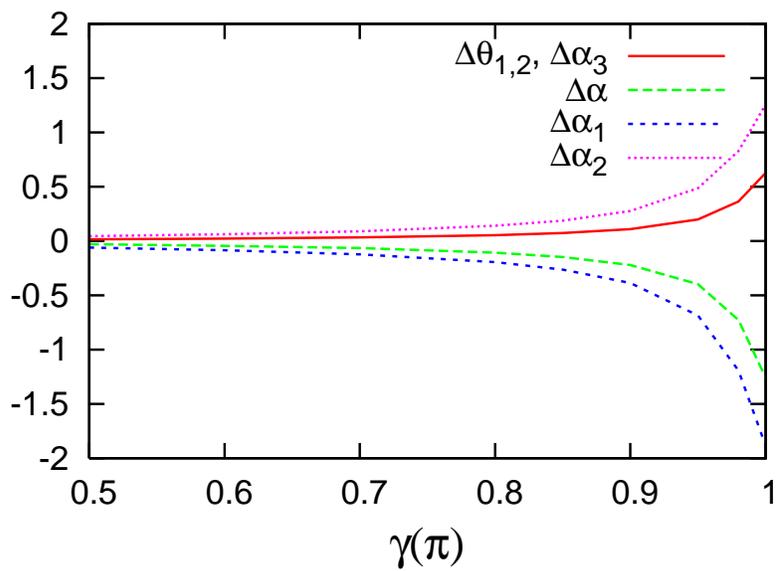}
\end{center}
\caption{ $\Delta \theta_{\phi_i}$ vs. $\gamma (\pi)$, in case b. } 
\label{CPphase2}
\end{figure}

\begin{figure}
\begin{center}
\includegraphics[height=3.2 in]{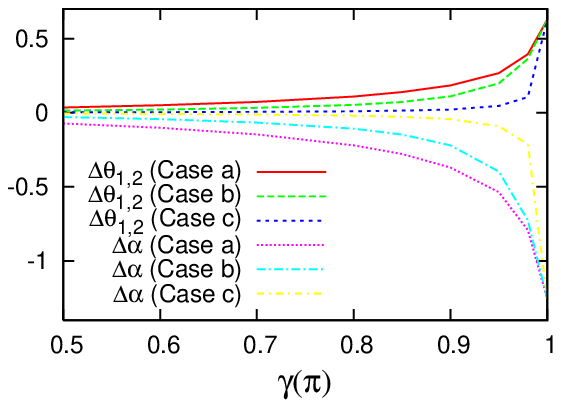}
\end{center}
\caption{ $\Delta \theta_{1,2}$, $\Delta_\alpha$ vs. $\gamma (\pi)$, in cases a, b and c. } 
\label{CPphase}
\end{figure}

Let us consider a parameter subspace:
$100|m_{S_1S_2}|^2 > m_{SS_1}^2 \sim 0.1 m_{SS_2}^2 > 10
|m_{S_1S_2}|^2$. Assume that 
$\langle S \rangle / \langle S_{1,2,3}\rangle$ is small ($\sim 0.01$) in the false vacuum and large ($\sim
0.1$) in the true vacuum. In the broken phase, the
$m_{SS_1}^2$ and $m_{SS_2}^2$ terms will dominate over the
$m_{S_1S_2}^2$ term in the tree-level potential and thus favor
small $\beta_1$ and $\beta_2$.
In the symmetric phase, the dominant terms
change to $m_{SS_2}^2$ and $m_{S_1S_2}^2$ due to the sharp
decrease of the VEV of $S$. For $\gamma \sim \pi$, $\beta_1$ is 
pushed to a large value in this region. A large $\Delta\beta_1$ is then generated. 
We will follow this line to generate large $\Delta \theta_{\phi_i}$ in our analyses.
In Figure~\ref{CPphase1}, we illustrate the $\gamma$-dependence 
of $\beta_1$ and $\beta_2$, in the false and true vacua respectively. 
The relevant $\gamma$-dependence of $\Delta \theta_{\phi_i}$ is illustrated in
Figure~\ref{CPphase2}. For both of them, the parameter values are 
from case b (see Table~\ref{parameter}).

Besides the baryon asymmetry, it would be desirable to also
account for the CDM.
In the scenario of the lightest neutralino $\chi_1^0$ as a CDM candidate, the CDM generation has a
strong dependence on the mass ($m_{\chi_1^0}$) and composition of $\chi_1^0$. 
Since $m_{\chi_1^0}$ is sensitive to the soft mass terms of $H_d^0$,
$H_u^0$ and $S$ due to their roles in determining $\tan\beta$,
we will consider EWBG with respect to $\gamma$ in three cases first (cases a, b and c) which are specified 
by different $m_{H_d^0}^2$ values (see Table~\ref{parameter}), and then discuss the continuous 
dependence of EWBG and the $\chi_1^0$ relic density on $m_{H_d^0}^2$ for a given $\gamma$ value. 
In Figure~\ref{CPphase} we show the $\gamma$ dependence of $\Delta \theta_1=\Delta \theta_2$ and $\Delta \alpha$ 
in cases a, b and c (since the baryon asymmetry generation mainly couples to these three phases). 
From this figure we see that for a given $\gamma$ value a larger $m_{H_d^0}^2$ value gives a smaller phase variation crossing the wall. 
This is because a large $m_{H_d^0}^2$ value make $|S|$ small in both vacua, which further leads to a small 
$\Delta |S|$ (or small $\Delta \beta_i$ due to Eq.~(\ref{variation})). As expected from the previous discussion, the large $|\Delta\theta_{\phi_i}|$ appears in the 
large $\gamma$ region.  For $\gamma=\pi$ there exists a 
$Z_2$ discrete symmetry in the neutral Higgs potential which might cause a dilution between 
matter and anti-matter bubbles for SCPV-induced EWBG (the explicit $CP$ breaking from the fermion sector 
avoids problems with cosmological domain walls~\cite{Vilenkin:1984ib}).  The vacua degeneracy is lifted  
as long as $\gamma$ deviates from  $\pi$. Because the nucleation rates of these two vacua are exponentially biased 
by the difference of their energy densities, a slight shift from $\pi$ for $\gamma$ can greatly diminish the dilution. In this article therefore we will 
choose $\pi/2 \le \gamma < \pi$.

\subsection{New Contributions to Electric Dipole Moments}

In this subsection, let us discuss possible new contributions to the EDMs in the SCPV-induced baryogenesis (also see~\cite{letter}). 
According to the discussions in the previous subsection, no $CP$ violation occurs at 
zero temperature for $\gamma=0, \pi$, so there is no EDM problem.

For $\gamma \neq 0, \pi$, the Higgs fields may obtain nonzero phases at zero temperature. However, their contributions 
to the EDMs are highly suppressed. Among the
four gauge independent phases, $\beta_3$ and $\beta_4$ are zero
due to the $A_h$ and $A_{\lambda}$ terms, so there will be no phases
entering the $h$ and $A_h$ terms after we do the field redefinitions. The
only phases are associated with the $m_{S S_1}^2$
$m_{S S_2}^2$ and $m_{S_1 S_2}^2$ terms. (The phases will enter the
fermion Yukawa couplings also, but they only lead to overall phases in the mass matrices which do not affect the CKM matrix.) 
Therefore, the $CP$ violation will only affect the Higgs sector.

The Feynman diagrams contributing to the electron and quark EDMs
which are mediated by the Higgs fields involve two vertices
containing the Yukawa couplings, and include a mass insertion
on the fermion line to flip the chirality. The EDMs will therefore be
proportional to the cube of Yukawa coupling, making it very
small. The EDMs are also proportional to the $CP$ violation term which comes from the secluded sector and couples very weakly to the ordinary fermions.
In particular, if any of the three soft mass terms  $m_{S S_1}^2$ $m_{S S_2}^2$
and $m_{S_1 S_2}^2$ vanish, there will be no tree-level $CP$
violation. Also, for the parameters that we consider, the imaginary part of
$m_{S_1 S_2}^2$ is very small. 
Combining
all these effects, new EDM contributions from the Higgs sector are
very small. Our numerical estimate shows that they are 6 to 7
orders smaller than the experimental upper limit.

\section{Wall Thickness and Velocity}

In the early Universe the first order EWPT is realized by
nucleating bubbles of the broken phase. To calculate the baryon
asymmetry, understanding the dynamical behavior of these bubbles,
such as the nucleation rate, expansion velocity, and bubble wall profile,
is important but analytically difficult. In this section, we will focus
on the wall profile and wall velocity.

\subsection{Wall Thickness}

\label{Wall Thickness}

The kinetic Lagrangian for the complex scalar fields can be
written as
\begin{eqnarray}  
{\cal L} \sim \sum_i \left( \frac{1}{2} \partial_{\mu} \phi_i
\partial^{\mu} \phi_i + \frac {\phi_i^2}{2}\partial_{\mu} \theta_{\phi_i}
\partial^{\mu} \theta_{\phi_i} \right )~,~\, \label{Eaction}
\end{eqnarray}
where $\phi_i$ and $\theta_{\phi_i}$ collectively denote the magnitudes and
phases of the Higgs fields, respectively. For the unitary gauge, we are left with 
10 dynamical equations (with respect to 6 magnitudes and 4 gauge-invariant phase variables).
Once the stationary solutions to these 10 equations are found, we can obtain the
relevant physical information on the bubble wall, such as the wall
thickness and velocity. However, these classical field equations
are extremely complicated, 
so we have to solve them numerically.

A useful numerical method was devised in~\cite{MQS,P.John,CMS}.
The basic idea~\cite{MQS} is that to solve the classical field equations one has to
find the field configuration for which
\begin{eqnarray}
S_A = \int_{-\infty}^{+\infty}dz\left[\sum_iE_{\phi_i}(z)^2
+\frac{1}{T_c^2}\sum_j E_{\beta_j}(z)^2\right] \equiv 0 ~,~\, \label{action}
\end{eqnarray}
where $E_{\phi_i}(z)$ and $E_{\beta_j}(z)$ are the
relevant field equations under the planar wall approximation. 
($E_{\beta_j}(z)$ is rescaled by a factor $\frac{1}{T_c}$, so that $E^i_{magnitude}(z)$ and $E^j_{phase}(z)$
have the same dimensions.) This can be achieved by searching for the absolute
minimum of the action $S_A$. We need to find an appropriate
ansatz satisfying the boundary conditions at infinity as a first step. The kink
ansatz is especially suitable since 
it satisfies the boundary conditions very well and has a smooth, gradually changing
behavior as one moves from negative infinity to positive infinity.
Therefore, we will take the kink ansatz for the independent
field variables
\begin{eqnarray}
\phi_i(z) &=&
\frac{\langle\phi_i\rangle_t+\langle\phi_i\rangle_f}{2} +
\frac{\langle\phi_i\rangle_t-\langle\phi_i\rangle_f}{2}
\tanh(\frac{z}{\delta}) ~,~\, \nonumber \\
\beta_j(z) &=&
\frac{\langle\beta_j\rangle_t+\langle\beta_j\rangle_f}{2} +
\frac{\langle\beta_j\rangle_t-\langle\beta_j\rangle_f}{2}
\tanh(\frac{z}{\delta}) ~,~ j=1,2,4~\, \label{Kink Ansatz}
\end{eqnarray}
where the index ``$i$" runs over the six complex neutral Higgs
fields: $H_d^0$, $H_u^0$, $S$, $S_1$, $S_2$, $S_3$, and  ``t" and
``f" denote the VEVs in the true and false vacua, respectively.
In our convention the true vacuum is located 
at $z\to \infty$ and the false one is located at $z\to -\infty$. 
For simplicity, we assume that there are no off-sets which shift
the fields against each other.

However, this assumption does not hold for $\beta_3$. From the $A_h$ term 
in Eq.~(\ref{vphase}), it is easy to see that, as $\langle H_d^0\rangle$ 
and $\langle H_u^0\rangle$ become non-trivial crossing the bubble wall, 
$\langle \beta_3 \rangle$ will be suppressed to zero quickly. 
So a good ansatz for $\beta_3$ should be
\begin{eqnarray}
\beta_3(z)={\langle \beta_3 \rangle_f \over
2}[1-\tanh({{z+a}\over{\delta}})]~.~
\end{eqnarray}
This is an anti-kink function and $a\geq \delta$. For simplicity, we set
$a=\delta$, so all kink ansatzs are mediated by only
one parameter -- the wall thickness $\delta$. This discussion does not 
contradict the assumption $\Delta \beta_3=\beta(\infty)-\beta(-\infty)=0$
in Subsection~\ref{CP subsection2}, since there 
we are analyzing the phase change crossing the wall while $\beta_3 \sim 0$ 
inside the wall.

\begin{figure}
\begin{center}
\includegraphics[height=3.2 in]{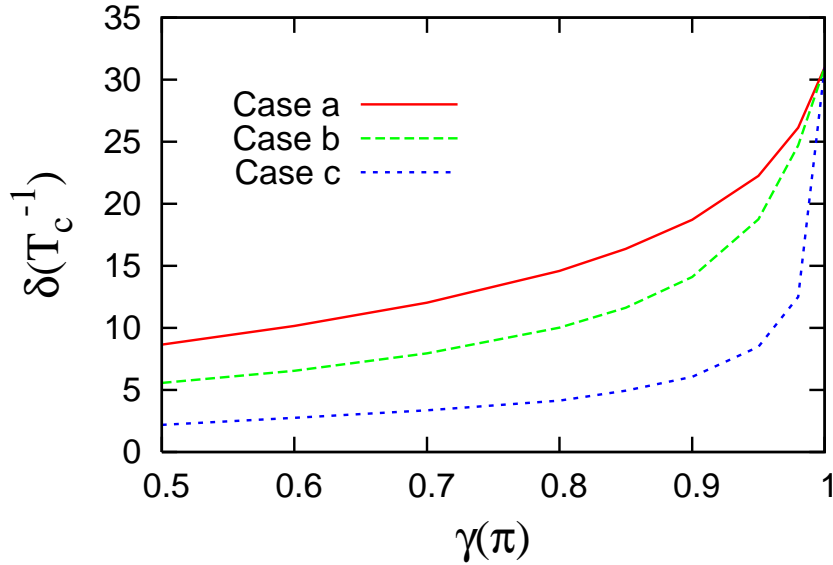}
\end{center}
\caption{Wall thickness $\delta$ vs. $\gamma (\pi)$.}
\label{wallthickness}
\end{figure}

Figure~\ref{wallthickness} describes the dependence of the wall
thickness on $\gamma$ ($\frac{\pi}{2} \le \gamma < \pi$) by
minimizing the action $S_A$ over $\delta$ for the parameter
sets in Table~\ref{parameter}. Again, the three curves
correspond to the three different $m^2_{H_d^0}$ values: 2, 4 and
6. These plots tell us that the wall thickness becomes larger as
$\gamma$ increases and reaches its peak value around $\gamma = \pi$.
In the action $S_A$ the kinetic
terms and potential slope terms are proportional to $\delta^{-3}$
and $\delta$, respectively. The wall thickness is a result of the balance
between them. According to
the discussions in Section~\ref{Generation of Large $CP$
Violation}, large $\gamma$ induces large magnitudes for $\Delta\beta_j$ or 
$\Delta\theta_{\phi_i}$, which can increase the weight of the
kinetic terms in $S_A$. The wall thickness therefore is an approximately 
monotonic increasing function with respect to $\gamma$. Numerical
simulation shows that the wall thickness varies in the range $(3 -
30) T_c^{-1}$.

\subsection{Wall Velocity}
The velocity of the bubble wall also plays a crucial role in EWBG. 
As a technical requirement,
the wall velocity cannot be smaller than 0.01. Otherwise, the
effect of EW sphalerons and Higgs-mediated (in the thin wall
regime) decay processes cannot be neglected while the $CP$-violating
current is propagating in the false vacuum, and the new relaxation
effect of EW sphalerons will appear before the wall recaptures
the left-chiral particles composing the flux. 

The wall velocity of a new-born bubble is mainly mediated by three kinds of
forces: 1) the driving force ($F_d$) responsible for the bubbles'
expansion; 2) the shrinking force ($F_s$) due to the surface
tension of the bubbles ($T_s$); and 3) the friction coming from
the plasma ($F_f$). Though some progress has been reported in the MSSM recently~\cite{MJS},
a systematic study on the dynamics of the bubble wall is still absent in the sMSSM.
In the following, we will argue that the wall velocity in these models 
should be comparable to or smaller than that in MSSM or NMSSM under the thin wall approximation~\cite{Linde}\footnote{This ``thin wall''
is different from the one defined in next section in  
that the comparison scale is the bubble radius rather than the
mean free path of the scattered particles.}. 

Consider two bubbles from the sMSSM and MSSM, respectively, and assume that the energy density differences across the wall, $\varepsilon=V_{out}-V_{in}$, are
the same for both bubbles. In this section, a prime will be
used to denote the quantities from the sMSSM. The driving force per area acting on the walls is
$F_d=\varepsilon$ and the shrinking force is $F_s=2T_s/r$, so we can
approximately estimate the critical radii for both kinds of
bubbles by employing the equation $F_d=F_s$, which gives
\begin{eqnarray}
r_c=2T_s/\varepsilon,   \\
r_c'=2T_s'/\varepsilon.
\end{eqnarray}
For $T'_s=aT_s$, we have $r'_c=ar_c$. Here $a \gg 1$ since the
large VEVs of the Higgs fields in the secluded sector will contribute
significantly to $T'_s$ according to the second term of
Eq.~(\ref{Eaction}).

Now suppose that both bubbles sweep a distance $\Delta r$, {\em i.e.},
the bubble radii respectively increase to $r_c+\Delta r$ and
$r'_c+\Delta r$, and that the wall tensions $T_s$ and $T'_s$ are left
invariant during this process; then the shrinking forces due to
the surface tension will become smaller. The net force for both bubbles are changed
from zero to
\begin{eqnarray}
\Delta F=F_d-F_s=\varepsilon-{{2T_s}\over{r_c+\Delta r}}, \\
\Delta F'=F'_d-F'_s=\varepsilon-{{2T'_s}\over{r'_c+\Delta r}}, 
\end{eqnarray}
and
\begin{eqnarray}
\Delta F- \Delta F' = 2T_s({{-1} \over{r_c+\Delta r}}+{1
\over{r_c+\Delta r/a}})>0.
\end{eqnarray}
That is, in the sMSSM the shrinking force decreases
more slowly if there is no other physics involved 
(note that the driving forces stay unchanged during
this process).

On the other hand, due to thermal scattering from particles in the
hot plasma, the bubble wall will suffer friction or thermal
pressure $F_f$ during its expansion, where $F_f$ is determined by the
thermal environment and the wall velocity relative to
the plasma. Now suppose that bubbles in both models expand at the
same velocities, microscopically this thermal pressure is mainly induced by 
momentum transfer of the scattered particles, particularly those with 
\begin{eqnarray}
m(\infty) \gg m(-\infty),  \frac{k_i.k_f}{|k_i||k_f|} < 0.
\end{eqnarray}
Besides the ones in the MSSM, the sMSSM contains extra light 
degrees of freedom, e.g., the singlets in the Higgs sector and the additional exotic 
particle species. We may well expect that in the sMSSM the thermal pressure is not smaller than
that in the MSSM given $T_c=T_c'$ and $v_w=v_w'$.  
Therefore, the wall
velocity in the sMSSM should be 
non-relativistic. In this article, we will set $v_w=0.05$.

\section{Non-local Electroweak Baryogenesis}

\label{NEWBG}

In EWBG scenarios, the non-local baryogenesis
or charge transport mechanism~\cite{CKN1} refers to the case where
particles have $CP$-violating interactions with a bubble wall.
These $CP$-violating interactions cause an asymmetry in some $CP$-odd
quantum numbers or transport charges other than $B$ carried
by particle currents into the false vacuum. In front of the
expanding bubble wall, where the thermal equilibrium is broken to
some extent, the asymmetry of transport charges induces 
a non-vanishing number density for left-chiral particles carrying EW charges. 
This asymmetry is subsequently converted by the $B+L$-violating EW
sphaleron processes, such as $t_{\overline{R}} t_{\overline{R}}
b_{\overline{R}} \tau_{\overline{R}} \leftrightarrow 0$ and
$t_{\overline{R}} b_{\overline{R}} b_{\overline{R}}
\nu_{\overline{R}} \leftrightarrow 0$, into an asymmetry in baryon
number. Some instant later the wall sweeps over the region. 
The generated net $n_B$ is then frozen in the bubble due to the
exponentially suppressed EW sphaleron effects.
There are two typical regimes relevant to non-local baryogenesis, 
depending on the particles under consideration: the thin wall regime for $\delta/\Delta < 1$ and
the thick wall region for $\delta/\Delta \geq 1$, where $\delta$ is the
wall thickness and $\Delta$ is the particle's mean free path or propagation length.

For the thin wall regime~\cite{JPT, Joyce:1994zn}, the physical
picture is relatively simple since we may neglect the damping
effect on the $CP$-violating source caused by the self-energy
thermal correction.
The reflectivity and
transmittivity of particle scattering can be calculated in 
quantum mechanics, where the fermions interacting with the
bubble wall are treated as free particles with space-dependent
masses, and the $CP$-violating source can be modeled as a
delta function current.

For the thick wall regime, the effects of additional decay
processes in the wall cannot be neglected, and hence modeling
the $CP$-violating source will be very complicated. Generally, two
parallel approaches are available: the semiclassical force method
\cite{JPT1,JPT2} and spontaneous baryogenesis~\cite{CKN2}. The
first one is based on the fact that, under the WKB approximation
($p\gg \delta^{-1}$), a $CP$-violating mass term in the
Lagrangian can induce a $CP$-violating shift in gradients of the
dispersion relation at first order. This $CP$-violating source
then appears in the diffusion equations, interpreted as one kind
of ``semi-classical" source. This approach has been applied to the
MSSM~\cite{HJS} and the non-minimal NMSSM~\cite{SJHMGS}. 
However, this approach neglects the contributions from the non-WKB region,
which may greatly enhance the produced baryon asymmetry. The
second approach is a little like the one used in the thin wall
regime: the $CP$-violating source is the particle current reflected
by the bubble wall, but the damping and multiple-scattering effects 
are included.

With $CP$-violating sources produced, the particles carrying 
the $CP$-odd transport charges will undergo a diffusion process outside the bubble. 
Dynamical baryogenesis requires a slight
departure from thermal equilibrium in front of the wall,
implying that not all particle ``decay'' processes ({\em i.e.}, those which affect the charge number density) 
are fast or adiabatic compared to the expanding velocity of the bubble wall. If
\begin{eqnarray}
\frac{D\Gamma}{v_w^2} > 1
\end{eqnarray}
with $\Gamma$ being the decay rate and $D$ the associated
diffusion constant, the processes will be considered as fast,
and vice versa. In our calculations, almost all processes included
in the diffusion equations are fast, otherwise they will not have
enough time to significantly affect the
charge number density. The only exception is the EW sphaleron
process. We assume that they are slow during the diffusion and can be neglected in
mediating the charge number density. We can therefore
deal with the diffusion of the particles carrying transport charges and the generation of
the net baryon number separately. We must then: (1) solve the diffusion
equations of the particles carrying the transport charges, and find 
the number density of all left-chiral particles $n_L$ which carry EW gauge charges; 
(2) insert $n_L$ into the baryon 
diffusion equation, to find the generated baryon asymmetry.
The description of the particle transport in front of the bubble
wall by the diffusion equations is well-justified because $v_w
\ll v_s$. Here $v_s=1/\sqrt{3}$ is the sound speed in the
plasma.

In the following, we will  study the roles in EWBG played by different particles in
the sMSSM. We will calculate the contributions from leptons in the thin wall regime, 
and the contributions from squarks, charginos and quarks in the
thick wall regime by assuming the spontaneous baryogenesis
mechanism.

\subsection{The Lepton Contribution -- Thin Wall Regime}
\label{thin wall regime}

In the supersymmetric extension of the SM, leptons generally have a mean
free path in the thermal plasma much larger than the bubble wall
thickness. To consider their contribution to EWBG, we
can work in the relatively simple thin wall regime~\cite{JPT}.In the thin wall regime, the scattering of particles by the bubble wall is described
in terms of freely propagating particles with space-dependent masses, with the Lagrangian
\begin{eqnarray}
{\cal L} \: =\: {\overline\psi}_Li\gamma^\mu\partial_\mu \psi_L +
{\overline\psi}_R i\gamma^\mu\partial_\mu \psi_R -
m(z){\overline\psi}_L\psi_R \: - \: m^*(z){\overline\psi}_R\psi_L
\: ~,~\, \label{lagrangian}
\end{eqnarray}
with the mass term determined by the Yukawa coupling. For charged leptons
it is 
\begin{eqnarray}
m_l(z) = y_l v_1(z)e^{i\theta_1(z)} ~.~\,
\label{cpmass1}
\end{eqnarray}
Since the Yukawa coupling $y_{\tau} \gg y_{e,\mu}$, we will only study the contributions from
$\tau$ leptons.

The $CP$-violating current in this case is produced by $\tau_R\rightarrow \tau_L$, where 
the $CP$-odd transport charge is axial $\tau$ number 
\begin{eqnarray}
{\mathcal Q}_{\tau_{L,R}}^{\tau} = {\rm Diag}(1,-1).
\end{eqnarray} 
Explicitly solving the Dirac equation and only
considering the contribution from the non-WKB momentum subspace
($k_i^z<\delta^{-1}$) as a conservative approximation, one obtains
the $CP$-odd reflection asymmetry~\cite{JPT, Joyce:1994zn}
\begin{eqnarray}
\Delta R_{\tau_R\rightarrow \tau_L} (k_i, k_f) &=& 4t_{\phi}
(1-t_{\phi}^2)\int_{-\infty}^{\infty} dz ~\cos(2k_i^zz) {\rm
Im}\left[m_{\tau}(z){{m_{\tau}(\infty)^*}\over{|m_{\tau}(\infty)|}}\right] \nonumber \\
&=&{-\sqrt{\pi}\over 2}
m_{\tau}(\infty)^2\delta \Delta \theta_1{e^{-{(\delta
k_i^z)}^2}\over k_i^z}, \label{ref asym}
\end{eqnarray}
where $t_{\phi}=\tanh\phi$ and
$\tanh(2\phi)=|m_{\tau}(\infty)|/|k_i^z|$. In the second line, the kink ansatz
defined by Eq.~(\ref{Kink Ansatz}) and the assumption of $\Delta\theta_1/2$ as  perturbative (it turns out that $\frac{1}{2} \Delta\theta_1
\simlt 0.3$ in the parameter region we are interested in) have been used.

The axial $\tau$ current $j_{\tau_L}$ induced by the
scattering of leptons is~\cite{JPT, Joyce:1994zn}
\begin{eqnarray}
j_{\tau_L}= \int_{k_i^z>m_{\tau}}{{d^3k_i}\over{(2\pi)^3}}{{k_f^z}
\over {E}}(f_f^\tau(k_i^z)-f_t^\tau(-k_i^z))\Delta
R_{\tau_R\rightarrow \tau_L} (k_i, k_f){\mathcal Q}_{\tau_L}^{\tau} ~.~\, \label{118}
\end{eqnarray}
Here $f_t$ and $f_f$ in Eq.~(\ref{118}) are the Fermi-Dirac
distributions in the wall frame for the particles localized in and
outside the bubble, respectively,  given by
\begin{eqnarray}
f^\tau(k_i^z)={{\gamma_w^2(1-v_w\frac{k_i^z}{E})}
\over\displaystyle {e^{\gamma_w(E-v_wk_i^z)/T_c}+1}}~.~\,
\label{114}
\end{eqnarray}
The chemical potential in the exponential factor is suppressed due
to the large value of the critical temperature $T_c$. Similarly,
due to the exponential suppression for $E>T_c$, we may only consider the contribution from the small
energy region $E<T_c$ and make a perturbative expansion with
respect to $\frac{E}{T_c}$, approximately yielding the formula for 
the injected $CP$-violating current
\begin{eqnarray}
j_{\tau_L}={{ -m_{\tau}(\infty)^2 v_w \delta \Delta\theta_1 }\over
{(2\pi)^2}}h(\delta,T_c)~,~\, \label{chiral current}
\end{eqnarray}
where
\begin{eqnarray}
h(\delta, T_c) = \int_0^{T_c} dk_{i\bot} \int_0^{1/\delta}
dk_i^z{{k_\bot }\over{\sqrt{k_\bot^2+(k_i^z)^2}}}
(({{\gamma_wk_i^z}\over{T_c}}-{{k_i^z}\over{\sqrt{k_\bot^2+(k_i^z)^2}}})
+ O[(\frac{E}{T_c})^3])  \label{R}
\end{eqnarray}
is a function of the wall thickness and critical temperature. 
The first term in Eq.~(\ref{R}) exactly leads to Eq.~(25) in Ref.~\cite{Joyce:1994zn}
up to an ansatz-dependent factor after $h(\delta,T_c)$ is
integrated out. The second term, which will suppress the
injected chiral flux, is neglected in Ref.~\cite{Joyce:1994zn}.
Note that the $CP$-violating current $j_{\tau_L}$ is proportional to
$m_{\tau}(\infty)^2$. Given that the associated calculations can
be applied to the other two families of leptons as well, we are
justified to only consider the $\tau$.

While propagating in the false vacuum,  for a wall velocity that is not too large 
the behavior of the scattered leptons can be approximately described by diffusion
processes. The injected current
will lead to a ``diffusion tail" of the particles in front of the
moving wall. If this diffusion tail or persistence length $\xi$ is
much larger than the width of the wall, it will be a good
approximation to describe the injected current as a delta-function-like source.  
While the current is propagating in the
false vacuum, leptons will undergo some Higgs-mediated or chiral
mixing decay processes which may have a negative effect on the
$CP$-violating current amplitude. A reasonable assumption is that
the characteristic decay time for leptons is much longer than the
time it takes for the wall to recapture them. We can then neglect
lepton decays or chiral mixing processes and only consider the
left-chiral leptons propagating in the false vacuum. Furthermore,
we also need to make an assumption that the EW sphaleron processes
are too slow to alter the density in front of the wall
significantly and thus can be neglected except for their role in
generating baryons. These two assumptions require
$\frac{D\Gamma}{v_w^2} < 1$ for the associated processes. For
$\tau$ leptons, they work well if $v_w
> 0.01$~\cite{JPT}. Neglecting these decay processes, the
diffusion equation in the bubble wall frame can be rewritten as~\cite{JPT}
\begin{eqnarray}
D_{\tau_L} n''_{\tau_L}(z)-v_w n'_{\tau_L}(z)+J_{\tau_L}=0
~,~\,    
\end{eqnarray}
with
\begin{eqnarray}
n'_{\tau_L} &=& \partial n_{\tau_L}(z)/\partial z~,~ \nonumber \\
J_{\tau_L} &=& -\xi_{\tau_L} j_{\tau_L} \delta'(z) ~,~
\label{propagating equation}
\end{eqnarray}
where the persistence length $\xi_{\tau_L}$ is given by $\xi_{\tau_L}\sim
6D_{\tau_L}\langle v_{\tau_L}\rangle$~\cite{JPT}, with
$D_{\tau_L} \sim 100 ~T_c^{-1}$ being the related diffusion constant~\cite{JPT} and $\langle v_{\tau_L} \rangle$
being the average velocity of the injected current relative to the
wall; $\delta'(z) = \partial \delta (z)/\partial z$ and $z$ is related to the coordinates in 
the plasma frame by $z=r + v_w t$. Solving
Eq.~(\ref{propagating equation}) by combining with the boundary
conditions
\begin{eqnarray}
\lim_{|z|\rightarrow\infty}n_{\tau_L}(z)=0~,~\,
\end{eqnarray}
one obtains
\begin{equation}
n_{\tau_L}(z) = \left\{ \begin{array}{ll}
            \mathcal{C}_{\tau_L} e^{{v_w \over D_{\tau_L}}(z)}  & \ \ \ \ z<0 \\
          0 & \ \ \ \ z>0
         \end{array} \right. \ ~,~\,
\end{equation}
with
\begin{eqnarray}
\mathcal{C}_{\tau_L}=6\langle v_{\tau_L} \rangle j_{\tau_L}.
\end{eqnarray}

The baryon generation is described by the diffusion equation
\begin{eqnarray}
D_L n_B''-v_wn_B'-n_f\Gamma_{ws}n_L=0. \label{106}
\end{eqnarray}
Here $n_L$ is the net number density of left-chiral particles carrying
EW charges. Only such particles are involved in the EW sphaleron processes
and hence able to source the baryon generation. $D_L$ is effective diffusion constant. 
In the thin wall region, the main contribution to $n_L$
is from the scattering of $\tau$ leptons, so we have
$n_L=n_{\tau_L}$ and $D_L=D_{\tau_L}$. Under the approximation ($\Theta(-z)$ is step function)
\begin{eqnarray}
\Gamma_{ws}(z)&=&\Theta(-z)\Gamma_{ws}(-\infty) \nonumber\\
&=&6 \kappa \alpha_w^5 T_c \Theta(-z),
\end{eqnarray}
with $\kappa \sim 21$ for $T_c>T_{EW}\sim 100$
GeV~\cite{Bodeker:1999gx,Moore:2000mx}, we find for $z\geq 0$
\begin{eqnarray}
n_B= \frac{- n_f D_{\tau_L} \Gamma_{ws}
\mathcal{C}_{\tau_L}}{v_w^2}. \label{108}
\end{eqnarray}
Here the boundary condition $n_B(-\infty)=0$ has been used. Then
with $s={2\pi^2g_{*}T_c^3\over 45}$ the baryon asymmetry is solved
to be
\begin{eqnarray}
{n_B \over s} &=& {540\gamma_w^3\langle v_{\tau_L} \rangle
D_{\tau_L}{m_{\tau}(\infty)^2\delta
\Delta\theta_1h(\delta,T_c)\Gamma_{ws}}\over{(2\pi)^4 v_w
g_{*}T_c^3}}. \label{baryon asymmetry}
\end{eqnarray}

Eq.~(\ref{baryon asymmetry}) tells us that in the thin wall regime
of the non-local EWBG scenario, three model-dependent physical
quantities play critical roles: the variation of the $CP$ phase crossing the wall $\Delta \theta_1$, the wall thickness $\delta$, and
its velocity $v_w$. $\Delta \theta_1$ will bias the symmetry
between the dynamical behaviors of particles and anti-particles.  
For the perturbative approximation in the non-WKB momentum region,
{\it i.e.}, both $\Delta \theta_1$ and momentum are not large
($\Delta \theta_1/2<1$ and $k_i^z<\delta^{-1}$), the
$CP$-violating current and baryon asymmetry are proportional to
$\Delta \theta_1$. 
It turns out that the contribution from the thin wall regime is  
non-trivial compared to the ones from the thick wall regime.  
The wall configuration is also essential. It measures the size of the region in which the
$CP$-violating current is generated, and it provides the $p_z$
upper limit or definition standard of non-WKB particles. From Eqs.
(\ref{chiral current}) and (\ref{baryon asymmetry}), 
the $CP$-violating current and baryon asymmetry are proportional
to the wall thickness, and are also affected by the wall thickness
through the integration upper limit on $p_z$. In addition, $1/v_w$ measures the time for 
which the EW sphaleron processes continue before the particles in the $CP$-violating
current are recaptured by the bubble wall, so it is natural that the
baryon produced asymmetry is inversely proportional to $v_w$.

\subsection{The Squark, Chargino and Quark Contributions -- Thick Wall Regime}

In this subsection we will calculate the baryon asymmetry generated by 
squarks, charginos and quarks by employing the
spontaneous baryogenesis mechanism. 
Compared to the thin wall regime, there are two new
effects which are important for our understanding of 
the interactions between particles and the bubble wall: \\

(1) The effect of thermal scattering. This can lead to a
self-energy correction to the particle propagators and hence
suppress the $CP$-violating source (due to the shortened life-time
of the particles), and will be taken into account by including the
imaginary part of the fermion self-energy in the dispersion
relation, {\em i.e.},
\begin{eqnarray}
[\omega({\bf k}) + i \tilde  \gamma]^2={\bf k}^2+m^2, \label{dispersion}
\end{eqnarray}
where $\tilde \gamma$ is the associated damping rate. This immediately gives
\begin{eqnarray}
|\Im(\bf k)|&=& \frac{1}{\Delta}, 
\end{eqnarray}
with 
\begin{eqnarray}
\Delta = \tau\left |\frac{ \Re({\bf k})}{\omega}\right |, \ \ \ \ \tau= \frac{1}{\tilde \gamma}.   \label{FP}
\end{eqnarray}
Since $\tau$ is the lifetime of the scattered particles, and $\Re({\bf k})/\omega$ is their 
velocity, $\Delta$ defines the propagation length of the scattered particles in the wall. 
In addition, from Eq.~(\ref{dispersion}), we have 
\begin{eqnarray}
\omega^2 &=& \Re(\bf k)^2 +m^2 + \tilde \gamma^2 (1-\frac{\omega^2}{ \Re(\bf k)^2}).
\end{eqnarray} 
We will work in the limit $\tilde \gamma^2 \ll \Re(\bf k)^2 +m^2$, which gives the relation
\begin{eqnarray}
 \omega^2 &=& \Re(\bf k)^2 +m^2. 
\end{eqnarray}
\\

(2) The effect of multiple scattering by the wall. This effect can be
interpreted as multiple insertions in a perturbation regime. 
Currently, there are two main insertion methods: mass insertion~\cite{mass insertion} and
Higgs insertion~\cite{CMQSW1,Rius:Sanz}: 

Mass insertion assumes that the tree-level physics is described
by kinetic energy terms and considers the space-dependent mass
terms as a perturbation, which can be denoted by
\begin{eqnarray}
{\cal L} = {\cal L}_{tree} + {\cal L}_{pert}=K-M,
\end{eqnarray}
where $K$ is kinetic energy terms and $M$ is mass terms. Since
tree-level physics in the mass insertion regime is a
Lorentz-invariant free theory of massless particles, this method
is effective only if $K\gg M$. However, due to the Boltzmann factor suppression for
the injected $CP$-asymmetric current, the main contribution to the
$CP$-violating source comes from the non-WKB momentum region, 
which usually makes multiple mass insertions unstable.

Compared with mass insertions, Higgs insertion is more
extensively assumed in the literature. Its starting point is to separate 
the mass terms into a free part and perturbative part, {\em i.e.},~\cite{CMQSW1}
\begin{eqnarray}
{\cal L} = {\cal L}_{tree} + {\cal L}_{pert}=(K-M^0(z))-\delta
M(z,z_i)~,~ \label{higgs insertion}
\end{eqnarray}
where $z$ denotes the point where the $CP$-violating current is 
calculated, and also the mass eigenstates are defined, and $z_i$ 
denotes the point where the scattering occurs. 
Using these local mass eigenstates as approximate asymptotic states, 
one can calculate the associated $S$ matrix and the $CP$-violating source. If these local 
mass eigenstates are not the gauge eigenstates in the false
vacuum, after the $CP$-violating source is found one needs to go
back to the gauge eigenstates by taking an inverse unitary
transformation to evaluate the diffusion equations in the false vacuum.
This method is justified for any scattered particle
species, but it requires that the Higgs field variations be small within a propagation length.
This implies $\Delta \ll \delta$. This relation adds an upper bound on the particle momentum
in the $z$ direction. Numerically, its effect is weak, 
since the contributions from the small momentum region to the $CP$-violating current 
are more important, due to Boltzmann factor suppression in the high momentum region.\\

We will use Higgs insertion method. For given momentum and energy, 
the $CP$-violating current at point ``$z$'' is mainly generated in the layer between $z$ and $z+\Delta$. 
Since $\Delta \ll \delta$, at the lowest order we can use a formalism similar to that in the thin wall regime 
to approximately calculate the current. Then the generated baryon asymmetry can be solved
by embedding this $CP$-violating source into the diffusion equations in the thick wall regime.
This approximation is adequate for our purpose, {\em i.e.}, for presenting typical features of EWBG 
in the sMSSM. \\

{\bf (I) The squark sector} \\

The mass square matrix of the up-type squark is
\begin{equation}
M_{\tilde u} =  \left( \begin{array}{c c}
M_{{\tilde u}_L^i}^2  & M_{{\tilde u}_{LR}^i}^2 \\
M_{{\tilde u}_{RL}^i}^2  & M_{{\tilde u}_R^i}^2  \\
  \end{array} \right),
\end{equation}
where the diagonal entries are defined by
\begin{eqnarray}
M_{{\tilde u}_L^i}^2 &=&m^2_{{\tilde Q}_i} + m^2_{u_i} +
\Delta_{{\tilde u}_L^i} + \Delta'_{{\tilde u}_L^i} ~,~\, \nonumber
\\ M_{{\tilde u}_R^i}^2 &=&m^2_{{\tilde u}_R^i} + m^2_{u_i} +
\Delta_{{\tilde u}_R^i} + \Delta'_{{\tilde u}_R^i} ~,~\, \nonumber \\
M_{{\tilde u}_{LR}^i}^2 &=&y_{u_i} (h s v_1 e^{i(\alpha + \theta_1)} - A_{h_{u_i}} v_2e^{-i \theta_2}) ~,~\, \nonumber \\
M_{{\tilde u}_{RL}^i}^2 &=&y_{u_i} (h s v_1 e^{-i(\alpha + \theta_1)} - A_{h_{u_i}} v_2e^{i \theta_2}) ~,~\, 
\end{eqnarray}
with
\begin{eqnarray}
m^2_{u_i} &=& y_{u_i}^2 v_1^2~,~\ \nonumber \\
 \Delta_{{\tilde u}_L^i} &=& (\frac{1}{2} -
\frac{2}{3} \sin^2\theta_W)
 \cos (2\beta) \frac{G^2}{2}(v_1^2+v_2^2)~,~\, \nonumber \\
\Delta_{{\tilde u}_R^i} &=& \frac {2}{3} \sin^2\theta_W
 \cos (2\beta) \frac{G^2}{2}(v_1^2+v_2^2)~,~\, \nonumber \\
\Delta'_{{\tilde u}_L^i}= \frac{1}{8} \Delta'_{{\tilde u}_R^i} &=&
\frac{-1}{4\sqrt{15}} g_{Z'}^2 ( v_1^2 - v_2^2 ) ~.~\,
\end{eqnarray} 
$M_{\tilde u}$ can be perturbatively expanded at ``$z$'' as 
 \begin{eqnarray}
M_{\tilde u}^0 &=& M_{\tilde u}(z) = \left( \begin{array}{c c}
M_{{\tilde u}_L^i}^2(z)  & M_{{\tilde u}_{LR}^i}^2(z) \\
M_{{\tilde u}_{RL}^i}^2(z)  & M_{{\tilde u}_R^i}^2(z)  \\
  \end{array} \right) 
\approx   \left( \begin{array}{c c}
m_{\tilde Q_i}^2  & 0  \\
0  & m_{{\tilde u}_R^i}^2  \\
\end{array} \right), \nonumber \\
\delta M^0_{\tilde u}  &=&  \left( \begin{array}{c c}
M_{{\tilde u}_L^i}^2(z_i) - M_{{\tilde u}_L^i}^2(z) & M_{{\tilde u}_{LR}^i}^2(z_i) -  M_{{\tilde u}_{LR}^i}^2(z) \\
M_{{\tilde u}_{RL}^i}^2(z_i) -  M_{{\tilde u}_{RL}^i}^2(z) &  M_{{\tilde u}_R^i}^2(z_i) -  M_{{\tilde u}_R^i}^2(z)  \\
  \end{array} \right).
\end{eqnarray}
In the first line, we have made a first order approximation for $M_{\tilde u}^0$.
In the second line, ``$z_i$'' denotes the space point where the particle scattering occurs.

According to the bilinear interactions, a $CP$-violating left-chiral squark current
$j_{\tilde u^i_L}$ is produced through $\tilde u^i_R\rightarrow \tilde u^i_L$, where
the $CP$-odd transport charge is axial top number
\begin{eqnarray}
{\mathcal Q}_{\tilde t_{L,R}}^t = {\rm Diag}(1,-1).
\end{eqnarray} 
Considering that this is mediated by the Yukawa couplings 
of the up-type quarks and that the Yukawa coupling of the top quark
$y_{t}=y_{u_3}$ is much larger than the other two, {\em i.e.},
$y_{u_1,u_2}$, we neglect the contributions from the first
two family squarks. After a field theory
calculation (for details, see Appendix~\ref{DR}), we find
\begin{eqnarray}
\Delta R_{\tilde t_{R} \rightarrow\tilde t_L}(k_i,k_f,z) &=&
\frac{-1}{4E_iE_f}   \frac{1}{v_i^z v_f^z}  \int_z^{z+\Delta(\tilde \gamma_{\tilde t} )} dz_1 dz_2
\sin((k_i^z-k_f^z)(z_1-z_2))
\nonumber \\
&& \{w_1(z_1) w_1(z_2)\sin(\theta_1'(z_1)-\theta_1'(z_2)) \nonumber
\\ &&+ w_2(z_1) w_2(z_2)\sin(-\theta_2(z_1)+\theta_2(z_2)) \nonumber
\\ &&
+w_1(z_1) w_2(z_2)\sin(\theta_1'(z_1)+\theta_2(z_2)) \nonumber
\\ &&+ w_2(z_1) w_1(z_2)\sin(-\theta_2(z_1)-\theta_1'(z_2)) \nonumber \\
&& - (z_1\to z) - (z_2 \to z)\},   \label{R1}
\end{eqnarray}
with 
\begin{eqnarray}
E_i^2=E_f^2&=& (k_i)^2+m_{\tilde t_R}^2=(k_f)^2+m_{\tilde
t_L}^2, \nonumber \\
v_i^z &=& \frac{k_i^z}{E_i}, \ \ \ \ 
v_f^z \ = \ \frac{k_f^z}{E_f}, \nonumber \\
w_1(z)&=& y_t h s(z) v_1(z), \nonumber \\
w_2(z)&=& - y_t A_{h_t} v_2(z), \nonumber \\
\theta'_1(z)&=& \alpha(z)+\theta_1(z),\nonumber \\
\tilde \gamma_{\tilde t} &=& \tilde \gamma_{\tilde t_R}+\tilde \gamma_{\tilde
t_L}.
\label{115}
\end{eqnarray}
Here the indices ``i'' and ``f'' denote the initial and final
states of the scattering, respectively; and $(z_i\to z)$ denotes 
terms similar to the first four ones in $\{\}$, but with $z_i$ replaced by $z$.
In this paper, we set the damping rates $\tilde \gamma_{\tilde
t_R}=\tilde \gamma_{\tilde t_L}=0.1T_c \sim \alpha_s T_c$ as an
approximation, since an analytical calculation for them is still absent. 
$\Delta R_{\tilde t_{R} \rightarrow\tilde t_L}(k_i,k_f,z)$ is sensitive 
to the $A_{h_t}$ soft parameter due to its dependence on $w_2$. 
This makes its contribution to EWBG very different for small and 
large $A_{h_t}$ (EWPT is not sensitive to $A_{h_t}$, since it enters 
the neutral Higgs effective potential only at loop level). Particularly, 
this contribution can be quadratically enhanced by a large $A_{h_t}$ 
through the $w_2w_2$ term.
We will discuss more on these issues in the next subsection. 
Using these results, the
space-dependent $CP$-violating current is solved to be
\begin{eqnarray}
j_{\tilde t_L}(z)&=& \int_{k_i^z>0}{{d^3k_i}\over{(2\pi)^3}}{{k_f^z} \over {E_f}}\left (f_z^{\tilde
t_R}(k_i^z)-f_{z+\Delta(\tilde \gamma_{\tilde t} )}^{\tilde t_R}(-k_i^z)\right )\Delta R_{\tilde t_{R}
\rightarrow\tilde t_L}(k_i,k_f,z){\mathcal Q}_{\tilde t_L}^t. \label{116}
\end{eqnarray}
Here $f_z^{\tilde t_R}(k_i^z)$ and $f_{z+\Delta}^{\tilde t_R}(-k_i^z)$ are
the thermal distributions of $\tilde t_R$ in the wall frame at ``$z$'' and ``$z+\Delta$'', respectively.
Compared to the one defined in Eq.~(\ref{114}), they have a factor 3 difference due to QCD color as well as a sign-flip 
for the second term in the denominator. \\

{\bf (II) The chargino sector} \\

The chargino mass matrix is 
\begin{eqnarray}
M_{\tilde \chi^{\pm}} =\left(\matrix{M_2 & u_2e^{-i\theta_2}
\cr u_1e^{-i\theta_1} & \mu e^{i\alpha} \cr}\right) ~,~
\,
\end{eqnarray}
with $u_i = g_2 v_i/\sqrt{2}$ and $\mu \equiv h s $.
It can be perturbatively expanded around $z$ as 
\begin{eqnarray}
M_{\tilde \chi^{\pm}}^0 &=& M_{\tilde \chi^{\pm}}(z), \nonumber \\
\delta M_{\tilde \chi^{\pm}} &=&\left(\matrix{0 &
u_2(z_i)e^{-i\theta_2(z_i)} - u_2(z)e^{-i\theta_2(z)} \cr u_1(z_i)e^{-i\theta_1(z_i)}-u_1(z)e^{-i\theta_1(z)} & \mu(z_i) e^{i\alpha(z_i)}- \mu(z) e^{i\alpha(z)}
\cr}\right).
\end{eqnarray}
For the parameter values that we are using, $|u_{1,2}(z)| \ll M_2, |\mu(z)|$, so we 
simply neglect the off-diagonal elements in  $M_{\tilde \chi^{\pm}}^0$.
In this sector, the $CP$-violating Higgsino current is produced through $\tilde W^c\rightarrow\tilde H^c$, where 
the $CP$-odd transport charge is vector Higgs number \footnote{In the chargino sector, the $CP$-odd transport charge can also be axial Higgs number. 
But this source usually is suppressed by a fast Higgsino violating
process $\Gamma_{\mu}$ corresponding to the $h \langle S \rangle \widetilde h_1
\widetilde h_2$ term in the Lagrangian~\cite{CMQSW1}. We will therefore not consider it here.}
\begin{eqnarray}
{\mathcal Q}_{\tilde W^+,\tilde H^+,\bar {\tilde {W}}^+,\bar {\tilde {H}}^-}^H= {\rm Diag}(0,1,0,1).
\end{eqnarray} 
After a field theory calculation (for details, see Appendix~\ref{DR}), we find
\begin{eqnarray}
\Delta R_{\tilde W^c\rightarrow\tilde H^c}(k_i,k_f,z)
&=&\frac{1}{E_iE_f}  \frac{1}{v_i^z v_f^z}  \int_z^{z+\Delta(\tilde \gamma_{\chi^c})} dz_1 dz_2
\sin((k_i^z-k_f^z)(z_1-z_2)) 
\nonumber \\
&& \{k_i^zk_f^z[u_1(z_1) u_1(z_2)\sin(\theta_1(z_1)-\theta_1(z_2))
\nonumber
\\ &&+ u_2(z_1) u_2(z_2)\sin(-\theta_2(z_1)+\theta_2(z_2)) ]+\nonumber
\\&&
M_2\mu(z)[u_1(z_1) u_2(z_2)\sin(\theta_1(z_1)+\theta_2(z_2)+\alpha(z))
\nonumber
\\ &&+ u_2(z_1) u_1(z_2)\sin(-\theta_2(z_1)-\theta_1(z_2)-\alpha(z)) ]
 \nonumber \\
&& - (z_1 \to z) - (z_2 \to z)\}, 
\label{R2}
\end{eqnarray}
where
\begin{eqnarray}
E_i^2=E_f^2&=&(k_i)^2+M_2^2=(k_f)^2+\mu(z)^2,  \nonumber \\ 
v_i^z &=& \frac{k_i^z}{E_i}, \ \ \ \ 
v_f^z  \ = \ \frac{k_f^z}{E_f}, \nonumber \\
\tilde \gamma_{\chi^c} &=&\tilde \gamma_{\tilde W^c}+\tilde \gamma_{\tilde
H^c}. 
\end{eqnarray}
For the damping rates, we will use the results in the MSSM, which are given by $\tilde \gamma_{\tilde
H^c}=0.025 T_c$ and $\tilde \gamma_{\tilde W^c}=0.065 T_c$~\cite{damping rates}. The $CP$-violating current is solved
to be
\begin{eqnarray}
j_{\tilde H^c}(z) &=&\int_{k_i^z>0}{{d^3k_i}\over{(2\pi)^3}}{{k_f^z} \over {E_f}}    \nonumber 
\\ &&  \left (f_z^{\tilde
W^c}(k_i^z)-f_{z+\Delta(\tilde \gamma_{\chi^c})}^{\tilde W^c}(-k_i^z)\right)   \Delta R_{\tilde W^c
\rightarrow\tilde H^c}(k_i,k_f,z)
{\mathcal Q}_{\tilde H^+,\bar {\tilde {H}}^-}^H, 
\end{eqnarray}
where $f_z^{\tilde W^c}(k_i^z)$ and $f_{z+\Delta}^{\tilde W^c}(-k_i^z)$ are
the thermal distributions of the charged gaugino $\tilde W^c$ in
the wall frame, which have a factor 2 difference from that defined
in Eq.~(\ref{114}) due to $\tilde W^c=\tilde W^c_L+\tilde W^c_R$.\\

{\bf (III) The quark sector} \\

The top quark  mass is 
\begin{eqnarray}
m_t=  h_t v_2e^{i\theta_2} ~.~\,
\label{cpmass}
\end{eqnarray}
which can be perturbatively expanded around $z$ as 
\begin{eqnarray}
m_t^0 &=&  m_t(z), \nonumber \\
\delta m_t &=&  h_t v_2(z_i)e^{i\theta_2(z_i)} - h_t v_2(z)e^{i\theta_2(z)}.
\end{eqnarray}
In this sector, the $CP$-violating current is produced by $t_R \to t_L$, 
where the $CP$-odd transport charge is axial top number
\begin{eqnarray}
{\mathcal Q}_{t_{L,R}}^t = {\rm Diag}(1,-1).
\end{eqnarray} 
Then (for details, see Appendix~\ref{DR})
\begin{eqnarray}
\Delta R_{t_R \rightarrow t_L}(k_i,k_f,z)
&=&\frac{1}{E_iE_f}    \frac{1}{v_i^z v_f^z}  \int_z^{\Delta(\tilde \gamma_{t} )} dz_1 dz_2
\sin((k_i^z-k_f^z)(z_1-z_2))   
\nonumber \\
&& \{h_t^2k_i^zk_f^z v_2(z_1) v_2(z_2)\sin(\theta_2(z_1)-\theta_2(z_2)) \nonumber \\
&& - (z_1\to z) - (z_2 \to z)\},  \label{R3}
\end{eqnarray}
where
\begin{eqnarray}
E_i^2=E_f^2&=&(k_i)^2 +|m_t^0(z)|^2=(k_f)^2 + |m_t^0(z)|^2, \nonumber \\
v_i^z &=& \frac{k_i^z}{E_i}, \ \ \ \
v_f^z \ = \ \frac{k_f^z}{E_f}, \nonumber \\
\tilde \gamma_{t} &=&\tilde \gamma_{t_R}+\tilde \gamma_{t_L}.
\end{eqnarray}
Again, we approximate the damping rates by $\tilde \gamma_{t_L}=\tilde \gamma_{t_R}=0.1 T_c \sim \alpha_s T_c $. 
The trace in the $S$ matrices of the chargino and top quark scattering gives an extra factor ``4'', so the factor ``4'' in the 
denominator of Eq.~(\ref{810}) disappears in both Eq.~(\ref{R2}) and Eq.~(\ref{R3}).
The $CP$-violating current is solved to be
\begin{eqnarray}
j_{t_L}(z)&=&\int_{k_i^z>\Lambda_k^{t_L}}{{d^3k_i}\over{(2\pi)^3}}{{k_f^z} \over {E_f}}   \nonumber \\ && \left(f_z^{t_R}(k_i^z)-
\frac{k_i^z}{\sqrt{{k_i^z}^2-\Delta m_t^2}}
f_{z+\Delta(\tilde \gamma_{t} )}^{t_R}(-k_i^z)\right) 
 \Delta R_{t_R
\rightarrow t_L}(k_i,k_f,z) {\mathcal Q}_{t_L}^t, \label{121}
\end{eqnarray}
where $f_z^{t_R}(k_i^z)$ and $f_{z+\Delta}^{t_R}(-k_i^z)$ are
the thermal distributions of the right chiral top quark in
the wall frame, which have a factor 3 difference from that defined
in Eq.~(\ref{114}) due to color, 
\begin{eqnarray}
\Delta m_t^2 = m_t^2(z+\Delta(\tilde \gamma_{t} ))- m_t^2(z)
\end{eqnarray} 
is the difference of the quark mass 
square at $z$ and $z+\Delta(\tilde \gamma_{t} )$, and
\begin{eqnarray}
\Lambda_k^{t_L}= \sqrt{m_t^2(\infty) - m_t^2(z)}  \label{ct}
\end{eqnarray}
is the momentum cutoff. 
The factor in front of $f_t^{t_L}(-k_i^z)$ in Eq.~(\ref{121}) is due to the large
mass difference of the scattered particles inside and outside the bubble. For 
$\tau$ leptons, stops and charginos, this difference is small compared to $T_c$, so this factor 
is approximately equal to 1.\\

Before moving to the calculation of the produced baryon asymmetry, we have 
some comments on the $CP$-violating currents obtained from the lowest Higgs insertion. 
First, according to the discussions in Subsection~\ref{CP subsection2}, the single Higgs 
phases $\theta_{\phi_i}$ are not physical since their values depend on two 
decoupled degrees of freedom: $A$ and $B$. Indeed, though working in the $\theta_{\phi_i}$ basis, 
from Eq.~(\ref{R1}), Eq.~(\ref{R2}) and Eq.~(\ref{R3}) we see that all of these currents 
depend on some special linear combinations of $\theta_{\phi_i}$, where their dependences 
on $A$ and $B$ are cancelled and they are only sensitive to $\beta_i(z)$. 
Second, for the $CP$-violating currents from the stop and the chargino sectors, 
the terms $\propto \omega_1\omega_2$ and $\propto u_1u_2$ will be reduced to the $v_1\partial v_2 - v_2\partial v_1$
type sources in the MSSM if $\theta_{1,2}(z)=0$ and $\alpha(z)=$ constant. 
Though in this limit these contributions are trivial for $\partial \tan \beta =0$, 
they are not in Eq.~(\ref{R1}) and Eq.~(\ref{R2}) due to the space-dependence of $\theta_{1,2}$ in the wall. 
Third, the terms $\propto v_{i}v_{i}$ in Eq.~(\ref{R1}) and Eq.~(\ref{R2}) are zero in the limit $\theta_{1,2}(z)=0$ and $\alpha(z)=$ constant.
They therefore never appear in the MSSM currents and denote a new-type of $CP$-violating current
in the stop and chargino sector. This is also the only type of $CP$-violating current at the leading order in the quark sector. 
The vector Higgsino current of the $v_1\partial v_2 + v_2 \partial v_1$ type 
does not appear in our calculations since it denotes a resummation effect from higher order 
Higgs insertions~\cite{Rius:Sanz, CMQSW1, CQSW2}.  Finally, we emphasize that all of these features are results of SCPV-driven EWBG, 
so they are not sensitive to the concrete embeddings of the sMSSM. (For different embeddings, the $U(1)_Y$ and $U(1)'$ gauge charges 
in Eq.~(\ref{charges}) and hence the formulas in Eq.~(\ref{gauge eigenstates}) are different. However, we did not
use these results when deriving the formulas of the $CP$-violating currents.) 

The field theory calculations used in the thick wall regime cannot be directly applied to the thin wall regime. 
In the thin wall regime, the main contribution is from an interference effect between a $CP$ and a non-$CP$ term 
(see the Appendix of~\cite{Joyce:1994zn}). The non-$CP$ term is induced by the effect that the fundamental solutions inside and outside 
the bubble are different due to the mixing between left- and right-chiral fermions. In the quantum mechanics calculations of the reflection and transmission 
coefficients, this effect is explicitly included. Its strength can be measured by the parameter $t_\phi$ which is defined in Subsection~\ref{thin wall regime}. 
In the thick wall regime, this effect is neglected in the field theory calculations. Instead we assume the same asymptotic states between the two sides 
of a layer inside the wall, which greatly simplifies the calculations. This assumption is justified by the fact that $t_\phi$ is approximately estimated to be
\begin{eqnarray}
t_\phi \simlt \frac{\Delta m(z)}{\overline \Delta m} \sim \frac{\Delta}{2\delta} \ll 1 
\end{eqnarray}
in the thick wall limit. Here $\Delta m(z) = m(z+\Delta)-m(z)$ is the mass difference of the incident particles between the two sides of the layer, 
$\overline \Delta m = m(\infty) - m (-\infty)$ is their mass variation crossing the wall,  and $\Delta$ is the thickness of the layer (see Eq.~(\ref{FP})). 
The main contribution in the thick wall regime therefore is from an interference effect of the $CP$-violating terms. Actually, 
if we apply the field theory calculation to leptons (similar to what we did for quarks), we will get their subleading contribution, which is given in the Appendix 
of~\cite{Joyce:1994zn}.\\

While these injected $CP$-violating Higgsino or squark currents
propagate in the false vacuum, the original thermal equilibrium
among different particle species is broken and these 
species may obtain nonvanishing net number densities due to 
the associated decay processes. In the thick wall
regime, the main decay processes include~\cite{mass insertion}:
(1) the Yukawa interaction $\Gamma_{y_T}$ corresponding to all
supersymmetric and soft trilinear interactions arising from the
$y_T H_2Q_LT_R$ term in the superpotential\footnote{As a first order approximation, we neglect the tau and bottom Yukawa interactions.  For relevant discussion, 
see~\cite{Chung:2009qs}.} ($Q_L$ is the
left-chiral superfield of the third family quarks and $T_R$ is the
right-chiral superfield of the top quark); (2) the Higgsino violating
process $\Gamma_{\mu}$ corresponding to the $h \langle S \rangle \widetilde h_1
\widetilde h_2$ term in the Lagrangian and supergauge interaction
$\Gamma_g$; (3) the axial top number violating process $\Gamma_m$ and the
Higgs-violating process $\Gamma_h$ corresponding to top quark mass
effects and Higgs self interactions, respectively; (4) the strong
sphaleron process $\Gamma_{ss}$~\cite{Mohapatra:1991bz} and weak sphaleron process
$\Gamma_{ws}$. $\Gamma_m$ and $\Gamma_h$ are suppressed in the false vacuum and
$\Gamma_{ws}$ is suppressed in the true vacuum (for a strong enough
first order EWPT); all of the others can occur both inside
and outside of the bubble. For simplicity, we neglect all processes 
with a singlet component 
involved in the current propagation. In the concrete calculations, further
simplifications are also made~\cite{mass insertion}. (a) We assume fast
enough $\Gamma_g$ which thermally equilibrates $n_q$ and
$n_{\tilde q}$; thus we can describe the system by the densities
$n_{Q_L}=n_{q_L}+n_{\tilde q_L}$, $n_{T_R}=n_{t_R}+n_{\tilde t_R}$, 
$n_{H_1}=n_{h_1}+n_{\tilde H_1}$ and $n_{H_2}=n_{h_2}+n_{\tilde
H_2}$. (b) We assume fast enough $\Gamma_{\mu}$; then we have
$n_{H_1}=n_{H_2}$ and hence are able to simplify the two
quantities $n_{H_1}$ and $n_{H_2}$ into one $n_H=n_{H_1}+n_{H_2}$.
(c) We assume slow $\Gamma_{ws}$, which allows us to ignore
leptons for the particle diffusion and the $n_L$ generation. Under
these approximations, the diffusion equations in the wall frame are\footnote{These equations are only valid in the limit of non-relativistic wall velocity. This is 
also true for Eq.~(\ref{propagating equation}). }~\cite{mass
insertion} 
\begin{eqnarray}
\label{nQ} v_\omega n'_{Q_L} &= & D_{Q_L} n''_{Q_L}-
\Gamma_{y_T}\left[\frac{n_{Q_L}}{k_{Q_L}}-\frac{n_{T_R}}{k_{T_R}}-
\frac{n_H}{k_H} \right]
-\Gamma_m\left[ \frac{n_{Q_L}}{k_{Q_L}}-\frac{n_{T_R}}{k_{T_R}} \right]\nonumber\\
&-& 6 \Gamma_{ss}
\left[2\,\frac{n_{Q_L}}{k_{Q_L}}-\frac{n_{T_R}}{k_{T_R}}+
9\,\frac{n_{Q_L}+n_{T_R}}{k_{B_R}} \right]+J_{\tilde t_L} + J_{t_L},  \\
\label{nT} v_\omega n'_{T_R}&= & D_{Q_L} n''_{T_R}+
\Gamma_{y_T}\left[\frac{n_{Q_L}}{k_{Q_L}}-\frac{n_{T_R}}{k_{T_R}}-
\frac{n_H}{k_H} \right]
+\Gamma_m\left[ \frac{n_{Q_L}}{k_{Q_L}}-\frac{n_{T_R}}{k_{T_R}} \right]\nonumber\\
&+& 3 \Gamma_{ss}
\left[2\,\frac{n_{Q_L}}{k_{Q_L}}-\frac{n_{T_R}}{k_{T_R}}+
9\,\frac{n_{Q_L}+n_{T_R}}{k_{B_R}} \right]-J_{\tilde t_L} - J_{t_L},
 \\
v_\omega n'_H&= & D_H n''_H+
\Gamma_{y_T}\left[\frac{n_{Q_L}}{k_{Q_L}}-\frac{n_{T_R}}{k_{T_R}}-
\frac{n_H}{k_H} \right] -\Gamma_h\,\frac{n_H}{k_H}+J_{\tilde H^c},
\label{nH}
\end{eqnarray}
with
\begin{eqnarray}
J_{\tilde t_L} (z) &=& -\partial_z j_{\tilde t_L} (z) \sim \frac{\tilde
\gamma_{\tilde t}}{\langle v_{\tilde t_L}\rangle} j_{\tilde t_L}(z), \nonumber \\
J_{t_L} (z) &=& -\partial_z j_{ t_L} (z) \sim \frac{ \tilde
\gamma_{ t} }{\langle v_{t_L}\rangle} j_{ t_L}(z), \nonumber \\
J_{\tilde H^c} (z) &=& -\partial_z j_{\tilde H^c} (z) \sim \frac{ \tilde
\gamma_{\chi^c}}{\langle v_{\tilde H^c}\rangle} j_{\tilde H^c}(z), \label{124}
\end{eqnarray}
where $\langle v_{\tilde t_L, t_L, \tilde H^c} \rangle$ is average velocity of the scattered particles 
and $\{k_{Q_L},k_{T_R},k_{B_R},k_H\}$ are statistical factors.

For light particles compared to $T_c$, their statistical factor is 2
for bosons and 1 for fermions. If these particles are
very heavy, then their statistical factors are suppressed 
exponentially. Explicitly, this suppression factor is (e.g., see~\cite{KT})
\begin{eqnarray}
12\Big(\frac{m}{2\pi T_c}\Big)^{3/2}\exp\left (\frac{-m}{T_c}\right ), \label{301}
\end{eqnarray}
with $m$ being the particle mass. In this paper, we assume that all
charginos and neutralinos are light for simplicity. As for squark
masses, we take $m^2_{\tilde Q_3}=m^2_{\tilde u_R^3}=8$ while keeping
the other squark soft mass squares to be 25 (see
Table~\ref{parameter}). This soft mass pattern is
phenomenologically favored due to the large beta functions for
$m^2_{\tilde Q_3}$ and $m^2_{\tilde u_R^3}$. According to Eq.~(\ref{301}),
we immediately find\footnote{As an approximation, we use their
soft masses as the squarks' physical masses.}
\begin{eqnarray}
k_{Q_L}\approx 10,\quad k_{T_R} \approx 5,\quad k_{B_R}\approx
3,\quad k_H\approx 12,
\end{eqnarray}
and
\begin{eqnarray}
\bar D&=& \frac{5}{28}D_{Q_L} +\frac{23}{28}D_H, \nonumber \\
\bar\Gamma
&=&\frac{23}{336}(\Gamma_m+\Gamma_h), \nonumber \\
\bar J&=&\frac{23}{28}(J_{\tilde t_L}+J_{t_L}+J_{\tilde H^c}).
\end{eqnarray}
Here $\bar D$, $\bar \Gamma$, and $\bar J$ are the effective diffusion constant, decay rate and $CP$-violating current (for their definition, see~\cite{mass
insertion}).
It has been implicitly assumed that the other Higgs singlets
are not thermally equilibrated with the EW charged ones. Using
$D_H \sim 110/T_c$ and $D_{Q_L} \sim 6/T_c$~\cite{JPT}, one 
finds $\bar D \sim 91/T_c$. This result further
constrains the wall velocity: slow weak sphaleron processes
require
\begin{eqnarray}
 \bar D \Gamma_{ws}/v_w^2 < 1,
\end{eqnarray}
or
\begin{eqnarray}
v_w > 0.02.
\end{eqnarray}
As for $\bar \Gamma$, using the estimation for
$\Gamma_m(z)+\Gamma_h(z)$ with $\tan\beta>1$ in~\cite{mass
insertion}, we have
\begin{eqnarray}
\bar \Gamma(\infty) \approx 10^{-2} T_c.
\end{eqnarray}

To solve Eq.~(\ref{nQ}-\ref{nH}), we further assume that
$\Gamma_{ss}$ and $\Gamma_{y_T}$ are large, and approximate
\begin{eqnarray}
\Theta(z)\bar J(z) &\rightarrow&  \bar J(z), \nonumber \\
\Theta(z)\bar{\Gamma}(\infty) &\rightarrow&
\bar{\Gamma}(z),\label{119}
\end{eqnarray}
following~\cite{CMQSW1,mass insertion}. Then using the boundary conditions $n_L(\pm\infty)=0$,
we find
\begin{eqnarray}
n_L=\sum_i (n_{q_L^i}+n_{\tilde{q}_L^i}) = \mathcal {C}_{Q_L} e^{z\,v_\omega/\bar D}
\end{eqnarray}
for $z<0$, where
\begin{eqnarray}
\mathcal
{C}_{Q_L}=\frac{5}{46}\frac{1}{\bar D \lambda_+}\ \int_0^\infty d\zeta\
\bar J(\zeta)\ e^{-\zeta \lambda_+},
\label{128}
\end{eqnarray}
with 
\begin{equation}
\lambda_+= \frac{1}{2\bar D}\left\{v_\omega+ \sqrt{v_\omega^2+4
\bar \Gamma(\infty) \bar D} \right\} \approx 10^{-2}T_c. 
\end{equation}
This solution is the same as the MSSM one obtained in~\cite{CMQSW1,mass insertion} except the coefficient, 
which is caused by the difference of the $\{k_{Q_L},k_{T_R},k_{B_R},k_H\}$ values.
Solving the diffusion equation for $n_B$, which has exactly
the same form as Eq.~(\ref{106}) with $D_L \rightarrow \bar D$ and
$\mathcal{C}_L\rightarrow \mathcal{C}_{Q_L}$, we obtain the baryon
number density for the $z\geq 0$ region
\begin{eqnarray}
n_B= \frac{-n_f\bar D \Gamma_{ws} \mathcal{C}_{Q_L}}{v_w^2},
\label{109}
\end{eqnarray}
and finally
\begin{equation}
\frac{n_B}{s}=- \frac{45}{2\pi^2g_{*}T_c^3}\frac{15
\Gamma_{ws}}{46v_w^2 \lambda_+}   \int_0^\infty d\zeta\
\bar J(\zeta)\ e^{-\zeta \lambda_+} . \label{129}
\end{equation}

\subsection{Numerical Results of Non-local EWBG}

\begin{figure}
\begin{center}
\includegraphics[height=3.2 in]{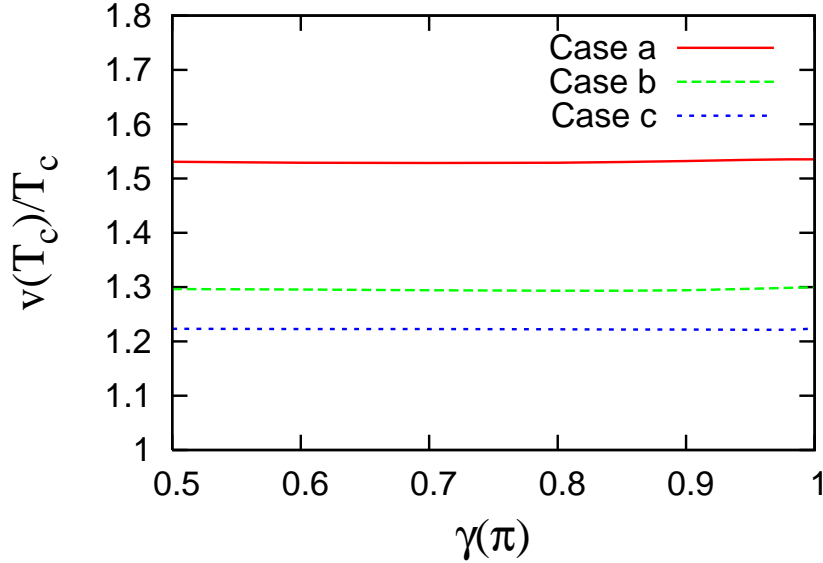}
\end{center}
\caption{$\frac{v(T_c)}{T_c}$ vs. $\gamma (\pi)$ for cases a, b, and c defined in Table~\ref{parameter}.}
\label{phi3}
\end{figure}

\begin{figure}
\begin{center}
\includegraphics[height=3.2 in]{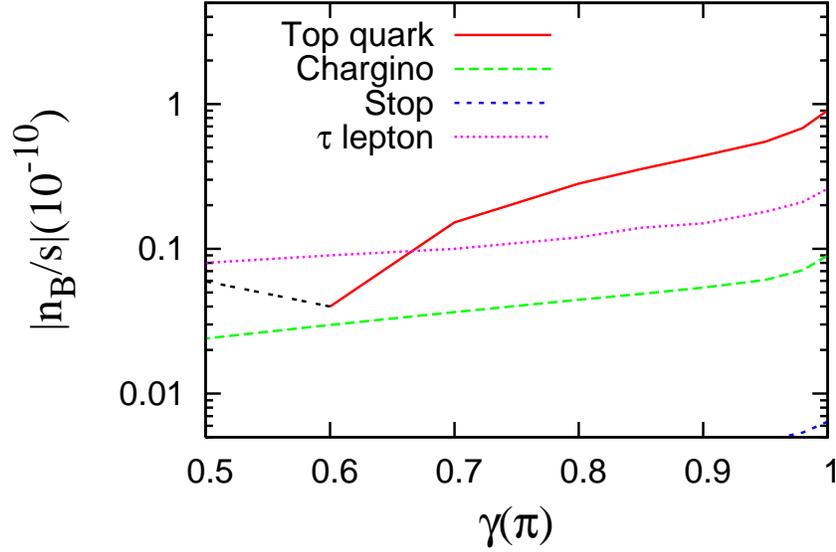}
\end{center}
\caption{$\frac{n_B}{s}$ vs. $\gamma (\pi)$ for $v_w=0.05$, in case a. There is a sign flip for $n_B/s$ in the region denoted by black-dashed line for the top quark curve. }
\label{fig101}
\end{figure}

\begin{figure}
\begin{center}
\includegraphics[height=3.2 in]{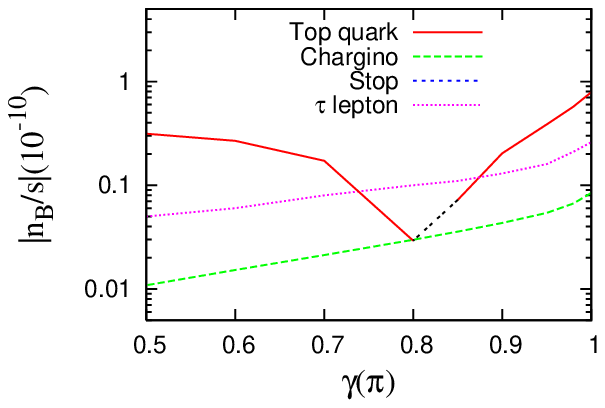}
\end{center}
\caption{$\frac{n_B}{s}$ vs. $\gamma (\pi)$ for $v_w=0.05$, in case b. There is a sign flip for $n_B/s$ in the region denoted by black-dashed line for the top quark curve. The stop contribution is smaller than the scale of this figure.}
\label{fig102}
\end{figure}

\begin{figure}
\begin{center}
\includegraphics[height=3.2 in]{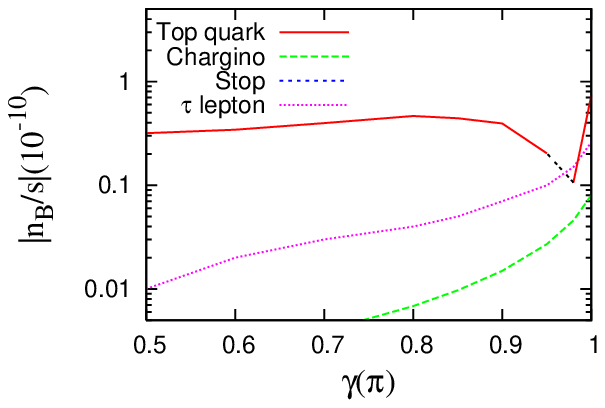}
\end{center}
\caption{$\frac{n_B}{s}$ vs. $\gamma (\pi)$ for $v_w=0.05$, in case c. There is a sign flip for $n_B/s$ in the region denoted by black-dashed line for the top quark curve. The stop contribution is smaller than the scale of this figure.}
\label{fig103}
\end{figure}

According to the previous sections, the explicit $CP$
phase, $\gamma$, plays a crucial role in mediating the 
EWBG. Both the spontaneous $CP$ phases and the bubble
wall physics show a strong dependence on it. There should be
a similar $\gamma$ dependence, therefore, for the generated baryon
asymmetry.

Let us first consider the
$\gamma$ dependence of $v(T_c)/T_c$, the measure of the EWPT strength. Some related issues on EWBG have been
discussed in Section~\ref{EWPT} where no ECPV is implicitly assumed. As expected, $v(T_c)/T_c$
shows a strong dependence on the soft SUSY breaking parameters
$m^2_{\tilde Q_3}$, $m^2_{\tilde u_R^3}$ and $A_h$. In
Figure~\ref{phi3}, we illustrate the $\gamma$ dependence of
$v(T_c)/T_c$. The weak $\gamma$ dependence is because $\gamma$ is 
from the secluded sector of the neutral Higgs potential, which has no direct coupling to
$H_d^0$ and $H_u^0$. For all of the three cases a, b and c, the first order EWPT is strong enough for implementing the 
EWBG.

In Figure~\ref{fig101}-\ref{fig103}, we show the $\gamma$
dependence of the baryon asymmetry $|n_B/s|$ produced 
by $\tau$ leptons, top squarks, charginos and top quarks\footnote{We did not consider the contribution from neutralinos. Since the relevant $CP$-violating sources and the interactions with the bubble wall are similar for neutralinos and charginos, their contributions are expected to be comparable with those from charginos. For relevant discussions in the MSSM, e.g., see~\cite{CMQSW1}.}. We set $v_w=0.05$. The produced baryon asymmetry is sensitive to $\gamma$. This is easy to understand since both the magnitude of SCPV and the bubble physics 
are sensitive to $\gamma$.
Within theoretical uncertainties, the observed value $n_B/s = ( 8.82\pm 0.23)\times 10 ^{-11}$ can be explained for
a large range of $\gamma$ values. Top quarks play a significant role in all three cases because of  
the top Yukawa coupling enhancement. In addition, due to the large change of  the top quark mass 
crossing the bubble wall, the cancellation between $f_z^{t_R}(k_i^z)$ and $f_{z+\Delta}^{t_R}(-k_i^z)$ in Eq.~(\ref{121}) is relatively weak.  
The stop contribution depends on the magnitude of the soft parameter $A_{h_t}$.
For a small $A_{h_t}$, it is dominated by the $w_1w_1\sim s^2 v_1^2$ term in Eq.~(\ref{R1}). 
As $m_{H_d^0}^2$ increases, $v_1$ and $s$ usually becomes smaller and smaller.
The stop contribution is then quartically suppressed. This explains why the stop contributions
are smaller in case b and c than in case a, and also explains why the stop contribution is small 
for a small $m_{\chi_1^0}$ in Figure~\ref{figure6} of the next section 
(for all figures in this article, we chose $A_{h_t}=0.1$). For a large $A_{h_t}$, the stop 
contribution is dominated by the $w_2w_2 \sim A_{h_t}^2 v_2^2$ term in Eq.~(\ref{R1}). 
It is not sensitive to the soft parameter $m_{H_d^0}^2$ and can be comparable with or even larger than
the chargino contribution. Numerical results show that the stop contribution becomes comparable with the chargino one as
$A_{h_t}$ increases to $\sim T_c$ and with the top quark one as $A_{h_t}$ increases to $\sim4T_c$. 
As pointed out in the previous section, the $v_2v_2$-type $CP$-violating current is 
absent at the leading order in the MSSM. Both the large contribution by top quarks and 
the enhancement effect of the stop contribution by a large $A_{h_t}$ therefore are absent in the MSSM.

In addition to $\gamma$, there are two other important factors which
can influence the produced baryon asymmetry. The first is 
$v_w$. One can see this from Eq.~(\ref{baryon asymmetry}) and Eq.~(\ref{109}).
A larger baryon asymmetry can be
generated if the bubble wall is expanding at a more
non-relativistic velocity; and vice versa. The second is the stop masses. 
In this paper we assume the
stop soft mass $|m_{\tilde Q_3}|=|m_{\tilde u_R^3}|$ to be 
heavier than $T_c$, even though they are lighter than those of the
other squarks and sleptons. If they (or more precisely, the stop
physical masses) are light compared to $T_c$, then the associated
statistical factors will be modified to~\cite{mass insertion}
\begin{eqnarray}
k_{Q_L}\approx 18,\quad k_{T_R} \approx 9,\quad k_{B_R}\approx
3,\quad k_H\approx 12,
\end{eqnarray}
which gives an additional factor 3 in the numerator of
Eq.~(\ref{129}).

\section{Neutralino CDM and Non-local EWBG}

\begin{figure}
\begin{center}
\includegraphics[height=3.2 in]{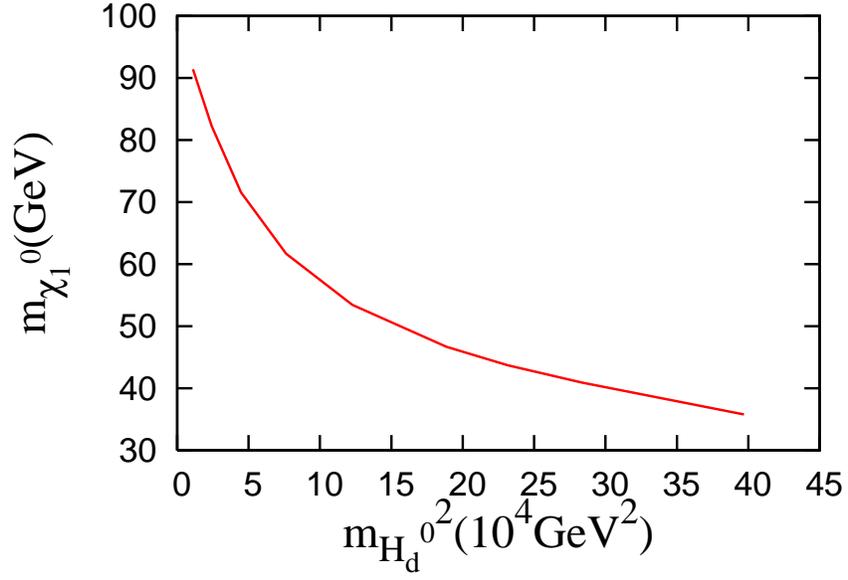}
\end{center}
\caption{$m_{\chi_1^0}$ vs. $m_{H_d^0}^2$.}
\label{figure1}
\end{figure}

\begin{figure}
\begin{center}
\includegraphics[height=3.2 in]{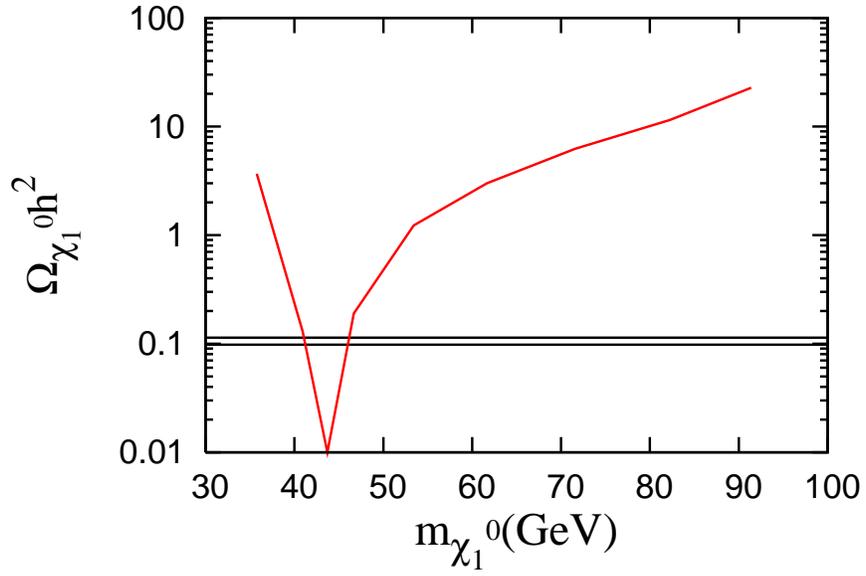}
\end{center}
\caption{ $\Omega_{\chi_1^0}h^2$ vs. $m_{\chi_1^0}$. The two black parallel lines give the upper and lower bounds on the DM relic density measured by astrophysical and cosmological probes~\cite{Amsler:2008zz}. }
\label{figure2}
\end{figure}

\begin{figure}
\begin{center}
\includegraphics[height=3.2 in]{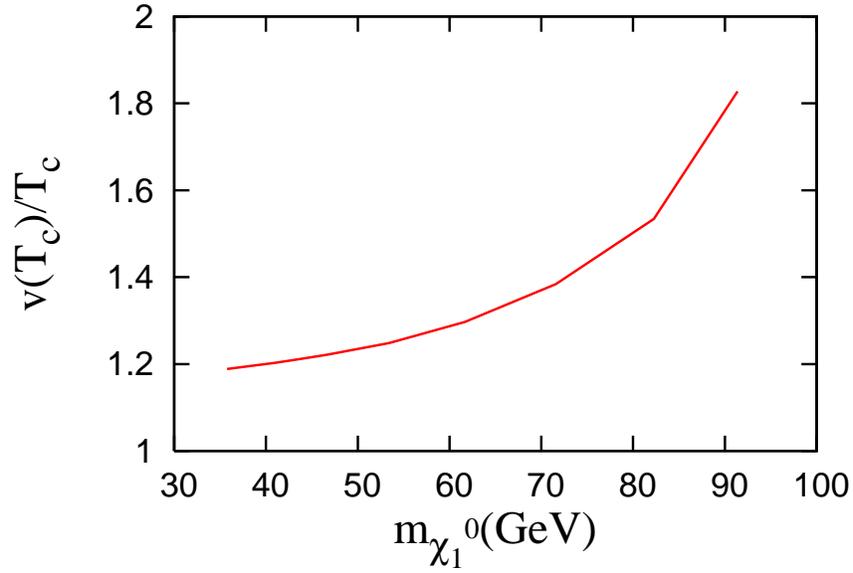}
\end{center}
\caption{$\frac{v(T_c)}{T_c}$ vs. $m_{\chi_1^0}$.}
\label{figure3}
\end{figure}

\begin{figure}
\begin{center}
\includegraphics[height=3.2 in]{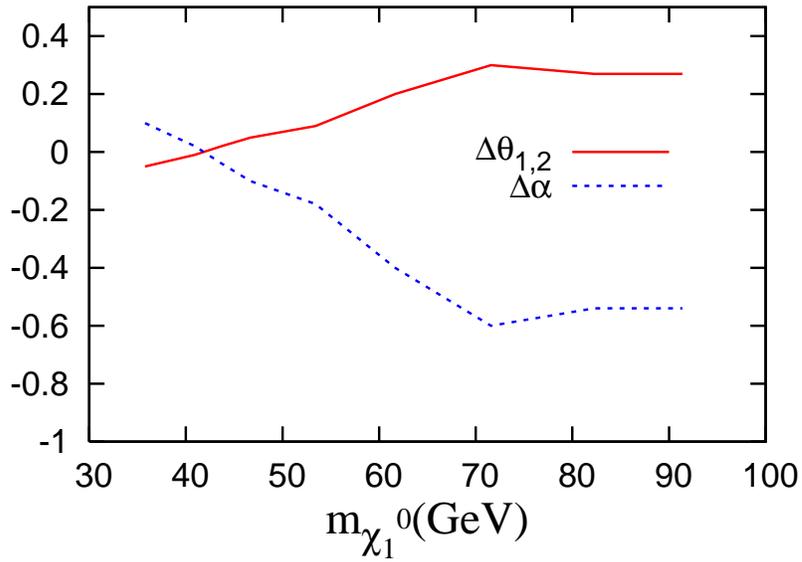}
\end{center}
\caption{$\Delta \theta_{1,2}$ and $\Delta \alpha$ vs. $m_{\chi_1^0}$.}
\label{figure4}
\end{figure}

\begin{figure}
\begin{center}
\includegraphics[height=3.2 in]{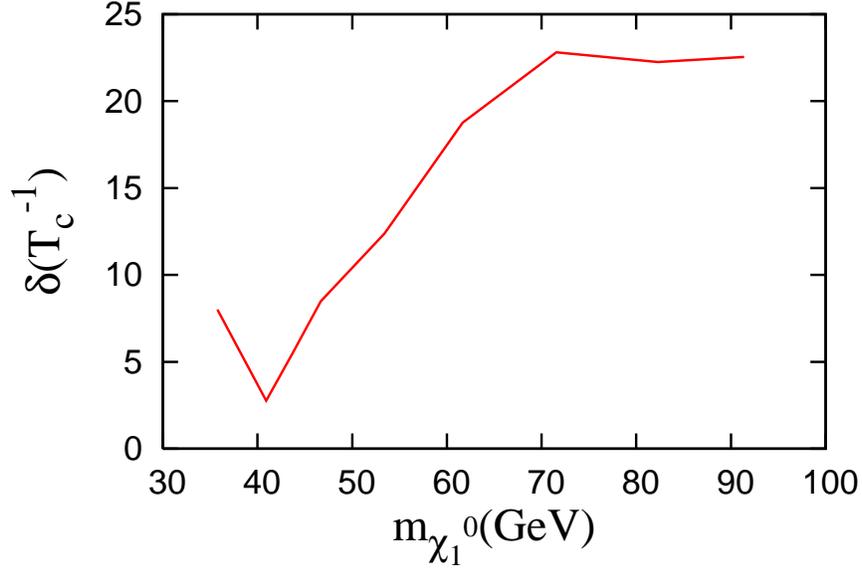}
\end{center}
\caption{$\delta$ vs. $m_{\chi_1^0}$.}
\label{figure5}
\end{figure}

\begin{figure}
\begin{center}
\includegraphics[height=3.2 in]{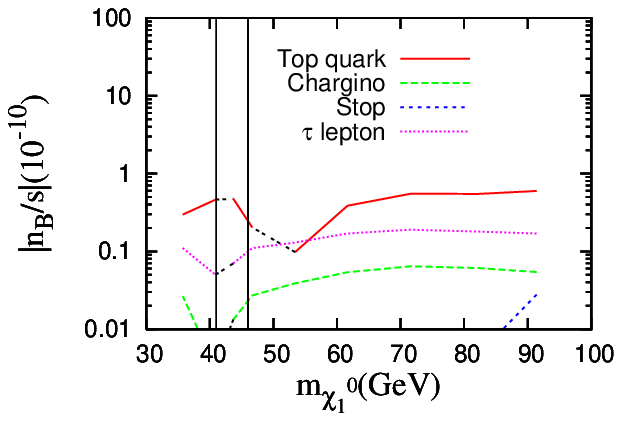}
\end{center}
\caption{$\frac{n_B}{s}$ vs. $m_{\chi_1^0}$. There is a sign flip for $n_B/s$ in the region denoted by black-dashed lines. The two black solid lines 
specify the $m_{\chi^0_1}$ values for which the observed dark matter relic density can be explained by $\chi_1^0$.}
\label{figure6}
\end{figure}

Besides the baryon asymmetry, another mystery is dark matter (DM), 
which has been measured precisely by astrophysical and cosmological probes~\cite{Amsler:2008zz}: 
\begin{eqnarray}
\Omega_{\rm DM}h^2=0.106 \pm 0.008.
\end{eqnarray}
Given that supersymmetry with $R$-parity conservation could have a neutral LSP as a good candidate 
for the latter, e.g., the lightest neutralino $\chi_1^0$, one naturally wants to know whether or not these two cosmological problems 
can be solved in the same framework.

To answer this question, we will not scan the whole space, since it is large. Instead, 
we vary parameters to which the neutralino DM relic density or $m_{\chi^0_1}$ is sensitive. 
From the discussions in the previous sections, $m_{H^0_d}^2$ can be such a parameter.
Explicitly, we use parameter values given in Table~\ref{parameter}, except that we vary $m_{H^0_d}^2$ between $10^4 \sim 5\times 10^5$ GeV$^2$
and set $\gamma= 0.95 \pi$. 
The relation between $m_{\chi^0_i}$ and $m_{H^0_d}^2$ is illustrated in Figure~\ref{figure1}. As $m_{H^0_d}^2$ increases, $m_{\chi^0_i}$ 
decreases from more than 90 GeV to 35 GeV. This is because $m_{\chi_1^0} \sim h v_1$ in the limit of large $\langle S_{1,2,3}\rangle$ 
and $v_1 \ll v_2 \sim \langle S \rangle$ (see Eqs.~(\ref{901}-\ref{903})) while a larger $m_{H^0_d}^2$ implies a smaller $v_1$. This feature is important for the DM relic density, since the $Z$ pole region is covered. 
As is well-known, the relic density of neutralino CDM is usually overproduced. One way to reduce the abundance 
is from the $Z$-pole region, where we have $m_Z \sim 2 m_{\chi_1^0}$ and the annihilation of 
two $\chi_1^0$ through a $Z$ boson is greatly enhanced. In this region, therefore, we can easily obtain an acceptable value for the
$\chi_1^0$ relic density.
As a first-order approximation, we calculate the $\chi_1^0$ relic density today, only counting in the processes mediated by gauge bosons 
(for general discussions on the relevant calculations, e.g., see~\cite{Bertone:2004pz}). 
The results  are shown in Figure~\ref{figure2}.

Let us come back to EWBG for this parameter region. With $m_{\chi_1^0}$ (or equivalently $m_{H^0_d}^2$) as a varying parameter, the strength of the EWPT,
the magnitude of $CP$-violation, the thickness of the bubble wall and the produced baryon asymmetry by different particle species are shown in 
Figures~\ref{figure3}-\ref{figure6}.  From Figure~\ref{figure3}, it is easy to see that the measure of the EWPT strength $v(T_c)/T_c$ 
is always larger than 1.1.  As  $m_{\chi_1^0}$ increases or equivalently $m_{H^0_d}^2$ decreases, its value increases. This can be understood 
because a smaller scaled $m_{H^0_d}^2$ (the real $m_{H^0_d}^2$ value is also smaller) usually leads to a larger scaled $v_1(T_c)$ in the true vacuum, 
while keeping the scaled $v_2(T_c)$ and $T_c$ almost invariant ($T_c \sim 1.4$ in our simulation).  In Figure~\ref{figure4}, we see that  $\Delta \theta_{1,2}$ and $\Delta \alpha$ are close to zero around 
$m_{\chi_1^0} \approx 40$ GeV, which explains the small wall thickness in Figure~\ref{figure5} (recall that a large phase variation crossing the wall usually leads to 
a small wall thickness (see discussion in Subsection~\ref{Wall Thickness})) and a sign flip of $n_B/s$ for the same $m_{\chi_1^0}$ value in Figure~\ref{figure5}. Another sign flip for the top curve in Figure~\ref{figure5}
is related to the ones shown in Figures~\ref{fig101}-\ref{fig103}, which is a result of balance between the contributions from the particle reflection and transmission.  
Finally, for most region of $m_{\chi^0_i}\sim 35-90$ GeV a strong enough first order EWPT and a reasonable baryon asymmetry can be obtained, 
and in the $Z$ pole region the right $\chi_1^0$ relic density can be produced (see Figure~\ref{figure6}). The two cosmological mysteries, baryon asymmetry and DM, can therefore be explained simultaneously. 

%
%
%
%
%
%
%
%

\section{Some Other Cosmological Effects}

\subsection{Superconducting Cosmic Strings }

There may exist phase transitions in the early Universe due
to  gauge symmetry breaking, and then topological
defects, such as monopoles and/or cosmic strings, etc, can be
produced~\cite{Vilenkin:1984ib, AVES-I}. If the 
gauge group $G$ is broken down to
a subgroup $H$, we will have monopoles and cosmic strings if
the homotopy group  $\pi_2 (G/H) \not=I$ and 
$\pi_1 (G/H) =Z\not=I$, respectively, where $Z$ is the group of integers. In particular, 
 the cosmic strings may be superconducting because 
of the presence of a charged field condensate in the core
of the string~\cite{Witten:1984eb}.  The gauge symmetry in the 
simplest cosmic string model is 
$U(1)\times U(1)'$ where the $U(1)'$ gauge symmetry
is broken via the Higgs mechanism~\cite{Witten:1984eb}. 
Note that $\pi_1 (U(1)') =Z$, so we can
have a topologically stable cosmic string solution. Although the
$U(1)$ symmetry is respected by the vacuum, it can be broken in the 
string core and then give rise to a bosonic condensate. Moreover,
there may exist fermionic superconducting strings that have
fermionic zero modes in its core~\cite{Witten:1984eb}. These 
fermions can have non-zero $U(1)$ charges, and arise from 
Yukawa couplings to the scalar fields.
In the non-supersymmetric models, we only need to introduce two
complex scalar fields: one is charged under $U(1)$ while the other
is charged under $U(1)'$~\cite{Witten:1984eb}. However, 
in order to have the bosonic
superconductivity for cosmic strings in the supersymmetric 
$U(1)\times U(1)'$ model, one needs to introduce at least five 
chiral superfields~\cite{Morris:1995wd}.
Interestingly, one can realize the fermionic superconductivity in
the simplest supersymmetric Abelian Higgs models~\cite{Davis:1997bs}.

In the sMSSM, we have the fermionic superconductivity, and  supersymmetry
is broken in the core of the cosmic strings (supersymmetry is not
broken outside of the string core) which is similar to
that in Ref.~\cite{Davis:1997bs}. However, 
we do not have the bosonic superconductivity since all
the Higgs fields are charged under the $U(1)'$ gauge symmetry.
Because the $U(1)'$ gauge symmetry is broken around the TeV 
scale,  the string tension $T_{\rm S}$ should be around
the TeV scale as well. 
The gravitational effect of cosmic strings with
$T_{\rm S}^2/M^2_{\rm Pl} \sim 10^{-30}$ is extremely weak and is not within foreseeable detection capabilities. Interestingly,
we can search for the signatures of particle emissions,
such as positrons or other decaying particles produced
where the current on the string quenches. Similar
to the discussions in Ref.~\cite{Ferrer:2005xva}, which also 
has strings around 1 TeV, the sMSSM might explain the observed 511 keV emission from the 
electron-positron annihilation in the Galactic 
bulge~\cite{Milne:2001zs, Jean:2003ci, Knodlseder:2003sv}.

\subsection{Gravitational Waves}

There are a few experiments under consideration that might
detect for the first time a stochastic background of GWs, for example, the planned
Laser Interferometer Space Antenna (LISA)~\cite{LISA}, 
and the Big Bang Observer (BBO)~\cite{BBO}, which is the 
follow-on mission to LISA.
One of the well-motivated and convincing sources for GWs 
is inflation, and the discovery of a relic gravitational 
background could be a smoking gun signal from inflation.
Another source of the stochastic GWs are strong first-order
phase transitions. In contrast with the inflationary spectrum, 
the spectrum from a phase transition is not flat and has 
a characteristic peak related to the temperature at which
the phase transition took place. 
Especially, LISA and BBO will have fairly
good sensitivity at the frequencies that coincide with 
the redshifted spectrum of GWs produced during an electroweak
phase transition at temperatures around 100 GeV.
Thus, the GW observations will
give us information about EWBG since 
the strong first-order electroweak
phase transition is a necessary requirement.

There are two distinct mechanisms that can generate the GWs 
during a first-order electroweak phase transition:
the bubble collisions~\cite{Kosowsky:1991ua, Kosowsky:1992rz,
Kosowsky:1992vn, Kamionkowski:1993fg} and the turbulent 
motion of the primordial 
plasma~\cite{Kosowsky:2000rq, Kosowsky:2001xp, Dolgov:2002ra,
Caprini:2006jb, Gogoberidze:2007an}. 
The GWs from the electroweak phase transition have been
studied in a model independent way in 
Refs.~\cite{Nicolis:2003tg} and~\cite{Grojean:2006bp}. 
The results can be presented as a function of
two main parameters: the typical size of the colliding
bubbles and the available energy. It was concluded that
a sufficiently strong electroweak phase transition could
lead to an observable GW signal at LISA. However,
for a definite model, the bubble size and the available
energy are correlated. Thus, it is still very important to 
study the concrete models in details. 

It was found previously that in the MSSM the produced 
amount of GWs is orders of magnitude below the LISA sensitivity. 
The situation in the NMSSM seems much more promising 
because the trilinear term $SH_d H_u$ can naturally lead to a much stronger 
electroweak phase transition~\cite{Apreda:2001us}. 
Recently, it was pointed out that in 
Ref.~\cite{Apreda:2001us} a rough approximation was used 
to calculate the bubble configurations, which
overestimated the strength of the phase transition
and the GW signal~\cite{Huber:2007vva}. Moreover, 
in Refs.~\cite{Huber:2007vva, Huber:2008hg} the nMSSM, 
which solves  the $\mu$ problem and domain wall
problem simultaneously, was considered. It was shown that
the GW signals can be detected at BBO for part of parameter
space, but cannot be observed at 
LISA~\cite{Huber:2007vva, Huber:2008hg}.

In the sMSSM the strong first-order electroweak phase 
transitions  arises from the trilinear term $SH_d H_u$
in the superpotential,
which is similar to the NMSSM and nMSSM. The GW signals might therefore be detectable 
at BBO for part of the parameter space, but cannot be 
observed at LISA~\cite{Huber:2007vva, Huber:2008hg}. 
As a side comment, the phase transition
for the $U(1)'$ gauge symmetry breaking may not be
strong first order since the coupling
$\lambda$ in the trilinear term $\lambda S_1 S_2 S_3$
is small, and one might not be able to observe the GWs from
the $U(1)'$ gauge symmetry breaking.

%
%
%
%
%
%
%
%

\section{Discussion and Conclusions}

In this paper we discussed EWBG and its correlation with the neutralino CDM in the sMSSM. We first 
constructed two anomaly-free models. In Model I, we embedded the
$SU(3)_C\times SU(2)_L \times U(1)_Y \times U(1)'$ gauge symmetry
into a larger gauge symmetry, $E_6$, whose representations are
anomaly-free. We considered three families of SM fermions and one
pair of Higgs doublets arising from three $E_6$ fundamental
representation ${\bf 27}$s. To include the SM singlet Higgs
particles, achieve gauge coupling unification and cancel the
$U(1)'$ anomaly, we assumed that three pairs of vector-like SM
singlets and one pair of vector-like Higgs doublets from two pairs
of ${\bf 27}$ and ${\bf {27^*}}$ are light while the other
particles in the two pairs of ${\bf 27}$ and ${\bf {27^*}}$ are absent
or very heavy. In Model II,
we calculated the anomaly-free conditions, and added the minimal
number of exotic particles to cancel the $U(1)'$ anomalies. We
required that all the particles have rational $U(1)'$ charges.
Although we had minimal exotic particles, we lost the gauge
coupling unification. We also presented the general superpotential
and the supersymmetry breaking soft terms in both models.

We discussed the one-loop effective potential at finite
temperature in the 't~Hooft-Landau gauge and in the
$\overline{MS}$-scheme.
We showed that there
exists a strong enough first order EWPT because of the large
trilinear term $A_h h S H_u H_d$ in the tree-level Higgs
potential. Unlike the MSSM, the stop masses can be
very heavy. The EWPT features in both models are
quite similar because the exotic particles' contributions to the
one-loop effective potential at finite temperature are suppressed due to their heavy masses.
We therefore restricted the detailed analysis to Model I.

In the early Universe the first order EWPT is realized by
nucleating bubbles of the broken phase. Crossing the bubble walls, the VEVs of the Higgs fields (including their magnitudes and phases) are space-dependent. 
Because the dynamical behaviors of these bubbles,
such as wall profile and expansion velocity, can have important 
influences on the production of the baryon asymmetry, we also 
studied their physics in detail. Numerical results show that the bubble wall thickness
varies from $3 \sim 30~T_c^{-1}$ as a monotonically
increasing function of the phase changes of the Higgs fields. We also argued that the
wall velocity in the sMSSM cannot be larger than that
in the MSSM under the same phase transition condition and thus
should be non-relativistic. This fact implies that  EW sphaleron processes 
have more time to occur and hence will enhance the final baryon asymmetry.

We also discussed possible $CP$ violation
introduced by the extended Higgs sector. Unlike the MSSM, where
there is no $CP$-violation at tree level in the Higgs sector, 
the $CP$ symmetry can be broken spontaneously as well as explicitly. 
With loop corrections included, there may coexist vacua at finite temperature with broken and unbroken EW symmetry. 
The values of these spontaneous $CP$ phases usually are different in these vacua. 
In our work, the SCPV provides
a direct source for baryogenesis while its magnitude is mediated by an
explicit one from the secluded sector.
These new $CP$ sources do not introduce significant new contributions to EDMs. After proper field redefinitions, the
$CP$-violation phases will only appear in the Higgs mass matrix
through soft masses associated with singlet components. Numerical results show that, for typical
parameter values, their contributions to EDMs will be about six or
seven orders smaller than the experimental upper limits. These contributions disappear completely in the
limit with a trivial explicit $CP$ phase, where the spontaneous $CP$ phases are trivial in the true vacuum but not where 
EWBG occurs.

Subsequently, we systematically studied non-local EWBG in both the thin wall and thick wall
regimes. We calculated the contribution from leptons in the thin wall regime, and the ones from
squarks, charginos and quarks in the thick wall regime in terms of the profile of bubble wall. For leptons and quarks, the induced $CP$-violating
currents outside the bubbles are proportional to Yukawa couplings squared, so we only
considered the contributions from $\tau$ leptons and top quarks. 
We found that the $CP$-violating currents for stops and charginos are very different from those obtained
in the MSSM. Due to the space-dependence of the relevant $CP$ phases, they do not require a variation of $\tan\beta$ 
in the bubble wall to have a non-trivial structure at the lowest order of Higgs insertion. 
In addition, there exists a new type of $CP$-violating current at the leading order which can contribute to the 
generation of the baryon asymmetry. The $CP$-violating current of the new type has important influence on the EWBG.  
First, the stop contribution can be quadratically enhanced by a large $A_{h_t}$. Second, in addition to 
$\tau$ leptons, top squarks and charginos, top quarks can also play a significant role in the EWBG. 
We emphasize that all of these features are results of the SCPV-driven EWBG, 
so they are not sensitive to the concrete embeddings of the sMSSM.
Numerical results show that the produced baryon asymmetry 
is large enough to explain the cosmological observations.

After that, we studied the correlation between EWBG and neutralino CDM in the sMSSM. 
Though we did not scan the whole parameter
space, we found that there exists a region where a strong enough first order EWPT, large $CP$ phase variations crossing the bubble wall, 
a reasonable baryon asymmetry,  
as well as an acceptable neutralino LSP relic density can be achieved simultaneously. 
We also commented on possible cosmological signals of the model: superconducting cosmic strings and GWs. 
Particle emission from the decays of cosmic strings and GWs from EWPT
could be observed within the foreseeable future.

The secluded $U(1)'$-breaking sector plays an essential role. It induces ECPV and SCPV at tree level in the Higgs sector, 
helping avoid significant new contributions to the EDMs which may hinder a successful EWBG in supersymmetric models, and also influences the 
bubble wall physics. 
It provides more degrees of freedom to drive the realization of EWBG, and makes the EWBG physics in the sMSSM very different from those in the
MSSM and NMSSM. These differences make it interesting to further study the relevant collider implications for the LHC~\cite{lowmass}, which we will leave to future exploration.

\section*{Acknowledgments}

We thank Michael Ramsey-Musolf and Carlos Wagner for 
fruitful discussions. The work of Paul Langacker is supported by the IBM Einstein 
Fellowship and by NSF grant PHY-0503584. The work of Tianjun Li is supported in part 
by the Natural Science Foundation of China 
under grant No. 10821504, 
by the DOE grant DE-FG03-95-Er-40917, 
and by the Mitchell-Heep Chair in 
High Energy Physics. 
The work of Tao Liu is supported by the Fermi-McCormick Fellowship 
and by the DOE through Grant No. DE-FG02- 90ER40560. This work was completed 
at the Aspen Center for Physics.

\newpage

\appendix

\section{Mass Matrices for the Particles in Model I}

\setcounter{equation}{0}

\renewcommand{\theequation}{\Alph{section}.\arabic{equation}}

For the Higgs mass matrices, here we only present the results in the limit $m_{S_1
S_2}^2=0$, where there is no $CP$ violation. It is not difficult to
extend them to the general case with non-zero $m_{S_1 S_2}^2$ and
$CP$ violation, as considered in the main text. Let us define
\begin{eqnarray}
v_1=\langle H_d^0 \rangle ~,~ v_2=\langle H_u^0\rangle ~,~ \tan\beta={{v_2}\over v_1}
~,~\,
\end{eqnarray}
\begin{eqnarray}
s=\langle S\rangle ~,~  s_i = \langle S_i\rangle 
~.~\,
\end{eqnarray}
Then the mass-square matrix for the $CP$-odd neutral Higgs particles in
the basis $\{H_d^{0i} = Im(H_d^0), H_u^{0i}, S^{0i},
S_1^{0i}, S_2^{0i}, S_3^{0i}\}$ is
\begin{eqnarray}
M_{A^{0}}^2 =\left(\matrix{ O_{A^{0}} & C_{A^{0}} \cr C_{A^{0}}^T&
S_{A^{0}} \cr}\right) ~,~ \,
\end{eqnarray}
where
\begin{eqnarray}
O_{A^{0}} = \left(\matrix{\beta_{H_d}^2  & A_h h s & A_h h v_2 \cr
A_h h s & \beta_{H_u}^2  & A_h h v_1 \cr A_h h v_2 & A_h h v_1 &
\beta_{S}^2 \cr }\right) ,\,
\end{eqnarray}
\begin{eqnarray}
S_{A^{0}}= \left(\matrix{ \beta_{S_1}^2 & A_{\lambda} \lambda s_3
& A_{\lambda} \lambda s_2 \cr A_{\lambda} \lambda s_3 &
\beta_{S_2}^2& A_{\lambda} \lambda s_1 \cr A_{\lambda} \lambda s_2
& A_{\lambda} \lambda s_1 & \beta_{S_3}^2  \cr}\right) ,\,
\end{eqnarray}
\begin{eqnarray}
C_{A^{0}}= \left(\matrix{0 & 0& 0\cr 0 & 0 & 0\cr -m^2_{S S_1} &
-m^2_{S S_2} & 0 \cr }\right) ,\,
\end{eqnarray}
and
\begin{eqnarray}
\beta_{H_d}^2 = m_{H_d}^2 +h^2 (v_2^2 +s^2) +{{G^2}\over 4}
(v_1^2-v_2^2) +g_{Z'}^2 Q_{H_d} \Delta ~,~\,
\end{eqnarray}
\begin{eqnarray}
\beta_{H_u}^2 = m_{H_u}^2 + h^2 (v_1^2 +s^2) +{{G^2}\over 4}
(v_2^2-v_1^2) +g_{Z'}^2 Q_{H_u} \Delta  ~,~\,
\end{eqnarray}
\begin{eqnarray}
\beta_{S}^2 = m_{S}^2 + h^2 (v_1^2 +v_2^2) +g_{Z'}^2 Q_{S} \Delta
~,~\,
\end{eqnarray}
\begin{eqnarray}
\beta_{S_i}^2 = m_{S_i}^2 + \lambda^2 \sum_{j\ne i}s_j^2  + g_{Z'}^2
Q_{S_i} \Delta ~,~\,
\end{eqnarray}
where
\begin{eqnarray}
\Delta \equiv Q_S s^2 + Q_{H_d} v_1^2 + Q_{H_u} v_2^2 + \sum_{i=1}^3 Q_{S_i} s_i^2
~.~\,
\end{eqnarray}

Similarly, in the basis $\{H_d^{0r} = Re(H_d^0),
H_u^{0r}, S^{0r}, S_1^{0r}, S_2^{0r}, S_3^{0r}\}$,  the
mass-square matrix for the $CP$-even neutral Higgs particles is
\begin{eqnarray}
M_{H^{0}}^2 =\left(\matrix{ O_{H^{0}} & C_{H^{0}} \cr C_{H^{0}}^T&
S_{H^{0}} \cr}\right) ~,~ \,
\end{eqnarray}
where
\begin{eqnarray}
O_{H^{0}}= \left(\matrix{ \kappa_{H_d}^2  & \kappa_{H_d, H_u} &
\kappa_{H_d, S} \cr \kappa_{H_d, H_u} & \kappa_{H_u}^2  &
\kappa_{H_u, S}  \cr \kappa_{H_d, S}  & \kappa_{H_u, S}  &
\kappa_S^2  \cr }\right) ,\,
\end{eqnarray}
\begin{eqnarray}
S_{H^{0}}= \left(\matrix{ \kappa_{S_1}^2 & \kappa_{S_1, S_2}&
\kappa_{S_1, S_3}  \cr \kappa_{S_1, S_2} & \kappa_{S_2}^2 &
\kappa_{S_2, S_3}\cr \kappa_{S_1, S_3}  & \kappa_{S_2, S_3} &
\kappa_{S_3}^2 \cr}\right) ,\,
\end{eqnarray}
\begin{eqnarray}
C_{H^{0}}= \left(\matrix{ \kappa_{H_d, S_1} & \kappa_{H_d, S_2} &
\kappa_{H_d, S_3} \cr \kappa_{H_u, S_1} & \kappa_{H_u, S_2} &
\kappa_{H_u, S_3} \cr \kappa_{S, S_1} +m^2_{S, S_1} & \kappa_{S,
S_2}+m^2_{S, S_2} & \kappa_{S, S_3} \cr}\right) ,\,
\end{eqnarray}
and
\begin{eqnarray}
\kappa_{H_d}^2 = 2 \left( {{G^2} \over {4}} + g_{Z'}^2 Q_{H_d}^2
\right) v_1^2 + m_{H_d}^2 +h^2 (v_2^2 +s^2) +{{G^2}\over 4}
(v_1^2-v_2^2) +g_{Z'}^2 Q_{H_d} \Delta ~,~\,
\end{eqnarray}
\begin{eqnarray}
\kappa_{H_u}^2 = 2 \left( {{G^2} \over {4}} + g_{Z'}^2 Q_{H_u}^2
\right) v_2^2 + m_{H_u}^2 +h^2 (v_1^2 +s^2) +{{G^2}\over 4}
(v_2^2-v_1^2) +g_{Z'}^2 Q_{H_u} \Delta ~,~\,
\end{eqnarray}
\begin{eqnarray}
\kappa_S^2 = 2 g_{Z'}^2 Q_S^2 s^2 +  m_{S}^2 +
h^2 (v_1^2 +v_2^2) + g_{Z'}^2 Q_{S} \Delta ~,~\,
\end{eqnarray}
\begin{eqnarray}
\kappa_{S_i}^2 = 2 g_{Z'}^2 Q_{S_i}^2 s_i^2 + m_{S_i}^2 +
\lambda^2  \sum_{j\ne i}s_j^2 + g_{Z'}^2 Q_{S_i} \Delta ~,~\,
\end{eqnarray}
\begin{eqnarray}
\kappa_{H_d, H_u} = 2 \left( h^2 - {{G^2} \over {4}} + g_{Z'}^2
Q_{H_d} Q_{H_u} \right) v_1 v_2 - A_h h s ~,~\,
\end{eqnarray}
\begin{eqnarray}
\kappa_{H_i, S} = 2 \left( h^2 + g_{Z'}^2 Q_{H_i} Q_S \right) v_i
s - |\epsilon_{i j}| A_h h v_j ~,~\,
\end{eqnarray}
\begin{eqnarray}
\kappa_{H_i, S_j} = 2 g_{Z'}^2 Q_{H_i} Q_{S_j} v_i s_j ~,~
\kappa_{S, S_i} = 2 g_{Z'}^2 Q_S Q_{S_i} s s_i ~,~\,
\end{eqnarray}
\begin{eqnarray}
\kappa_{S_i, S_j} = 2 (\lambda^2 + g_{Z'}^2 Q_{S_i} Q_{S_j}) s_i
s_j -|\epsilon_{ijk}| A_{\lambda} \lambda s_k ~.~\,
\end{eqnarray}

The charged Higgs mass is
\begin{eqnarray}
M_{H^{\pm}}^2 = M_W^2 + {{2 A_h h s}\over\displaystyle
{\sin 2\beta}} - h^2 (v_1^2 +v_2^2) ~,~\,
\end{eqnarray}
where $M_W^2={g_2^2} (v_1^2 +v_2^2)/2$.

In the basis $\{ {\tilde B}^{\prime}, \tilde B, \tilde W_3^0,
\tilde H_d^0, \tilde H_u^0, \tilde S, \tilde S_1, \tilde S_2,
\tilde S_3\}$, the neutralino mass matrix is
\begin{eqnarray}
M_{\tilde \chi^{0}} =\left(\matrix{ M_{\tilde \chi^{0}}(6, 6) &
M_{\tilde \chi^{0}}(6, 3) \cr M_{\tilde \chi^{0}}(6, 3)^T&
M_{\tilde \chi^{0}}(3, 3) \cr}\right) ~,~ \, \label{901}
\end{eqnarray}
where
\begin{eqnarray}
M_{\tilde \chi^{0}} (6, 6)= \left(\matrix{M_1^{\prime}
&0&0&\Gamma_{H_d^0}^*&\Gamma_{H_u^0}^*&\Gamma_{S}^*\cr 0&M_1&0& -{1\over
\sqrt 2} g_1 H_d^{0*} & {1\over \sqrt 2} g_1 H_u^{0*} &0\cr 0&0&M_2& {1\over
\sqrt 2} g_2 H_d^{0*} & -{1\over \sqrt 2} g_2 H_u^{0*}&0\cr \Gamma^*_{H_d^0} &
-{1\over \sqrt 2} g_1 H_d^{0*}
 & {1\over \sqrt 2} g_2 H_d^{0*}
&0& h S& hH_u^{0} \cr \Gamma^{*}_{H_u^0} & {1\over \sqrt 2} g_1 H_u^{0*} &
-{1\over \sqrt 2} g_2 H_u^{0*} & h S &0& h H_d^{0} \cr \Gamma^*_S &0&0& h H_u^0
&  h H_d^0 &0 \cr}\right) ,\, \label{902}
\end{eqnarray}
and
\begin{eqnarray}
M_{\tilde \chi^{0}} (3, 3)= \left(\matrix{ 0& \lambda S_3&
\lambda S_2 \cr \lambda S_3 &0& \lambda S_1 \cr \lambda S_2 &
\lambda S_1 & 0  \cr}\right) ,\,     \label{903}
\end{eqnarray}
where $\Gamma_{\phi} \equiv \sqrt 2 g_{Z'} Q_{\phi} 
\phi$; and $M_1^{\prime}$, $M_1$ and $M_2$ are the gaugino masses for
$U(1)^{\prime}$, $U(1)$ and $SU(2)_L$, respectively. The first row
of $M_{\tilde \chi^{0}} (6, 3)$ is given by $\left( \Gamma^*_{S_1}
\ \  \Gamma^*_{S_2}  \ \ \Gamma^*_{S_3} \right)$, while the other
entries are zero.

The chargino mass matrix  is
\begin{eqnarray}
M_{\tilde \chi^{\pm}} =\left(\matrix{M_2 & \frac{g_2}{\sqrt 2}  H_u^{0*}
\cr \frac{g_2 }{\sqrt 2}H_d^{0*} & h S\cr}\right) ~.~ \,
\end{eqnarray}
In contrast to the Higgs mass matrix, the neutralino and chargino mass matrices 
as well as the other ones in the following, are written in a general background with 
$CP$ violation.

In the basis $(B_{\mu}, W_{\mu}^3, Z'_{\mu})$, the mass matrix for
the neutral gauge bosons is
\begin{equation}
 M_{B, W^3, Z'}  =  \left( \begin{array}{c c c}
 M_{B, W^3, Z'}^{11}  & M_{B, W^3, Z'}^{12}
& M_{B, W^3, Z'}^{13} \\

M_{B, W^3, Z'}^{21}  &  M_{B, W^3, Z'}^{22}
&  M_{B, W^3, Z'}^{23} \\
 M_{B, W^3, Z'}^{31}  &  M_{B, W^3, Z'}^{32} &
M_{B, W^3, Z'}^{33} \\
  \end{array} \right),
\label{eqgg}
\end{equation}
where
\begin{eqnarray}
M_{B, W^3, Z'}^{11} = {1\over 2} g_1^2 [|H_u^0|^2 + |H_d^0|^2] ~,~
\end{eqnarray}
\begin{eqnarray}
M_{B, W^3, Z'}^{12}=M_{B, W^3, Z'}^{21}  = - {1\over 2} g_1 g_2
[|H_u^0|^2 + |H_d^0|^2] ~,~
\end{eqnarray}
\begin{eqnarray}
M_{B, W^3, Z'}^{13} = M_{B, W^3, Z'}^{31} = g_1 g_{Z'} [ Q_{H_u}
|H_u^0|^2 - Q_{H_d} |H_d^0|^2] ~,~
\end{eqnarray}
\begin{eqnarray}
M_{B, W^3, Z'}^{22} = {1\over 2} g_2^2 [|H_u^0|^2 + |H_d^0|^2] ~,~
\end{eqnarray}
\begin{eqnarray}
M_{B, W^3, Z'}^{23} = M_{B, W^3, Z'}^{32} =
 g_2 g_{Z'} [ Q_{H_d} |H_d^0|^2 - Q_{H_u} |H_u^0|^2] ~,~
\end{eqnarray}
\begin{eqnarray}
M_{B, W^3, Z'}^{33} = 2 g_{Z'}^2 (Q_S^2 |S|^2 + Q_{H_u}^2 |H_u^0|^2
+ Q_{H_d}^2 |H_d^0|^2 +\sum_{i=1}^3 Q_{S_i}^2 |S_i|^2) ~.~
\end{eqnarray}

In addition, for a scalar $\phi$, we define
\begin{eqnarray}
\Delta_{\phi} \equiv [T_3^{\phi} - Q_{EM}^{\phi} \sin^2\theta_W]
 M_Z^2 \cos (2\beta)~,~\,
\end{eqnarray}
\begin{eqnarray}
\Delta'_{\phi} \equiv Q'_{\phi} g_{Z'}^2 [ Q_{H_u} |H_u^0|^2 +
Q_{H_d} |H_d^0|^2 +Q_S |S|^2 + \sum_{i=1}^3 Q_{S_i} |S_i|^2 ]
~.~\,
\end{eqnarray}
The up-type squark mass matrix is
\begin{equation}
 M_{\tilde u}  =  \left( \begin{array}{c c}
M_{{\tilde u}_L^i}^2  & h_{u_i}^* (h S H_d^0 - A_{h_{u_i}}^* H_u^{0 *})  \\
h_{u_i} (h^* S^* H_d^{0 *} - A_{h_{u_i}} H_u^{0 })   & M_{{\tilde u}_R^i}^2  \\
  \end{array} \right),
\end{equation}
where
\begin{eqnarray}
M_{{\tilde u}_L^i}^2 =m^2_{{\tilde Q}_i} + m^2_{u_i} +
\Delta_{{\tilde u}_L^i} + \Delta'_{{\tilde u}_L^i} ~,~\,
\end{eqnarray}
\begin{eqnarray}
M_{{\tilde u}_R^i}^2 =m^2_{{\tilde u}_R^i} + m^2_{u_i} +
\Delta_{{\tilde u}_R^i} + \Delta'_{{\tilde u}_R^i} ~,~\,
\end{eqnarray}
while the down-type matrix is obtained by $u \leftrightarrow d$. The mass matrices for the sleptons and exotic scalars are similar.

\section{$\Pi_{\Phi} (T)$ for Scalar Particles in Model I}

\setcounter{equation}{0}

The $\Pi_{\Phi} (T)$s for the Higgs doublets are
\begin{eqnarray}
\Pi_{H_d} (T) =
  {3\over 8} g_2^2 T^2 +{1\over {8}} g_1^2 T^2
+{1\over 2} g_{Z'}^2 Q_{H_d}^2 T^2 +\sum_{i=1}^3 \left(
{3\over 4} h_{d_i}^2 + {1\over 4} h_{e_i}^2\right) T^2
+{1\over 4} h^2 T^2~,~\,
\end{eqnarray}
\begin{eqnarray}
\Pi_{H_u} (T) =
  {3\over 8} g_2^2 T^2 +{1\over {8}} g_1^2 T^2
+{1\over 2} g_{Z'}^2 Q_{H_u}^2 T^2  +\sum_{i=1}^3 {3\over
4} h_{u_i}^2  T^2 + {1\over 4} h^2 T^2 ~.~\,
\end{eqnarray}
For the SM singlets:
\begin{eqnarray}
\Pi_{S} (T) &=& {1\over 2} g_{Z'}^2 Q_s^2 T^2 + {1 \over 4}
\alpha^2 T^2+ {1\over 2} h^2 T^2 + \sum_{i=1}^3 {3\over 4}
(\alpha_i^D)^2 T^2 \nonumber\\&& + \sum_{k=1}^2  {1\over 2}
(\alpha_k^{H^{\prime}})^2 T^2
 + \sum_{i=1}^3  {1\over 4} (\alpha_i^N)^2 T^2
~,~\,
\end{eqnarray}
\begin{eqnarray}
\Pi_{S_i} (T) = {1\over 2} g_{Z'}^2 Q_{S_i}^2 T^2 + {1\over 4}
{\lambda}^2 T^2 ~.~\,
\end{eqnarray}
For the longitudinal components of the
$SU(3)_C\times SU(2)_L\times U(1)_Y \times U(1)'$
 gauge bosons, $g$, $W$, $B$ and $Z'$:
\begin{eqnarray}
\Pi_{g_L} (T) = 6 g_3^2 T^2 ~,~\,
\Pi_{W_L} (T) = 6 g_2^2 T^2 ~,~\,
\end{eqnarray}
\begin{eqnarray}
\Pi_{B_L} (T) = 8 g_1^2 T^2 ~,~\,
\Pi_{Z'_L} (T) ={619 \over 2} g_{Z^{\prime}}^2 T^2  ~.~\,
\end{eqnarray}
For ${\tilde {\bar N}}_i$ and $S_L^i$:
\begin{eqnarray}
\Pi_{{\tilde {\bar N}}_i} (T) = {1\over 2} g_{Z^{\prime}}^2 Q_{N_i}^2 T^2+{1 \over 4
}(\alpha_i^N)^2 T^2 ~,~\,
\end{eqnarray}
\begin{eqnarray}
\Pi_{S_L^i} (T) = {1\over 2} g_{Z^{\prime}}^2 Q_{S_L^i}^2 T^2+{1
\over 4 }(\alpha_i^N)^2 T^2 ~.~\,
\end{eqnarray}
For $\tilde X$ and ${\tilde X}_3$:
\begin{eqnarray}
\Pi_{\tilde X} (T) = {1\over 2} g_{Z^{\prime}}^2 Q_{X}^2 T^2+{1 \over 4
}\alpha^2 T^2 ~,~\,
\end{eqnarray}
\begin{eqnarray}
\Pi_{{\tilde X}_3} (T) = {1\over 2} g_{Z^{\prime}}^2 Q_{X_3}^2 T^2+{1 \over 4
}\alpha^2 T^2 ~.~\,
\end{eqnarray}
For ${\tilde D}_i$ and ${\tilde {\bar D}}_i$:
\begin{eqnarray}
\Pi_{{\tilde D}_i} (T) = {2 \over 3} g_3^2 T^2 + {1 \over 18} g_1^2 T^2+
{1\over 2} g_{Z^{\prime}}^2 Q_{D_i}^2 T^2+{1 \over 4 }(\alpha_i^D)^2 T^2 ~,~\,
\end{eqnarray}
\begin{eqnarray}
\Pi_{{\tilde {\bar D}}_i} (T) = {2 \over 3} g_3^2 T^2 + {1 \over 18} g_1^2
T^2+ {1\over 2} g_{Z^{\prime}}^2 Q_{\bar{D_i}}^2 T^2+{1 \over 4
}(\alpha_i^D)^2 T^2 ~.~\,
\end{eqnarray}
For $H_u^{\prime k}$ and $H_d^{\prime k}$:
\begin{eqnarray}
\Pi_{H_u^{\prime k}} (T) = {3 \over 8} g_2^2 T^2 + {1 \over 8} g_1^2
T^2+ {1\over 2} g_{Z^{\prime}}^2 Q_{H_u^{\prime k}}^2 T^2
+{1 \over 4 }(\alpha_k^{H'})^2 T^2 ~,~\,
\end{eqnarray}
\begin{eqnarray}
\Pi_{H_d^{\prime k}} (T) = {3 \over 8} g_2^2 T^2 + {1 \over 8} g_1^2
T^2+ {1\over 2} g_{Z^{\prime}}^2 Q_{H_d^{\prime k}}^2 T^2
+{1 \over 4 }(\alpha_k^{H'})^2 T^2 ~.~\,
\end{eqnarray}
For $H_u^{\prime}$ and ${\bar H}_u^{\prime}$:
\begin{eqnarray}
\Pi_{H_u^{\prime}} (T) = {3 \over 8} g_2^2 T^2 + {1 \over 8} g_1^2
T^2+ {1\over 2} g_{Z^{\prime}}^2 Q_{H_u^{\prime }}^2 T^2  ~,~\,
\end{eqnarray}
\begin{eqnarray}
\Pi_{{\bar H}_u^{\prime}} (T) = {3 \over 8} g_2^2 T^2 + {1 \over 8} g_1^2
T^2+ {1\over 2} g_{Z^{\prime}}^2 Q_{{\bar H}_u^{\prime }}^2 T^2  ~.~\,
\end{eqnarray}
For the superpartners of the SM fermions:
\begin{eqnarray}
\Pi_{{\tilde Q}_i} (T) =
 {2\over 3} g_3^2 T^2 + {3\over 8} g_2^2 T^2 +{1\over {72}} g_1^2 T^2
+{1\over 2} g_{Z'}^2 {Q_{Q_i}}^2 T^2
 + {1\over 4} \left(h_{u_i}^2 + h_{d_i}^2\right) T^2~,~\,
\end{eqnarray}
\begin{eqnarray}
\Pi_{{\tilde u}_R^i} (T) =  {2\over 3} g_3^2 T^2 +{2\over {9}}
g_1^2 T^2 +{1\over 2} g_{Z'}^2 {Q_{{\bar u}_i}}^2 T^2+
{1\over 2} h_{u_i}^2 T^2~,~\,
\end{eqnarray}
\begin{eqnarray}
\Pi_{{\tilde d}_R^i} (T) =  {2\over 3} g_3^2 T^2 +{1\over {18}}
g_1^2 T^2 +{1\over 2} g_{Z'}^2  {Q_{{\bar d}_i}}^2   T^2  +
{1\over 2} h_{d_i}^2 T^2~,~\,
\end{eqnarray}
\begin{eqnarray}
\Pi_{{\tilde L}_i} (T) =
  {3\over 8} g_2^2 T^2 +{1\over {8}} g_1^2 T^2
+{1\over 2} g_{Z'}^2 {Q_{L_i}}^2 T^2
 + {1\over 4} h_{e_i}^2  T^2~,~\,
\end{eqnarray}
\begin{eqnarray}
\Pi_{{\tilde e}_R^i} (T) = {1\over {2}} g_1^2 T^2 +{1\over 2}
g_{Z'}^2 {Q_{{\bar e}_i}}^2 T^2 + {1\over 2} h_{e_i}^2 T^2~.~\,
\end{eqnarray}

\section{Derivation of $\Delta R$ in the Thick Wall Regime}

\label{DR}

\setcounter{equation}{0}

$\Delta R$ is the reflectivity asymmetry between the particle and antiparticle scatterings. 
It is crucial for the calculation of the $CP$-violating current in the EWBG.  
In the following, we will show how $\Delta R$ is calculated in the thick wall regime. 
We will explicitly calculate $\Delta R_{\tilde W^c\rightarrow\tilde H^c}(k_i,k_f,z)$ (see 
Eq.~(\ref{R2})) as an illustration. The calculations for $\Delta R_{\tilde t_{R} \rightarrow\tilde t_L}(k_i,k_f,z)$ and $\Delta R_{t_R \rightarrow t_L}(k_i,k_f,z)$ (see Eq.~(\ref{R1}) and Eq.~(\ref{R3})) are similar,  so we will not present their details.

The general formula for the probability calculation in quantum field theory is (eg. see~\cite{Peskin})
\begin{eqnarray}
P(\psi_i \to \psi_f) &=& \int \frac{d^3 p_f}{(2\pi)^3} \frac{1}{2E_f} | \langle p_f | \psi_i \rangle|^2   \label{801}\\   
&=& \int \frac{d^3 p_f}{(2\pi)^3} \frac{1}{2E_f}  \int \frac{d^3 p_i}{(2\pi)^3} \frac{\psi_i( p_i)}{\sqrt{2E_i}}  \int \frac{d^3 \tilde p_i}{(2 \pi)^3} \frac{\psi_i^*(\tilde p_i)}{\sqrt{2\tilde E_i}} 
  \langle p_f | p_i \rangle   \langle p_f | \tilde p_i \rangle^* . \nonumber    
\end{eqnarray}
Here the subscripts ``i''  and ``f'' denote initial and final states, respectively.
Given that in the model the Langrangian is approximately invariant under the translations in the $x$, $y$ and $t$ directions, 
it is natural to define  
\begin{eqnarray}
  \langle p_f | p_i \rangle  &=&   \langle p_f | i \int d^4x {\cal L} (z) |p_i \rangle \nonumber \\
     &=& iM(p_i \to p_f) (2\pi)^3 \delta(E_i-E_f) \delta(p_i^x-p_f^x) \delta(p_i^y-p_f^y),    \label{802}
\end{eqnarray}
with
\begin{eqnarray}
M &=&   \langle p_f^z | \int dz {\cal L} (z) |p_i^z \rangle .   \label{803}
\end{eqnarray}
Embedding Eq.~(\ref{802}) into Eq.~(\ref{801}), we find 
\begin{eqnarray}
P(\psi_i \to \psi_f) = \frac{1}{4E_iE_f} \frac{1}{v_i^z v_f^z} |M(k_i^z \to k_f^z)|^2,  \label{804}
\end{eqnarray}
where $k$, $E$ and $v$ are particle momentum, energy and velocity, respectively. For a given particle with mass $m$, 
all of the quantities are functions of $k_i^z$. 

For the scattering process  $\tilde W^c \rightarrow\tilde H^c$ in the chargino sector, $M(k_i^z,k_f^z, z)$ is 
\begin{eqnarray}
M(k_i^z,k_f^z, z) &=& \int_{z}^{z+\Delta(\tilde \gamma_{\tilde \chi^c})}  d z_1  \exp(i(k_i^z-k_f^z)z_1) \nonumber \\
&& (A(k_i^z,k_f^z, z_1) - A(k_i^z,k_f^z, z_1 \to z)),     \label{805}
\end{eqnarray}
with
\begin{eqnarray}
A(k_i^z,k_f^z, z_1) &=&  \bar u_{\tilde H^c} (k_f^z) [(u_1(z_1)\exp(-i\theta_1(z_1)-u_1(z)\exp(-i\theta_1(z))P_L   \nonumber \\ && +  (u_2(z_1)\exp(i\theta_2(z_1)-u_2(z)\exp(i\theta_2(z))P_R  ] u_{\tilde W^c} (k_i^z),  \label{806}
\end{eqnarray}
where $P_{L,R}$ denote chirality projection operators. Then,
\begin{eqnarray}
\Delta|M(k_i^z,k_f^z, z)|^2 &=& |M(k_i^z,k_f^z, z)|^2 -  (CP \ {\rm Conjugate}) \nonumber \\&=& \Delta {\cal A}(k_i^z,k_f^z,z_1,z_2) + \Delta {\cal A}(k_i^z,k_f^z,z_1 \to z,z_2\to z)  \nonumber\\
&& - \Delta {\cal A}(k_i^z,k_f^z,z_1 \to z,z_2) -\Delta {\cal A}(k_i^z,k_f^z,z_1,z_2 \to z),   \label{807}
\end{eqnarray}
with
\begin{eqnarray}
 \Delta {\cal A}(k_i^z,k_f^z,z_1,z_2) &=&  \int_{z}^{z+\Delta(\tilde \gamma_{\tilde \chi^c})}  d z_1 dz_2\exp(i(k_i^z-k_f^z)(z_1-z_2))  \nonumber \\
&& [A(k_i^z,k_f^z, z_1) A(k_i^z,k_f^z, z_2)^* -  (CP \ {\rm Conjugate})].  \label{808}
\end{eqnarray}
For the other $\Delta {\cal A}$s, only  ``$z_1$'' or ``$z_2$'' or both in the {\it second} line of Eq.~(\ref{808}) are replaced by ``$z$''.
It is easy to calculate
\begin{eqnarray}
 \Delta {\cal A}(k_i^z,k_f^z,z_1,z_2) &=&  \int_{z}^{z+\Delta(\tilde \gamma_{\tilde \chi^c})}  d z_1 dz_2  \sin((k_i^z-k_f^z)(z_1-z_2)) \nonumber \\
&& \{k_i^zk_f^z[u_1(z_1) u_1(z_2)\sin(\theta_1(z_1)-\theta_1(z_2))
\nonumber
\\ &&+ u_2(z_1) u_2(z_2)\sin(-\theta_2(z_1)+\theta_2(z_2)) ]+\nonumber
\\&&
M_2\mu(z)[u_1(z_1) u_2(z_2)\sin(\theta_1(z_1)+\theta_2(z_2)+\alpha(z))
\nonumber
\\ &&+ u_2(z_1) u_1(z_2)\sin(-\theta_2(z_1)-\theta_1(z_2)-\alpha(z)) ] \},  \label{809}
\end{eqnarray}
which implies 
\begin{eqnarray}
\Delta {\cal A}(k_i^z,k_f^z,z_1 \to z,z_2\to z) &\equiv& 0 .
\end{eqnarray}
Finally, embedding Eq.~(\ref{809}) into Eq.~(\ref{807}), we obtain Eq.~(\ref{R2}) according to the relation
\begin{eqnarray}
\Delta R_{\tilde W^c\rightarrow\tilde H^c}(k_i,k_f,z) =  \frac{1}{4E_iE_f} \frac{1}{v_i^z v_f^z} \Delta|M(k_i^z, k_f^z, z)|^2.    
\label{810}
\end{eqnarray}

\section{More Results at $T=0$ K in Model I}

\setcounter{equation}{0}

\begin{figure}
\begin{center}
\includegraphics[height=3.2 in]{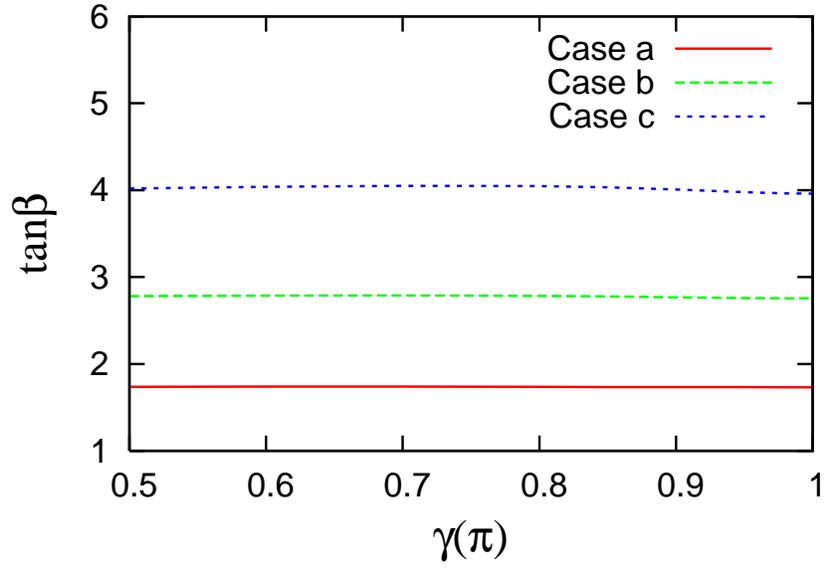}
\end{center}
\caption{$\tan\beta$ vs. $\gamma (\pi)$.}
\label{Afig1}
\end{figure}

\begin{figure}
\begin{center}
\includegraphics[height=3.2 in]{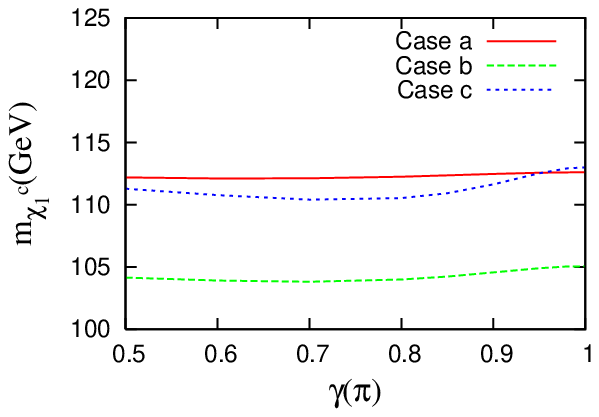}
\end{center}
\caption{$m_{\chi_1^c}$ vs. $\gamma (\pi)$.}
\label{Afig2}
\end{figure}

\begin{figure}
\begin{center}
\includegraphics[height=3.2 in]{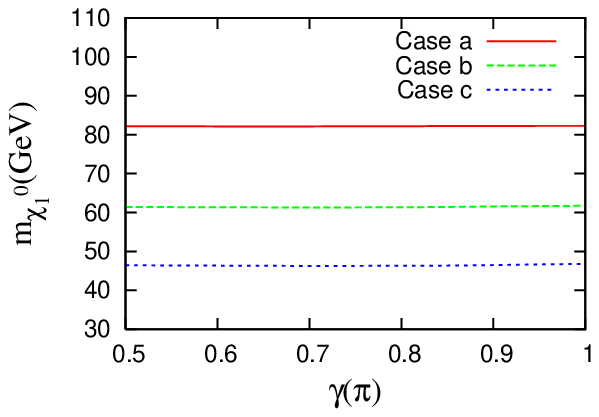}
\end{center}
\caption{$m_{\chi_1^0}$ vs. $\gamma (\pi)$.}
\label{Afig3}
\end{figure}

\begin{figure}
\begin{center}
\includegraphics[height=3.2 in]{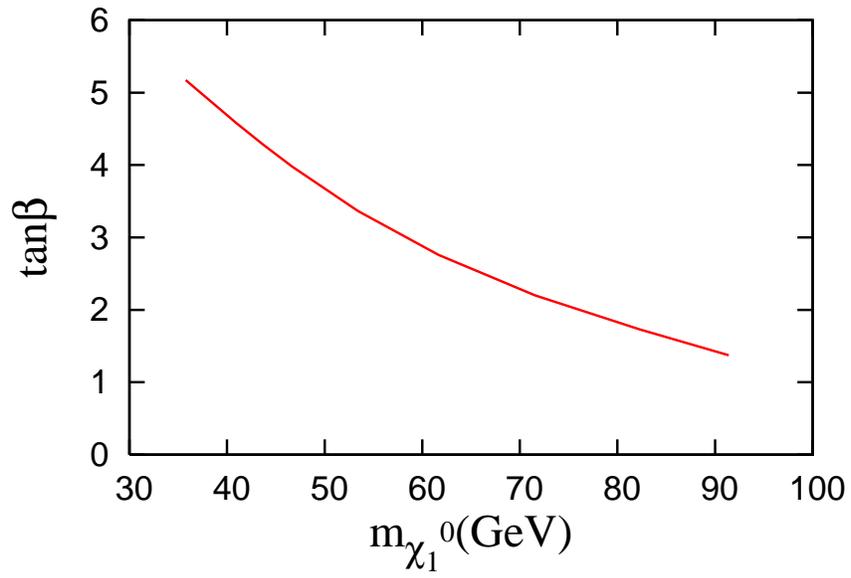}
\end{center}
\caption{$\tan\beta$ vs. $m_{\chi_1^0}$.}
\label{Afig4}
\end{figure}

\begin{figure}
\begin{center}
\includegraphics[height=3.2 in]{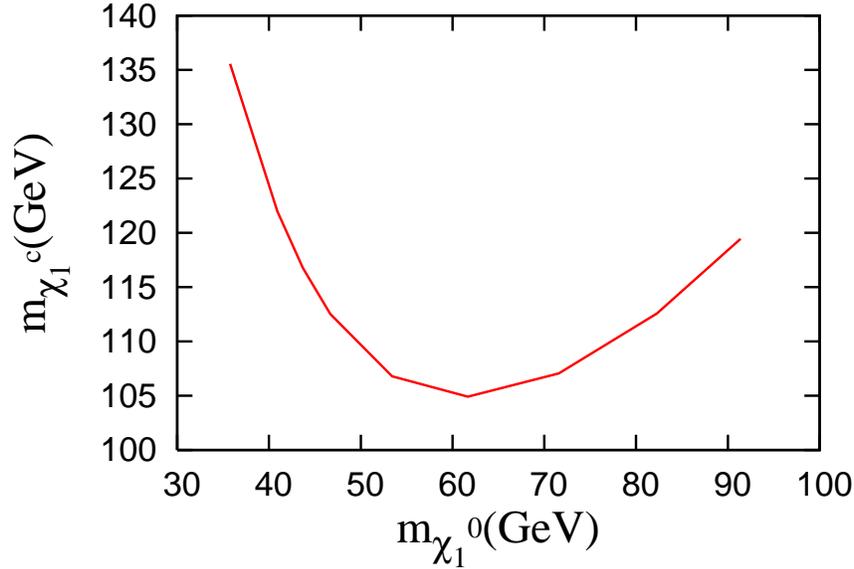}
\end{center}
\caption{$m_{\chi_1^c}$ vs. $m_{\chi_1^0}$.}
\label{Afig5}
\end{figure}

Additional $T=0 K$ results in the sMSSM with both ECPV and SCPV turned on 
are presented in Figure~\ref{Afig1} - Figure~\ref{Afig5}. These results include the $\gamma$ dependences of $\tan\beta$, 
$m_{\chi_1^c}$ and $m_{\chi_1^0}$ in cases a, b and c,
and the correlations between $\tan\beta$, $m_{\chi_1^c}$ and $m_{\chi_1^0}$ due to the variance of 
$m_{H_d^0}^2$. For discussions on the phenomenology in a similar framework, also see~\cite{Chiang:2008ud}.

\end{document}